\documentclass[a4paper]{article}

\usepackage[utf8]{inputenc}

\usepackage{lmodern}

\usepackage{amsmath}        
\usepackage{amsfonts}       
\usepackage{amssymb}        
\usepackage{amsthm}         
\usepackage{bbding}         
\usepackage{bbold}          
\usepackage{bm}             
\usepackage{graphicx}       
\usepackage{fancyvrb}       
\usepackage{xparse}         

\usepackage[style=numeric, maxnames=6, giveninits=true, backend=biber]{biblatex}
\usepackage[nottoc]{tocbibind} 
\usepackage{dcolumn}        
\usepackage{booktabs}       
\usepackage{paralist}       
\usepackage{xcolor}         
\usepackage{mathtools}      
\usepackage{textgreek}      
\usepackage{algorithm}
\usepackage{algorithmic}    
\usepackage{setspace}       
\usepackage{caption}        
\usepackage{subcaption}     
\usepackage{afterpage}

\usepackage[breaklinks=true,letterpaper=true,colorlinks,bookmarks=false]{hyperref}
\usepackage{cleveref}

\theoremstyle{plain}
\newtheorem{theorem}{Theorem}[section]
\newtheorem{lemma}[theorem]{Lemma}
\newtheorem*{lemma*}{Lemma}

\newtheorem*{theorem*}{Theorem}
\newtheorem{corollary}[theorem]{Corollary}
\newtheorem*{corollary*}{Corollary}
\newtheorem{proposition}[theorem]{Proposition}

\theoremstyle{plain}
\newtheorem{definition}[theorem]{Definition}
\newtheorem*{definition*}{Definition}

\theoremstyle{remark}
\newtheorem{remark}[theorem]{Remark}

\theoremstyle{definition}
\newtheorem{example}[theorem]{Example}
\newtheorem*{example*}{Example}

\newtheorem{problem}{Problem}
\newtheorem{assumption}[theorem]{Assumption}

\DefineVerbatimEnvironment{code}{Verbatim}{fontsize=\small, frame=single}

\newcommand{\R}{\mathbb{R}}
\newcommand{\N}{\mathbb{N}}
\newcommand{\E}{\mathbb{E}}

\renewcommand{\P}{\mathbb{P}}

\newcommand{\inv}[1]{#1^{-1}}
\newcommand{\norm}[2]{\|#1\|_{#2}}

\newcommand{\post}{\mu}
\newcommand{\prior}{\mu_0}
\newcommand{\fwop}{G}
\newcommand{\rv}[1]{#1}

\newcommand{\paramsp}{\mathcal{U}} 

\newcommand{\observsp}{\mathcal{Y}}
\newcommand{\onoise}{\rv{E}^o}

\newcommand{\noisesp}{\mathcal{E}}
\newcommand{\spextended}{\paramsp\times\mathcal{E}}

\newcommand{\kernel}{\kappa}
\newcommand{\kernoise}[1][]{\kernel_{\nsamp_{#1}}}

\newcommand{\extend}[1]{\widetilde{#1}}
\newcommand{\postextsl}[1]{\extend{\post}^\nsamp_{#1}}
\newcommand{\postextml}[1]{\extend{\post}^{\nsamp_{#1}}_{#1}}
\newcommand{\fwopext}{\extend{\fwop}}
\newcommand{\bsa}[1]{\mathcal{B}(#1)}
\newcommand{\kernoisearg}{(u,\cdot)}
\newcommand{\kernoiseargint}{(u,\td e)}
\newcommand{\extendarg}{(u,e)}
\newcommand{\nsamp}{S}
\newcommand{\npart}{N}
\newcommand{\margmeas}[1]{\post^m_{#1}}
\newcommand{\measxker}[2]{{#1}\otimes{#2}}
\newcommand{\priorxker}[1]{\measxker{\prior}{\kernel_{#1}}}
\newcommand{\td}{\text{d}}
\newcommand{\tp}{\beta}
\newcommand{\rnder}[2]{\frac{\td #1}{\td #2}}

\newcommand{\msp}{\mathcal{X}}
\newcommand{\mker}{P}
\newcommand{\mkerc}{P_{c}}
\newcommand{\mkere}{P_{e}}
\newcommand{\minvar}{\pi}
\newcommand{\minvarc}{\pi_{c}}
\newcommand{\minvare}{\pi_{e}}
\newcommand{\minvarex}{\minvar_\npart}
\newcommand{\minvarexc}{\minvar_{c,\npart}}
\newcommand{\minvarexe}{\minvar_{e,\npart}}
\newcommand{\mspex}{\msp^\npart}
\newcommand{\mkerex}{\mker_\npart}
\newcommand{\mkerexc}{\mker_{c,\npart}}
\newcommand{\mkerexe}{\mker_{e,\npart}}
\newcommand{\was}[2]{\mathcal{W}(#1,#2)}
\newcommand{\pp}{p}
\newcommand{\qq}{q}
\newcommand{\rr}{r}
\DeclareMathOperator*{\argmin}{arg\,min}
\newcommand{\km}{\overline{M}}

\newcommand{\mkerpm}{P^{\mathrm{PM}}}
\newcommand{\mkermc}{P^{\mathrm{MC}}}
\newcommand{\priorl}{\mu_{0-}}
\newcommand{\priorh}{\mu_{0+}}
\newcommand{\potbd}{\overline{\Phi}}
\newcommand{\propl}{Q_-}
\newcommand{\proph}{Q_+}
\newcommand{\minc}{\varepsilon_\star}

\renewcommand{\post}{\mu^\star}

\renewcommand{\margmeas}[1]{\mu^m_{#1}}

\renewcommand{\postextsl}[1]{\widetilde{\mu}^{\nsamp}_{#1}}

\renewcommand{\postextml}[1]{\widetilde{\mu}^{\nsamp_{#1}}_{#1}}

\newcommand{\postextat}[2]{\widetilde{\mu}^{#2}_{#1}}

\RenewDocumentCommand{\mkerc}{o}{%
  P_{c\IfValueT{#1}{,#1}}%
}

\RenewDocumentCommand{\mkere}{o}{%
  P_{e\IfValueT{#1}{,#1}}%
}

\renewcommand{\mkerpm}{P^{\mathrm{PMMH}}}
\renewcommand{\mkermc}{P^{\mathrm{MCWM}}}

\RenewDocumentCommand{\minvarc}{o}{%
  \mu_{c\IfValueT{#1}{,#1}}%
}

\RenewDocumentCommand{\minvare}{o}{%
  \mu_{e\IfValueT{#1}{,#1}}%
}

\newcommand{\minvarmc}[2]{\mu^{\mathrm{MCWM}}_{#1,#2}}

\renewcommand{\mkerex}{P_{\npart}}
\renewcommand{\minvarex}{\pi_{\npart}}

\renewcommand{\mkerexc}{P_{c,\npart}}
\renewcommand{\mkerexe}{P_{e,\npart}}

\renewcommand{\minvarexc}{\mu_{c,\npart}}
\renewcommand{\minvarexe}{\mu_{e,\npart}}

\NewDocumentCommand{\nstepc}{o}{%
  M_{c\IfValueT{#1}{,#1}}%
}

\NewDocumentCommand{\nstepe}{o}{%
  M_{e\IfValueT{#1}{,#1}}%
}

\NewDocumentCommand{\kerforward}{o}{%
  \overrightarrow{\kernel}%
  \IfValueT{#1}{_{#1}}%
}

\newcommand{\kerbackstage}[1]{\overleftarrow{\kernel}_{#1}}
\newcommand{\kerbackopt}[1]{\overleftarrow{\kernel}_{#1,\mathrm{opt}}}

\newcommand{\paramleb}{\lambda_{\paramsp}}
\newcommand{\noiseleb}{\lambda_{\noisesp}}

\newcommand{\minormeas}[1]{\varphi_{#1}}

\newcommand{\doeblinconst}{\delta_{\mathrm D}}
\newcommand{\costratio}{\theta}

\renewcommand{\minc}{\doeblinconst}
\renewcommand{\pp}{\costratio}

\newcommand{\likexact}{L^\star}
\newcommand{\likmarg}[1]{L_{#1}}

\newcommand{\abs}[1]{\left|{#1}\right|}

\begin{document}

\title{Reducing Complexity of Pseudo-Marginal Sequential Monte Carlo via Hierarchical Tempering
\\[0.5cm] \large{}}


\author{Josef Mart\'inek\thanks{Institute for Mathematics and Interdisciplinary Center for Scientific Computing (IWR), Heidelberg University, 69120, Heidelberg, Germany (martinek@math.uni-heidelberg.de, r.scheichl@uni-heidelberg.de)}, Julian Hofstadler\thanks{Department of Mathematical Sciences, University of Bath, Bath, U. K.
(jh4272@bath.ac.uk, a.m.g.cox@bath.ac.uk)}, Alexander M. G. Cox\footnotemark[2], Robert Scheichl\footnotemark[1]\\
{\tt\small }
}

\maketitle

\setlength{\emergencystretch}{100pt}
\begin{abstract}
The Bayesian approach to inverse problems provides a powerful framework for inferring unknown parameters from noisy observations. Among the computationally most challenging statistical inference problems are problems with intractable likelihoods, especially if no unbiased likelihood estimator is available. We focus on a class of such problems, where a biased likelihood estimator is obtained through a stochastic approximation of a forward map (\emph{stochastic solver}). Obtaining an empirical posterior approximation in applications is very expensive due to the repeated stochastic solver evaluations required by standard sampling methods. To address this computational bottleneck, we develop a novel algorithm based on a density-tempered marginalized Sequential Monte Carlo (SMC) sampler. Our approach (\emph{hierarchical SMC}) accelerates inference by leveraging a hierarchy of stochastic solvers with increasing accuracy both along the tempering sequence and within mutation stages of the SMC sampler. We establish, under additional assumptions, that the asymptotic speedup is proportional to the number of tempering stages~$T$. We demonstrate computational gains on several examples, including a nonlinear, high-contrast inverse problem arising in photoacoustic medical imaging. On this problem, we demonstrate a speedup by a factor of \(14.3\), reducing the inference wall-clock time from about \(4.6\) days to \(7.7\) hours on an NVIDIA A40 GPU.
\end{abstract}



\section{Introduction}

A challenging task in Bayesian statistics is to efficiently generate samples from complicated distributions such as posteriors $\post$ in (possibly infinite dimensional) inverse problems. Among the most popular methods are Markov chain Monte Carlo (MCMC) and Sequential Monte Carlo (SMC) algorithms. Both approaches require the density of $\post$ only up to a constant; however, being able to evaluate the likelihood function occurring in $\post$ is crucial. Problems involving intractable likelihood functions are particularly challenging from a computational perspective. If an unbiased estimator of the likelihood is available, the pseudo-marginal framework, see \cite{beaumont2003estimation, andrieu2009pseudo}, provides a method to overcome this issue. However, for many Bayesian inverse problems (BIPs) it is not even possible to provide an unbiased estimator for the likelihood.

The starting point of a BIP is a \textit{forward map}, denoted by $\fwop$, which depends on some parameter $u$. The related (classical) inverse problem is then to find $u$ which satisfies 
\[
\fwop (u) = y
\]
for some observed (in general noisy) quantity $y$. Inverse problems tend to be ill-posed, which means that there is possibly no solution, solutions are not unique, or that solutions are highly unstable with respect to small noise in the data $y$. The Bayesian approach is to put a prior distribution on $u$ and to use $\fwop$ in the likelihood function to derive a posterior $\post$. Under mild conditions, this leads to a well-posed problem, see \cite{stuart2010inverse}.

Sampling from $\post$ in the context of BIPs is typically difficult. Many problems are related to situations where $\fwop$ involves the solution of a differential equation \cite[Section 3]{stuart2010inverse} and we study such an example later. More precisely, this means evaluating $u \mapsto \fwop(u)$ requires the solution of the underlying differential equation with parameter $u$. The corresponding likelihood function is then defined as 
\begin{equation}\label{equ:likelihood_definition} 
u \mapsto
\exp\Bigl(- \Phi\bigl(\fwop(u)\bigr)\Bigr),
\end{equation}
for a suitable mapping $\Phi$ implicitly depending on the measured data $y$. In general, $\fwop$ cannot be computed analytically and needs to be approximated. Motivated by applications in nuclear engineering \cite{lartaud2025uncertainty}, radiation therapy \cite{cox2024bayesian}, or medical imaging \cite{kangasniemi2024stochastic} we focus on situations where a \textit{stochastic solver} $\fwopext$ is used, which may be any random quantity approximating $\fwop$.
\begin{example}\label{ex:stoch_solvers}
    Particular choices for $\fwopext$ that fit into our setting are:
    \begin{itemize}
        \item Monte Carlo estimators (e.g., based on i.i.d. samples; randomized quasi-Monte Carlo; multilevel or multifidelity techniques) \cite{giles2015multilevel, lemieux2009monte},
        \item Gaussian process surrogates \cite{rasmussen2006gaussian, stuart2018posterior}, or
        \item  probabilistic numerical methods \cite{hennig2022probabilistic}, such as randomized differential equation solvers in \cite{conrad2017statistical}.
    \end{itemize}
\end{example}

Using $\fwopext$ then 
yields the \textit{approximate posterior distribution} $\widetilde{\post}$ with likelihood function 
\begin{equation} \label{equ:approximate_likelihood}
u \mapsto \exp\left( -\Phi(\fwopext(u))\right),
\end{equation}
replacing the likelihood function in \eqref{equ:likelihood_definition}.
Note that even if $\fwopext$ is unbiased, i.e., if $\E [\fwopext(u)] = G(u)$ for any $u$, this does not yield an unbiased estimator for the likelihood in general. Thus, in practice, it is typically only possible to target the approximate posterior instead of $\post$ exactly.   

In terms of computational efficiency, there is a clear tradeoff. Namely, most stochastic solvers $\fwopext$ are tuned via an accuracy parameter, which we generically denote by $S \in (0, \infty)$. We associate large values of $S$ with a high accuracy that renders evaluations of $\fwopext$ and hence \eqref{equ:approximate_likelihood} expensive. Thus, standard, as well as pseudo-marginal MCMC or SMC methods suffer from two competing drawbacks. If $S$ is chosen large, Markov transitions become expensive and too few Markov steps are possible for constrained budgets, while for small $S$ the true posterior $\post$ is not sufficiently well targeted.  
A way to overcome this issue is to use a hierarchy of solvers with increasing accuracies  $(S_t)_{1 \leq t \leq T}$. In the context of SMC and BIPs this has been exploited, for instance, in \cite{beskos2017multilevel,jasra2017multilevel,latz2018multilevel,rohrbach2022multilevel}; see also \cite{dodwell2019multilevel, lykkegaard2023multilevel} and the references therein for connections to MCMC. In particular, \cite{latz2018multilevel} introduces a Multilevel Sequential$^2$ Monte Carlo method, where tempering is combined with a hierarchy of discretization levels to numerically solve the underlying PDE in $\fwop$ via finite element methods. 

Inspired by \cite{latz2018multilevel}, we develop and analyze a hierarchical density-tempered pseudo-marginal SMC algorithm; in contrast to \cite{latz2018multilevel}, we consider random solvers. BIPs using random forward models, as considered in this work, were previously studied in \cite{lie2018random}, where the authors noted the connection to pseudo-marginal methods. The work \cite{garegnani2021sampling} extends \cite{lie2018random} and numerically compares the performance of some pseudo-marginal MCMC methods for a model problem. Similar problems have recently been studied in \cite{lovbak2025investigation}, where the authors investigate the implications of the random forward model noise scale on the well-posedness of the problem. The methods considered in those works are (pseudo-)marginal MCMC methods, while we propose to accelerate pseudo-marginal SMC via hierarchical tempering.  

The pseudo-marginal approach, cf.~\cite{beaumont2003estimation,andrieu2009pseudo,andrieu2015convergence}, and the SMC approach, cf.~\cite{del2006sequential,chopin2020introduction}, are both well-established techniques in the context of challenging sampling problems. In \cite{duan2015density}, a density-tempered marginalized $\text{SMC}^2$ sampler for Bayesian inference in state-space models is developed. Practical aspects of making the $\text{SMC}^2$ algorithm adaptive have been studied in \cite{botha2023automatically}. Our analysis builds on the $\text{SMC}^2$ theory, but we do not require an unbiased estimator for the likelihood.
In the past few years, there has been growing interest in developing finite-sample complexity results for SMC samplers, such as in \cite{marion2023finite,fay2026complexity}.
 By allocating longer MCMC chains to the last mutation stage, the greedy waste-free algorithm of \cite{fay2026complexity} achieves  similar complexity reductions as we do via a different technique. Building on \cite{marion2023finite}, we theoretically analyze our hierarchical tempering algorithm and develop finite-sample guarantees. Additionally, using the coupling constructions in \cite{eberle2016reflection}, we also optimize the mutation stage of the SMC algorithm.

The intuition behind the hierarchical tempering approach used within our pseudo-marginal SMC sampler can be summarized as follows: Let $T \in \N$ be the number of tempering stages. The main idea is to employ a sequence of increasing solver accuracies $(S_t)_{1 \leq t \leq T}$ to significantly reduce the complexity of the SMC sampler when compared to a version that uses the highest accuracy $S_T$ throughout all tempering steps. We refer to the two approaches as the \textit{hierarchical} and the \textit{standard} algorithm, respectively. 
Cheap forward solvers reduce the complexity in the initial phase of the algorithm; the tempering mechanism, which simplifies the target, mitigates the potentially bad effect of using inexact stochastic solvers in the likelihood during that phase. At later stages, we work with a high precision to ensure sufficient quality of Markov transitions. Indeed, under suitable assumptions, the hierarchical algorithm with $S_t = \gamma^{t-T}S_T$ for $\gamma>1$ achieves the same (finite-sample) probabilistic error bound as the standard algorithm with $S_t = S_T$ for all $1\leq t \leq T$; see Theorem \ref{cor:st_vs_ad_complex}. 

Furthermore, we analyze the cost associated with both algorithms to achieve a given error threshold. 
The most expensive part of SMC samplers in applications is the evaluation of the likelihood \eqref{equ:approximate_likelihood}, which requires the stochastic solver. Assume for the moment that each SMC stage requires the same number of likelihood evaluations (typically proportional to the number of particles) and a likelihood evaluation in stage $t$ has a cost of $\mathcal{O}(S_t)$, which is for instance the case for Monte Carlo based stochastic solvers. Hence, setting again $S_t = \gamma^{t-T}S_T$ in the hierarchical SMC sampler, we expect 
\begin{equation}\label{equ:cost_ratio}
\frac{\text{cost standard algorithm}}{\text{cost hierarchical algorithm}} = \mathcal{O}(T).
\end{equation}
In Theorem \ref{cor:st_vs_ad_complex}, we prove that this is indeed the case, while achieving the same error bounds for both samplers. This aligns with numerical experiments where a significant speedup is achieved, while (numerically assessed) errors behave the same for both algorithms, compare to Section~\ref{sec:numerics}. 

\medskip
\noindent
 We summarize the main contributions of this work as follows:
\begin{enumerate}
    \item We provide a general framework for density-tempered pseudo-marginal SMC algorithms employing a hierarchy of increasingly accurate stochastic solvers -- see Section~\ref{sec:algorithm}. 
    \item We prove finite-sample error bounds for hierarchical pseudo-marginal SMC samplers -- cf. Theorem~\ref{th:error_est_ad}: for a suitably chosen (increasing) sequence of accuracies $(S_t)_{1 \leq t \leq T}$ the hierarchical algorithm achieves the same theoretical guarantees as the standard algorithm -- see Theorem~\ref{cor:st_vs_ad_complex}. 
    \item   We prove \eqref{equ:cost_ratio}. That is, in order to achieve the same error bound in the standard and hierarchical algorithm, the ratio of respective costs grows linearly in $T$ -- see Theorem \ref{cor:st_vs_ad_complex}.
    \item   We study how a hierarchy of Markov kernels improves mutation stages within SMC algorithms. We derive bounds for the mixing of the hierarchical mutation kernel which compute an optimal budget allocation strategy for hierarchical mutation stages -- see~Theorem~\ref{th:budget_opt}.   
    \item We apply our algorithm to a concrete Bayesian inverse problem arising in photoacoustic medical imaging. In that problem, our algorithm reduces the computational cost roughly by a factor of $14.3$ from $4.6$ days to $7.7$ hours, while (numerically computed) errors remain the same -- see Section~\ref{sec:numerics}. 
\end{enumerate}

The remainder of this paper is organized as follows. Section \ref{sec:bip} formalizes the Bayesian approach to inverse problems with stochastic forward solvers and establishes the framework for density-tempered pseudo-marginal SMC. Section \ref{sec:algorithm} details the hierarchical SMC algorithm. Section \ref{sec:acc_is} derives the core complexity bounds for the hierarchical acceleration along the tempering sequence. Section \ref{sec:mutation} provides convergence analysis and optimal budget allocation for hierarchical acceleration of the mutation. Finally, Section \ref{sec:numerics} empirically validates the theoretical cost gains through numerical experiments on a toy problem and a practical statistical inference problem. Auxiliary results and technical proofs are provided in the appendix.

\section{Bayesian inverse problems with stochastic forward solvers}\label{sec:bip}

We begin by setting up the Bayesian inverse problem (BIP) framework considered in this paper. Whenever a measure admits a density with respect to a reference measure specified in context, we use the same symbol for the measure and its density. The Bayesian approach to inverse problems assumes a prior distribution on an unknown parameter and uses measured data to update this prior knowledge. Bayes' Theorem then yields the posterior distribution, a conditional distribution of the parameter given the observed data. Let the parameter space $\paramsp$ be a Polish space equipped with its Borel $\sigma$-algebra $\bsa{\paramsp}$, and let $\rv{U}$ be a $\paramsp$-valued random variable, distributed according to the prior distribution $\prior$. Let $\observsp\subseteq\R^m$ be the observation space and let $\fwop:\paramsp\rightarrow\observsp$ be measurable. Then, the BIP 
considered in this paper is given as follows.

\begin{problem}\label{pr:BIP}
    Let $\rv{Y}=\fwop(\rv{U})+\onoise$, where the additive \emph{observation noise} $\onoise$ is independent of $U$ and satisfies $\onoise\sim\mathcal{N}(0,\Gamma)$ with $\Gamma\in\R^{m\times m}$ being a symmetric positive definite noise covariance matrix. Let $y\in\observsp$ be a realization of $\rv{Y}$; $y$ will be referred to as the \emph{data}. Find the \emph{posterior distribution} $\post$, i.e., the conditional distribution of $\rv{U}$ given the measured data $\rv{Y}=y$.
\end{problem}
Note that in Problem~\ref{pr:BIP} the distribution of $\onoise$ is known, but the realization is not. We interpret $\prior$ as prior knowledge about $U$ before observing the data $y\in\observsp$ and using it to update our belief. 

We define the \emph{potential} (or negative log-likelihood) $\Phi(u)=\frac{1}{2}\norm{y-\fwop(u)}{\inv{\Gamma}}^2=\frac{1}{2}(y-\fwop(u))^{\mathsf T}\inv{\Gamma}(y-\fwop(u))$. Since the realization  $y$ of $Y$ is fixed in Problem~\ref{pr:BIP}, the dependence of $\Phi$ on $y$ is suppressed. Since $\Phi$ is measurable and non-negative, $0 < \int_{\paramsp} \exp\bigl(-\Phi(u)\bigr)\,\prior(\td u) \leq 1$.
Hence, the posterior $\post$ is well-defined and its density is given by Bayes' formula as
\begin{equation}\label{eq:ip_bayes}
    \frac{\td\post}{\td\prior}(u) \propto \exp{\big(-\Phi(u)\big)}.
\end{equation}

The density $\td\post/\td\prior$ in~\eqref{eq:ip_bayes} contains a multiplicative constant that depends on $y$ but not on $u$. Our ultimate goal is to develop and analyze efficient algorithms to (approximately) sample from the posterior $\post$. Evaluating the likelihood $\exp{\big(-\Phi(u)\big)}$ analytically is often intractable and one relies on approximations. In this work, we focus on the situation where a random approximation of $\fwop(u)$ is employed.

Before we proceed, we introduce the following notation. If $(\mathcal{Z},\bsa{\mathcal{Z}})$ is a measurable space, then any Markov kernel $\kernel:\paramsp\times\bsa{\mathcal{Z}}\rightarrow[0,1]$ extends the measure $\prior$ from $\paramsp$ to $\paramsp\times\mathcal{Z}$ in the following standard way. The extended measure, denoted by $\prior\otimes\kernel$, is defined by
\begin{equation}\label{eq:extendmeas}
    (\prior\otimes\kernel)(A\times B)=\int_A\kernel(u,B)\prior(\td u),
\end{equation}
for $A\in\bsa{\paramsp}$ and $B\in\bsa{\mathcal{Z}}$.

\subsection{Approximate posteriors using randomized likelihoods}\label{sec:approx_post_def}

Various approaches for constructing approximations $\fwopext$ of $\fwop$ exist in practice (cf. Example \ref{ex:stoch_solvers}). In order to carry out our analysis, it is important to quantify the stochasticity introduced into the likelihood by using a random estimator $\fwopext$. We address this by incorporating an abstract noise model that allows us to measure the uncertainty 
in a general way and can be used in the theoretical analysis later. The noise model is not required explicitly in practice.

\begin{definition}[Stochastic solver and noise model]\label{def:ssolver}
Let $\noisesp$ denote the \emph{noise space}, a Polish space equipped with Borel $\sigma$-algebra $\bsa{\noisesp}$. Any jointly measurable map $\fwopext:\paramsp\times\noisesp\to\observsp$ that approximates the forward map $\fwop:\paramsp\rightarrow\observsp$ in some suitable, randomized way is called a \emph{stochastic solver}. Associated with this solver is a \emph{noise model} in the form of a Markov kernel
$\kernoise:\paramsp\times\bsa{\noisesp}\to[0,1]$, parametrized by the \emph{solver accuracy} $\nsamp>0$. 
\end{definition}

In Definition \ref{def:ssolver} we did not define any notion of convergence of the stochastic solver $\fwopext$ to the true forward map $\fwop$. Precise convergence assumptions are stated below, when needed for the analysis. The parameter $\nsamp>0$ represents the solver accuracy and computational cost, e.g., the number of samples in a Monte Carlo estimator, with larger $\nsamp$ typically corresponding to a more accurate and more expensive solver. The space $\noisesp$ represents the latent randomness introduced by the stochastic solver. For a fixed $u \in \paramsp$, the Markov kernel $\kernoise\kernoisearg$ governs the conditional distribution of the noise $e\in\noisesp$. Consequently, evaluating $\fwopext\extendarg$ yields a realization of the stochastic solver given a realization $e$ of the noise. Crucially, in the methods developed in this paper, $\fwopext$ never needs to be evaluated  at the same noise $e$ with two different $u_1$ and $u_2$; this might be intractable in practice.

\begin{example}[Additive noise model]\label{ex:estimator_for_G}
    Consider a random estimator $\rv{X}_u$ for $\fwop(u)$, $u\in\paramsp$, and let $E_u=X_u-\fwop(u)$. Then we write $\fwopext(u,E_u)=\fwop(u)+E_u$, where $E_u$ is distributed according to the probability measure $\kernoise(u,\cdot)$.
Crucially, only access to $X_u$ is needed in practice. While $\kernoise$ and $E_u$ are used in the theoretical analysis, they are not required for implementation of the developed methods. The law of $\fwopext(u,E_u)$ depends on $\nsamp$, but $\fwopext$ itself does not depend on $\nsamp$.
\end{example}

 The applications in Section \ref{sec:numerics} are all of additive type, as in Example \ref{ex:estimator_for_G}, and in addition the stochastic solver is unbiased, i.e., $\E[X_u]=\fwop(u)$ and thus $\E[E_u]=0$; see Section \ref{sec:numerics}. Many estimators, such as those in Example \ref{ex:stoch_solvers}, introduce additional bias such that $\E[E_u]\neq 0$. This bias might arise for example from discretization of some underlying geometry using a mesh with parameter $h>0$. The parameter $h$ then determines the bias of $X_u$ with respect to $\fwop(u)$. To analyze this bias, one can define a (deterministic) approximation $\fwop_h(u)$ of $\fwop(u)$ by $\fwop_h(u)=\E[X_u]$. The bias $\fwop_h(u)-\fwop(u)$ is then analyzed separately.

We now use the stochastic solver $\fwopext$ within the Bayesian framework introduced in Problem \ref{pr:BIP}. We denote the stochastic potential by $\extend{\Phi}\extendarg=\frac{1}{2}\norm{y-\extend{\fwop}\extendarg}{\inv{\Gamma}}^2$ and define the \emph{marginal approximation} $\margmeas{\nsamp}$ of the posterior $\post$ by its Radon--Nikod\'ym (RN) derivative
\begin{equation}\label{eq:margmeas_def}
    \frac{\td\margmeas{\nsamp}}{\td\prior}(u)\propto\int_\noisesp\exp\bigl(-\extend{\Phi}\extendarg\bigr)\kernoise\kernoiseargint.
\end{equation}
Clearly, $\margmeas{\nsamp}$ is well defined as a (deterministic) measure on $\paramsp$. Furthermore, $\margmeas{\nsamp}$ is a direct generalization of the marginal approximation $\mu^M_N$ from \cite{lie2018random}, since the distribution of $E_u\sim\kernoise(u,\cdot)$ depends on $u$. Therefore, we use the same terminology. Note that the RN derivative in \eqref{eq:margmeas_def} is given up to a (unique) normalizing constant that ensures $\margmeas{\nsamp}(\paramsp)=1$. The interpretation for $\margmeas{\nsamp}$ is as follows: Due to the integration w.r.t. $\kernoise$ we marginalize out the stochasticity of the forward solver, that is, for fixed $u\in\paramsp$ and $E_u\sim\kernoise(u,\cdot)$, the right-hand side in \eqref{eq:margmeas_def} is the expected value of $\exp\bigl(-\extend{\Phi}(u,E_u)\bigr)$.

Direct sampling from $\margmeas{\nsamp}$ is usually difficult or infeasible. However, the so-called pseudo-marginal framework can be used to overcome this issue; see \cite{andrieu2009pseudo,andrieu2015convergence}. Sampling using pseudo-marginal methods can be performed without the necessity to explicitly construct the noise model $\kernoise$, which is a crucial property, since $\kernoise$ is typically intractable in practice. Note that in general we have
\begin{equation*}
    \int_\noisesp\exp\bigl(-\extend{\Phi}\extendarg\bigr)\kernoise\kernoiseargint\neq\exp\Bigl(-\frac{1}{2}\norm{y-\fwop(u)}{\inv{\Gamma}}^2\Bigr).
\end{equation*}
This means that while sampling from $\margmeas{\nsamp}$ is possible using a pseudo-marginal method, the samples are not distributed according to the true posterior $\post$ and some error is introduced. In the case $\kernoise\kernoisearg=\kernoise(\cdot)$ this error has been analyzed in \cite{lie2018random}; see \cite[Theorem 3.1]{lie2018random} for the most general result. In practice, as the accuracy parameter $\nsamp$ increases, the marginal posterior $\margmeas{\nsamp}$ typically gets closer to the true posterior $\post$, but at an increasing cost to sample from $\margmeas{\nsamp}$.

Our main contribution is to propose and analyze a pseudo-marginal SMC algorithm that exploits a hierarchy of stochastic solvers with increasing accuracy. In particular, we propose a hierarchical tempering schedule for pseudo-marginal SMC that evolves through targets associated with different solver accuracies and terminates at the marginal posterior $\margmeas{\nsamp}$, with the final accuracy $\nsamp$ chosen such that the resulting empirical approximation is sufficiently accurate relative to the exact posterior $\post$. 

Pseudo-marginal SMC methods have been used in the context of sequential inference in state space models before; see \cite[Chapter 18]{chopin2020introduction} for an overview. In this setting, a particle filter can be used to obtain an unbiased likelihood estimator. A key difference in our case is that the stochastic solver yields a biased likelihood estimator, leading to the target marginal measure $\margmeas{\nsamp}$ being dependent on the solver accuracy $\nsamp$ of the likelihood estimator.

\subsection{Density-tempering for pseudo-marginal SMC}\label{sec:smc_fix_nsamp}
To formulate SMC in the pseudo-marginal setting, we require a sequence of target distributions defined on the augmented space $\spextended$ with the standard product $\sigma$-algebra. Our approach relies on tempering as a tool to bridge prior and posterior. Tempering has proven to be an efficient technique for transitioning from simple to complex distributions, see \cite[Chapter 17]{chopin2020introduction}, but it can become prohibitively expensive when both $S$ and $T$ are large.

We now describe the construction of the sequence of measures for density-tempered pseudo-marginal SMC. The construction is analogous to standard tempering; see \cite[Section 3.1]{latz2018multilevel}. Let $\nsamp>0$, $T\in\N$ and $0=\tp_0<\ldots<\tp_T=1$. For $t=0,\ldots,T$, we define $\postextsl{t}$ by its RN derivative
\begin{equation}\label{eq:inter_meas_sl}
    \frac{\td\postextsl{t}}{\td(\priorxker{\nsamp})}(u,e)\propto\exp\bigl(-\tp_t\extend{\Phi}\extendarg\bigr).
\end{equation}
For future use, we also define the marginal counterpart $\margmeas{t,\nsamp}$ of $\postextsl{t}$ by
\begin{equation}\label{eq:marg_inter_meas}
    \frac{\td\margmeas{t,\nsamp}}{\td\prior}(u)\propto\int_\noisesp\exp\bigl(-\tp_t\extend{\Phi}\extendarg\bigr)\kernoise\kernoiseargint.
\end{equation}
As a consequence of properties of RN derivatives, we obtain the intermediate tempering densities
\begin{equation}\label{eq:inter_is_density}
    \frac{\td\postextsl{t}}{\td\postextsl{t-1}}(u,e)\propto\exp\bigl((\tp_{t-1}-\tp_t)\extend{\Phi}\extendarg\bigr).
\end{equation}

From \eqref{eq:inter_meas_sl} it immediately follows that $\postextsl{t}(\cdot\times\noisesp)=\margmeas{t,\nsamp}(\cdot)$. This means that we can use the tempered sequence $(\postextsl{t})_{t=0}^T$ to sample from $\margmeas{T,\nsamp}=\margmeas{\nsamp}$. A tempering-based SMC sampler using a sequence of distributions on an extended space was developed in \cite{duan2015density} in the context of Bayesian inference in state-space models. In the next section, we extend this approach by leveraging a hierarchy of solvers with increasing accuracy for tempering, i.e., we additionally use $\nsamp=\nsamp_0,\ldots,\nsamp_T$, leading to $(\postextml{t})_{t=0}^T$.

\section{Hierarchical tempering}\label{sec:algorithm}

Let $T\in\N$, let $0=\tp_0<\ldots<\tp_T=1$ be a tempering schedule, and let $0<\nsamp_0\leq\ldots\leq\nsamp_T$ be the solver accuracies used along the tempering sequence. We allow the solver accuracy to depend on the tempering stage and consequently we target the extended measures $\postextml{t}$, $t=0,\ldots,T$, defined by
\begin{equation}\label{eq:inter_meas_ml}
    \frac{\td\postextml{t}}
    {\td(\priorxker{\nsamp_t})}(u,e)
    \propto
    \exp\bigl(-\tp_t\extend{\Phi}\extendarg\bigr).
\end{equation}
Their marginal distributions satisfy $\postextml{t}(\cdot\times\noisesp)=\margmeas{t,\nsamp_t}(\cdot)$. Thus, at stage $t$, the SMC sampler performs the change of measure from $\postextml{t-1}$ to $\postextml{t}$.
If $\nsamp_t=\nsamp_{t-1}$, this is the standard tempering update. If $\nsamp_t>\nsamp_{t-1}$, the update also changes the solver accuracy and is handled by one of the importance sampling (\textsf{IS}) methods specified in Algorithm~\ref{alg:gen_is}. If the noise kernels $\kernoise[t]$ admit known densities $k_{\nsamp_t}$ with respect to a common dominating measure $\nu$, we use the classical importance-sampling update, labeled \texttt{Classical} in Algorithm~\ref{alg:gen_is}(A). Otherwise, we use a method based on a generalized IS update, labeled \texttt{Generalized} in Algorithm~\ref{alg:gen_is}(B). More details on both IS methods are given in Section \ref{sec:acc_is}.

\begin{algorithm}[t]
\caption{Importance Sampling (\textsf{IS})}
\label{alg:gen_is}
\begin{algorithmic}
\STATE \textbf{Input:} Particle $(u^\prime,e^\prime)$, tempering
parameters $\tp_{t-1},\tp_t$, and solver accuracies
$\nsamp_{t-1},\nsamp_t$
\STATE \textbf{Output:} Particle
$(u,e)$ and unnormalized importance
weight $w$
\end{algorithmic}

\vspace{1ex}

\noindent
\begin{minipage}[t]{0.48\textwidth}
\textbf{(A) Classical}\vspace{0.5ex}
\begin{algorithmic}
\STATE
$(u,e)
\leftarrow
(u^\prime,e^\prime)$
\STATE
$\displaystyle
\begin{aligned}
w\leftarrow{}&
\exp\Big(
    (\tp_{t-1}-\tp_t)
    \extend{\Phi}(u^\prime,e^\prime)
\Big)
\frac{
    k_{\nsamp_t}(u^\prime,e^\prime)
}{
    k_{\nsamp_{t-1}}(u^\prime,e^\prime)
}
\end{aligned}$
\end{algorithmic}
\end{minipage}
\hfill
\begin{minipage}[t]{0.48\textwidth}
\textbf{(B) Generalized}\vspace{0.5ex}
\begin{algorithmic}
\STATE
$(u,e)
\leftarrow
(u^\prime,\extend{e})$ \ where \ $\extend{e}\sim\kernoise[t](u^\prime,\cdot)$\vspace{1ex}
\STATE
$w\leftarrow
\exp\Big(
    (\tp_{t-1}-\tp_t)
    \extend{\Phi}(u^\prime,e^\prime)
\Big)$
\end{algorithmic}
\end{minipage}
\end{algorithm}

After importance sampling and multinomial resampling (see \cite[Section 9.4]{chopin2020introduction}), we apply $\nstepc[t]$ transitions of a cheap kernel $\mkerc[t]$ based on an accuracy $0<\nsamp_{c,t}<\nsamp_t$, followed by $\nstepe[t]$ transitions of an expensive kernel $\mkere[t]$ based on the current accuracy $\nsamp_t$. Here and throughout, the subscripts $c$ and $e$ stand for cheap and expensive, respectively. The formulation of the resulting Algorithm~\ref{alg:pmsmc_specific} is inspired by \cite[Algorithm 17.3]{chopin2020introduction}. Note that we can obtain $u_{T}^n$ from the output particles $\extendarg^n_{T}$ by marginalizing out $e$. The \emph{standard SMC} sampler is recovered by taking
\[
    \nsamp_0=\cdots=\nsamp_T
    \qquad\text{and}\qquad
    \nstepc[t]=0,
    \quad t=1,\ldots,T.
\]
In Section~\ref{sec:acc_is}, we analyze the finite-sample complexity of \emph{hierarchical tempering} under the simplifying assumptions that $\nstepc[t]=0$ and that the solver accuracies in Algorithm \ref{alg:pmsmc_specific} form a geometric hierarchy,
$\nsamp_t=\gamma \nsamp_{t-1}$ with $\gamma >1$. The analysis additionally requires technical assumptions, including conditions on the densities of the intermediate measures $\postextml{t}$ and on the spectral gaps of the mutation kernels; these are stated precisely, together with a suitable error criterion and cost specification,  in Section~\ref{sec:acc_is}.
The main complexity result is the following.
\begin{theorem}[Asymptotic complexity gain]
\label{cor:T_cost_gain}
Suppose that the assumptions of Theorem~\ref{cor:st_vs_ad_complex}, and in particular condition~\eqref{eq:spec_gap_lim}, are satisfied for some fixed number of tempering stages $T\in\N$ and $\gamma>1$. Then
\[
\lim_{\varepsilon\searrow0}
\frac{\mathcal{C}_{\mathrm{S}}(\varepsilon)}{\mathcal{C}_{\mathrm{H}}(\varepsilon)}
\geq
(1-\gamma^{-1})T,
\]
where $\mathcal{C}_{\mathrm{S}}(\varepsilon)$ and $\mathcal{C}_{\mathrm{H}}(\varepsilon)$ denote the total computational costs of the standard and hierarchical SMC algorithms, respectively, for error tolerance $\varepsilon>0$.
\end{theorem}
The cost reduction stems from the geometric solver hierarchy: compared with standard SMC, the hierarchical algorithm uses cheaper solver accuracies at the earlier tempering stages. A more careful statement and a proof of Theorem~\ref{cor:T_cost_gain} are given in Theorem~\ref{cor:st_vs_ad_complex} in Section~\ref{sec:acc_is}.

\begin{algorithm}[t]
\caption{Hierarchical density-tempered marginalized SMC sampler}
\label{alg:pmsmc_specific}
\begin{algorithmic}[1]
\STATE \textbf{Input:} Number of particles $\npart$; tempering parameters
$0=\tp_0<\ldots<\tp_T=1$; 
\STATE \phantom{\textbf{Input:}} number of kernel transitions $\nstepc[t]$ and $\nstepe[t]$ for the cheap and expensive kernels $\mkerc[t]$ and $\mkere[t]$;
\STATE \phantom{\textbf{Input:}} solver accuracies
$0<\nsamp_0\leq\ldots\leq\nsamp_T$; cheap solver accuracies
$0<\nsamp_{c,t}<\nsamp_t$ for the kernels $\mkerc[t]$;
\STATE \phantom{\textbf{Input:}} an \textsf{IS} method from Algorithm~\ref{alg:gen_is} for stages with $\nsamp_t>\nsamp_{t-1}$.\vspace{0.5em}
\STATE \textbf{Output:} Unweighted particles
$\{(u,e)^n_T\}_{n=1}^{\npart}$.
\vspace{0.5em}
\STATE \textbf{Operations involving index $n$ are performed for
$n=1,\ldots,\npart$:}
\STATE $(u,e)^n_0\sim\prior\otimes\kernoise[0]$
\FOR{$t=1$ to $T$}
    \IF{$\nsamp_t>\nsamp_{t-1}$}
        \STATE
        $\bigl((u,e)^n_t,w^n_t\bigr)
        \leftarrow
        \textsf{IS}\bigl(
            (u,e)^n_{t-1},
            \tp_{t-1},
            \tp_t,
            \nsamp_{t-1},
            \nsamp_t
        \bigr)$
    \ELSE
        \STATE
        $w^n_t
        \leftarrow
        \exp\Bigl(
            (\tp_{t-1}-\tp_t)
            \extend{\Phi}\bigl((u,e)^n_{t-1}\bigr)
        \Bigr)$
        \STATE
        $(u,e)^n_t\leftarrow(u,e)^n_{t-1}$
    \ENDIF
    \STATE
    $W^n_t
    \leftarrow
    w^n_t\big/\sum_{m=1}^{\npart}w^m_t$
    \STATE
    $\bigl((u^\prime,e^\prime)^n_t\bigr)_{n=1}^{\npart}
    =
    \text{resample}\Bigl(
        (W^n_t)_{n=1}^{\npart},
        \bigl((u,e)^n_t\bigr)_{n=1}^{\npart}
    \Bigr)$
    \STATE
    $(u^{\mathrm c},e^{\mathrm c})^n_t
    \sim
    \mkerc[t]^{\nstepc[t]}
    \bigl((u^\prime,e^\prime)^n_t,\cdot\bigr)$
    \STATE
    $(u,e)^n_t
    \sim
    \mkere[t]^{\nstepe[t]}
    \bigl((u^{\mathrm c},e^{\mathrm c})^n_t,\cdot\bigr)$
\ENDFOR
\end{algorithmic}
\end{algorithm}

In Section~\ref{sec:mutation}, we study the hierarchical acceleration of mutation. We consider one SMC stage $t$ and optimize the allocation of the computational budget between cheap and expensive kernel transitions, $\nstepc[t]$ and $\nstepe[t]$, respectively. Algorithm~\ref{alg:pmsmc_specific} permits both mechanisms to be used simultaneously, although they are analyzed separately in this work. We demonstrate numerically in Section \ref{sec:numerics} that hierarchical tempering with \texttt{Generalized} IS consistently reduces cost across the tested regimes, while the additional savings from two-stage mutation depend on the mutation budget and the allocation to the cheap chain.

\begin{remark}
One could also employ a hierarchy of solver accuracies within a single mutation stage, thereby obtaining a hierarchy of cheap kernels before applying the target-invariant kernel. Such a construction could yield further complexity reduction in addition to that obtained in Section \ref{sec:acc_is}. We do not pursue this extension and instead restrict the analysis to one cheap and one expensive kernel, which leads to a simple and practically interpretable budget-allocation problem.
\end{remark}

\section{Hierarchical acceleration along the tempering sequence}\label{sec:acc_is}

\subsection{Complexity analysis assuming known densities}\label{sec:compl_anal}
We start by analyzing the complexity of the hierarchical SMC (Algorithm \ref{alg:pmsmc_specific}) using classical importance sampling (Algorithm \ref{alg:gen_is}(A)). To this end, we assume that the Markov kernels $\kernel_{\nsamp_t}$, $t=0,\ldots,T$, in \eqref{eq:inter_meas_ml} have densities w.r.t.~a fixed measure $\nu$ on $(\noisesp,\bsa{\noisesp})$, that is, $\kernoise[t](u,A)=\int_Ak_{\nsamp_t}(u,e)\nu(\td e)$ for some density $k_{\nsamp_t}:\paramsp\times\noisesp\rightarrow[0,\infty)$ and all $A\in\bsa{\noisesp}$. In addition, to carry out the complexity analysis we assume that the densities $k_{\nsamp_t}$ are known and $k_{\nsamp_t}>0$ on $\spextended$.

Under these assumptions the intermediate measures can be obtained from \eqref{eq:inter_meas_ml} via
\begin{equation}\label{eq:inter_meas_ml_density}
    \frac{\td\postextml{t}}{\td(\prior\otimes\nu)}(u,e)\propto\exp\bigl(-\tp_t\extend{\Phi}\extendarg\bigr)k_{\nsamp_t}(u,e).
\end{equation}
As in \eqref{eq:inter_is_density}, the intermediate tempering densities can be constructed via
\begin{equation}\label{eq:inter_is_density_ml}
    \frac{\td\postextml{t}}{\td\postextml{t-1}}(u,e)\propto\exp\bigl((\tp_{t-1}-\tp_t)\extend{\Phi}\extendarg\bigr)\frac{k_{\nsamp_t}(u,e)}{k_{\nsamp_{t-1}}(u,e)}.
\end{equation}
This is the change of measure implemented in Algorithm \ref{alg:gen_is}(A).

We now show that an SMC algorithm adapting the solver accuracy yields significant cost savings while achieving the same error guarantees as the standard algorithm. Namely, we compare the following two versions of Algorithm \ref{alg:pmsmc_specific} for a given target accuracy $\nsamp>0$:
\begin{itemize}
    \item Algorithm \ref{alg:pmsmc_specific} with $\nsamp_0 = \dots = \nsamp_T =\nsamp$ (\emph{standard SMC})
    \item 
    Algorithm~\ref{alg:pmsmc_specific} with $0<\nsamp_0<\ldots<\nsamp_T=\nsamp$ and classical IS
    in Algorithm~\ref{alg:gen_is}(A) (\emph{hierarchical SMC})
\end{itemize}
Note that for the purpose of the analysis we assume that $\nsamp_t>0$ is real-valued. In practice, one can use the \texttt{ceil} function to obtain integer values of $\nsamp_t$ when required.

Following the complexity viewpoint of \cite[Section 5]{marion2023finite}, we measure the
computational cost of Algorithm~\ref{alg:pmsmc_specific} by summing the costs of all Markov kernel transitions executed during the
mutation stages. Throughout this subsection, we consider the
configuration
\[
    \nstepc[t]=0,
    \qquad
    \nstepe[t]=\nstepe,
    \qquad
    t\in\{1,\ldots,T\}.
\]
Thus, at stage $t$, Algorithm~\ref{alg:pmsmc_specific} applies
$\nstepe$ transitions of the kernel $\mkere[t]$, which
admits $\postextml{t}$ as its invariant measure. We assume that
there exists a threshold $\nsamp_{\mathrm E}>0$ such that, for
every $t\in\{1,\ldots,T\}$ and every $\nsamp_t\geq\nsamp_{\mathrm E}$,
the kernel $\mkere[t]$ is geometrically ergodic in the sense of
\cite[AS3.]{marion2023finite} and reversible with respect to
$\postextml{t}$. For $\nsamp_t\geq\nsamp_{\mathrm E}$, the
corresponding $L^2$ spectral gap of $\mkere[t]$ is denoted by
$\rho_t(\nsamp_t)\in(0,1]$, where we emphasize its dependence on
$\nsamp_t$. The computational cost of one transition of
$\mkere[t]$ is assumed to be $\nsamp_t$. This is motivated by one
transition of the standard pseudo-marginal random-walk
Metropolis--Hastings kernel, which requires one likelihood
evaluation; see \cite{medina2016stability}. One likelihood evaluation is assumed to cost $\nsamp_t$ using a Monte Carlo estimator with a sample size of $\nsamp_t$.

Let $\mathcal{F}_b$ denote the space of measurable functions $f: \paramsp \rightarrow \R$ satisfying $|f(u)| \leq 1$ for any $u\in\paramsp$. For $(u,e) \in \spextended$ and $\nsamp_{t-1}, \nsamp_{t} > 0$, we define the normalizing constants and weights as
\begin{align*}
    z_{t,\nsamp_t} &= \int_{\spextended} \exp\bigl(-\tp_t\extend{\Phi}\extendarg\bigr)k_{\nsamp_t}(u,e)(\prior \otimes \nu)\big(\td(u,e)\big),\\
    w_{t,\nsamp_{t-1}, \nsamp_{t}}(u,e) &= \exp\bigl((\tp_{t-1}-\tp_t)\extend{\Phi}\extendarg\bigr)\frac{k_{\nsamp_{t}}(u,e)}{k_{\nsamp_{t-1}}(u,e)}.
\end{align*}
Note that in the weight $w_{t,\nsamp_{t-1}, \nsamp_{t}}(u,e)$ the term $\frac{k_{\nsamp_{t}}(u,e)}{k_{\nsamp_{t-1}}(u,e)}$ cancels out when $\nsamp_{t-1} = \nsamp_{t}$.
\begin{assumption}
For $\gamma\geq 1$ fixed, we make the following assumptions on the sequence of intermediate measures. There exist $W_\gamma>0$ and $Z_\gamma>0$ such that for all $t\in\{1,\ldots, T\}$ we have
\begin{align}
    \sup_{\nsamp>0}\sup_{(u,e)\in\spextended} \Big( w_{t,\nsamp,\gamma\nsamp}(u,e) \Big)&\leq W_\gamma\,,\tag{A1}\label{eq:assump_W_ad}\\
    \sup_{\nsamp>0}\frac{z_{t-1,\nsamp}}{z_{t,\gamma\nsamp}}&\leq Z_\gamma\,. \tag{A2} \label{eq:assump_Z_ad}
\end{align}
Furthermore, there exist $C_{\mathrm B}>0$ and $\nsamp_{\mathrm B}>0$ such that for all $\nsamp>\nsamp_{\mathrm B}$ and $f\in\mathcal{F}_b$ we have
\begin{equation}
    \Big|\int_\paramsp f(u)\margmeas{\nsamp}(\td u)-\int_\paramsp f(u)\post(\td u)\Big|\leq C_{\mathrm B}\nsamp^{-1}\,. \tag{A3} \label{eq:assump_CB_ad}
\end{equation}
With $\nsamp_{\mathrm E}$ the ergodicity threshold defined above we set
$\displaystyle\varepsilon_{m,\gamma}=\min\left(\frac{1}{3W_\gamma Z_\gamma}, \frac{2C_{\mathrm B}}{\gamma^{T-1}\nsamp_{\mathrm E}}, \frac{2C_{\mathrm B}}{\nsamp_{\mathrm B}}\right)$.
\end{assumption}
Assumptions~(\ref{eq:assump_W_ad}--\ref{eq:assump_Z_ad}) control how much the intermediate distributions can change with increasing solver accuracy, while Assumption \eqref{eq:assump_CB_ad} controls the error of the marginal posterior approximation.

The following theorem provides a finite-sample error bound for both standard and hierarchical SMC, extending results from \cite{marion2023finite}. The result for the standard algorithm follows  by simply setting $\gamma=1$.

\begin{theorem}\label{th:error_est_ad}
    Let $T\in\N$, $0=\tp_0<\ldots<\tp_T=1$ and $\gamma \geq 1$, and suppose (\ref{eq:assump_W_ad}--
    \ref{eq:assump_CB_ad}) hold. For any $\varepsilon>0$ with $\varepsilon \le \varepsilon_{m,\gamma}$, let the target accuracy, the number of SMC particles and the number of kernel transitions be
    \begin{equation}
        \nsamp_T=2C_{\mathrm B} \varepsilon^{-1}\,, \quad
        \npart= \Big\lceil 2\log(128 T)\varepsilon^{-2} \Big\rceil \quad \text{and} \quad
        \nstepe[\gamma]=\bigg\lceil \frac{\log(16\npart T)}{\min_{t=1,\ldots,T}\rho_t(\nsamp^\gamma_t)} \bigg\rceil\,,\label{eq:def_param_ad}
    \end{equation}
    respectively. The sequence of intermediate accuracies is defined by $\nsamp^\gamma_t = \gamma^{t-T}\nsamp_T$ for $t \in \{0, \dots, T\}$. 
    Then, the SMC algorithm defined in Algorithm \ref{alg:pmsmc_specific} admits for any $f\in\mathcal{F}_b$ the error estimate
    \begin{equation}\label{eq:smc_error_ad}
        \P\bigg(\Big| \frac{1}{\npart}\sum_{n=1}^\npart f(u_{T}^n)-\int_\paramsp f(u)\post(\td u)\Big|\leq\varepsilon\bigg)\geq\frac{3}{4}.
    \end{equation}
\end{theorem}

\begin{remark}
    For simplicity, the error bound established in Theorem \ref{th:error_est_ad} guarantees an accuracy of $\varepsilon$ with probability of $3/4$. This can be extended to $1-\delta$ for $\delta \in (0, 1)$, either by modifying the constants in the finite-sample argument to depend on $\delta$, or by utilizing a median-of-means estimator over independent runs of the algorithm; see \cite{marion2023finite} and \cite{fay2026complexity}.
\end{remark}

\begin{proof}[Proof of Theorem \ref{th:error_est_ad}]
    Throughout the proof we write $\nstepe=\nstepe[\gamma]$. Let $\hat{\mu}_{\nsamp_T}(f) := \frac{1}{\npart}\sum_{n=1}^\npart f(u_{T}^n)$. By the triangle inequality:
    \begin{equation}\label{aux:err_dec1}
        \Big|\hat{\mu}_{\nsamp_T}(f) - \int_\paramsp f(u)\post(\td u)\Big| \leq \Big|\hat{\mu}_{\nsamp_T}(f) - \int_\paramsp f(u)\margmeas{\nsamp_T}(\td u)\Big| + \Big|\int_\paramsp f(u)\margmeas{\nsamp_T}(\td u) - \int_\paramsp f(u)\post(\td u)\Big|.
    \end{equation}
    
    Using \eqref{eq:assump_CB_ad} and \eqref{eq:def_param_ad} yields
    \begin{equation}\label{aux:det_err1}
        \Big|\int_\paramsp f(u)\margmeas{\nsamp_T}(\td u) - \int_\paramsp f(u)\post(\td u)\Big| \leq C_{\mathrm B} \nsamp_T^{-1} = \frac{\varepsilon}{2}.
    \end{equation}
    
    To bound the other term in \eqref{aux:err_dec1}, define the extension $\extend{f}: \mathcal{U} \times \mathcal{E} \to \mathbb{R}$ by $\extend{f}(u,e) = f(u)$. Because $f \in \mathcal{F}_b$, it follows that $|\extend{f}(u,e)| \le 1$ for any $(u,e) \in \mathcal{U} \times \mathcal{E}$. Furthermore, $\frac{1}{\npart}\sum_{n=1}^\npart \extend{f}(u, e)_{T}^n = \hat{\mu}_{\nsamp_T}(f)$ and
    $$\int_{\mathcal{U} \times \mathcal{E}} \extend{f}(u, e) \postextml{T}(\text{d} u, \text{d} e) = \int_{\mathcal{U}} f(u) \mu_{\nsamp_T}^m(\text{d} u).$$
    
    We want to apply \cite[Corollary 3.1]{marion2023finite} to the measure $\postextml{T}$ and the test function $\extend{f}$ with tolerance $\varepsilon/2$. Let us verify the assumptions. 
    By \eqref{eq:assump_W_ad}, since $\nsamp^\gamma_t = \gamma\nsamp^\gamma_{t-1}$ for all $t \in \{1, \dots, T\}$, the importance weights for the hierarchical algorithm are uniformly bounded by $W_\gamma$. In combination with \eqref{eq:assump_Z_ad}, we obtain that the assumption \cite[AS1]{marion2023finite} is satisfied. The particle count $\npart$ defined in \eqref{eq:def_param_ad} again satisfies \cite[Corollary 3.1, Assumption 1]{marion2023finite}. Analogously, the chain lengths $\nstepe$ defined in \eqref{eq:def_param_ad} satisfy \cite[Corollary 3.1, Assumption~2]{marion2023finite}. Thus, we obtain by \cite[Corollary 3.1]{marion2023finite} that
    \begin{equation*}
        \P\bigg(\Big|\hat{\mu}_{\nsamp_T}(f) - \int_\paramsp f(u)\margmeas{\nsamp_T}(\td u)\Big| \leq \frac{\varepsilon}{2}\bigg) \geq \frac{3}{4},
    \end{equation*}
    which in combination with \eqref{aux:det_err1} proves \eqref{eq:smc_error_ad}.
\end{proof}

The following theorem quantifies the cost gain of the hierarchical algorithm.

\begin{theorem}\label{cor:st_vs_ad_complex}
    Let $T\in\N$ and $0=\tp_0<\ldots<\tp_T=1$. Suppose (\ref{eq:assump_W_ad}--\ref{eq:assump_CB_ad}) hold for $\gamma=\gamma_{\mathrm{H}}>1$ and $\gamma=1$. Suppose further that there exists a $\rho^* >0$ independent of $T$ such that $\liminf_{\nsamp \to \infty} \rho_t(\nsamp) \geq \rho^* > 0$ for all $t \in \{1, \dots, T\}$.
    
    For any $\varepsilon>0$ with $\varepsilon \le \min(\varepsilon_{m,\gamma_{\mathrm{H}}}\,,\,\varepsilon_{m,1})$, let the target accuracy, the number of SMC particles and the number of kernel transitions for \emph{hierarchical SMC} ($\gamma=\gamma_{\mathrm{H}}$) and for \emph{standard SMC} ($\gamma=1$) be chosen as in \eqref{eq:def_param_ad} with intermediate accuracies defined by $\nsamp^{\gamma_{\mathrm{H}}}_t = \gamma_{\mathrm{H}}^{t-T}\nsamp_T$ and by $\nsamp^{1}_t = \nsamp_T$, respectively. 
Then, the hierarchical and standard SMC algorithm both admit for any $f\in\mathcal{F}_b$ the same error estimate in~\eqref{eq:smc_error_ad} and the asymptotic computational gain through hierarchical tempering satisfies
\begin{equation*}
        \liminf_{\varepsilon \searrow 0} \frac{\mathcal{C}_{\mathrm{S}}}{\mathcal{C}_{\mathrm{H}}} \geq T \rho^* \frac{1 - \gamma_{\mathrm{H}}^{-1}}{1 - \gamma_{\mathrm{H}}^{-T}},
    \end{equation*}
where $\mathcal{C}_{\mathrm{H}}$ and $\mathcal{C}_{\mathrm{S}}$ denote the computational costs of hierarchical SMC and standard SMC, respectively. 
If, in addition, for all $t \in \{1, \dots, T\}$ we have
    \begin{equation}\label{eq:spec_gap_lim}
        \lim_{\nsamp \to \infty} \rho_t(\nsamp) = \rho^*_t > 0,
    \end{equation}
     then 
    \begin{equation}\label{eq:comp_gain}
        \lim_{\varepsilon \searrow 0} \frac{\mathcal{C}_{\mathrm{S}}}{\mathcal{C}_{\mathrm{H}}} = T \frac{1 - \gamma_{\mathrm{H}}^{-1}}{1 - \gamma_{\mathrm{H}}^{-T}}.
    \end{equation}
\end{theorem}

\begin{proof}
    Throughout the proof we write $\nsamp_t=\nsamp^{\gamma_{\mathrm{H}}}_{t}$, and use $\min_t$ instead of $\min_{t \in \{1, \dots, T\}}$. Under the stated assumptions the error bounds follow as in Theorem \ref{th:error_est_ad} for $\gamma=\gamma_{\mathrm{H}}$ and $\gamma=1$.
    Using 
    \[
    \sum_{t=1}^T \gamma_{\mathrm{H}}^{t-T} = \frac{1 - \gamma_{\mathrm{H}}^{-T}}{1 - \gamma_{\mathrm{H}}^{-1}} =: K^{-1}
    \]
    yields
    \begin{equation}\label{eq:comp_gain_proof}
    \frac{\mathcal{C}_{\mathrm{S}}}{\mathcal{C}_{\mathrm{H}}}
         = \frac{\npart \nstepe[1] \nsamp_T T}{\npart \nstepe[\gamma_{\mathrm{H}}] \nsamp_T K^{-1}}
         = \frac{\nstepe[1]}{\nstepe[\gamma_{\mathrm{H}}]} T K.
    \end{equation} 
    Indicating the dependence on $\varepsilon>0$ explicitly, let $x(\varepsilon) = \frac{\log(16\npart(\varepsilon)T)}{\min_t \rho_t\big(\nsamp_T(\varepsilon)\big)}$ and $y(\varepsilon) = \frac{\log(16\npart(\varepsilon)T)}{\min_t \rho_t\big(\nsamp_t(\varepsilon)\big)}$. Applying the bound $z \leq \lceil z \rceil < z + 1$ yields
    \begin{equation*}
        \frac{x(\varepsilon)}{y(\varepsilon) + 1} < \frac{\nstepe[1]}{\nstepe[\gamma_{\mathrm{H}}]} < \frac{x(\varepsilon) + 1}{y(\varepsilon)}.
    \end{equation*}
Using the fact that $\rho_t(\nsamp) \leq 1$ for any $t, \nsamp$, we note that
\begin{equation*}
    \frac{x(\varepsilon)}{y(\varepsilon)+1} = 
\frac{\min_t \rho_t\big(\nsamp_t(\varepsilon)\big)}{\min_t \rho_t\big(\nsamp_T(\varepsilon)\big)} \cdot \frac{\log(16\npart(\varepsilon)T)}{\log(16\npart(\varepsilon)T)+\min_t \rho_t\big(\nsamp_t(\varepsilon)\big)}
        \geq 
{\min_t \rho_t\big(\nsamp_t(\varepsilon)\big)} \cdot \frac{\log(16\npart(\varepsilon)T)}{\log(16\npart(\varepsilon)T)+1} 
\end{equation*}
Since $\log(16\npart(\varepsilon)T)\to\infty$ as
$\varepsilon\searrow0$ and
$\liminf_{\nsamp\to\infty}\rho_t(\nsamp)\geq\rho^*>0$
by assumption, we obtain
\begin{align*}
    \liminf_{\varepsilon\searrow0}
    \frac{\mathcal{C}_{\mathrm{S}}}{\mathcal{C}_{\mathrm{H}}}
    \geq
    T K
    \liminf_{\varepsilon\searrow0}
    \frac{x(\varepsilon)}{y(\varepsilon)+1}\geq
    T K \rho^*.
\end{align*}

If the limits
$\lim_{\nsamp\to\infty}\rho_t(\nsamp)$ exist, we have
\[
    \lim_{\varepsilon\searrow0}
    \min_t\rho_t\bigl(\nsamp_t(\varepsilon)\bigr)
    =
    \lim_{\varepsilon\searrow0}
    \min_t\rho_t\bigl(\nsamp_T(\varepsilon)\bigr)
    =
    \min_t\rho_t^*.
\]
We therefore obtain
\begin{align*}
    \lim_{\varepsilon\searrow0}
    \frac{x(\varepsilon)}{y(\varepsilon)+1}
    =
    \lim_{\varepsilon\searrow0}
    \left(
        \frac{
            \min_t\rho_t\bigl(\nsamp_t(\varepsilon)\bigr)
        }{
            \min_t\rho_t\bigl(\nsamp_T(\varepsilon)\bigr)
        }
        \frac{
            \log(16\npart(\varepsilon)T)
        }{
            \log(16\npart(\varepsilon)T)
            +
            \min_t\rho_t\bigl(\nsamp_t(\varepsilon)\bigr)
        }
    \right)=1,
\end{align*}
and similarly
    $\lim_{\varepsilon\searrow0}
    \frac{x(\varepsilon)+1}{y(\varepsilon)}
    =1.$
    Thus, $\lim_{\varepsilon\searrow0} \frac{\nstepe[1]}{\nstepe[\gamma_{\mathrm{H}}]} = 1$ and substituting this into \eqref{eq:comp_gain_proof} leads to \eqref{eq:comp_gain}.
\end{proof}
Theorem \ref{cor:T_cost_gain} now follows immediately from Theorem \ref{cor:st_vs_ad_complex}. Note that in the standard algorithm, the maximum solver accuracy $\nsamp_T$ is used in the mutation stage across all $T$ tempering stages. By contrast, the geometric scaling $\nsamp_t = \gamma_{\mathrm{H}}^{t-T} \nsamp_T$ in the hierarchical algorithm ensures that the total cost is dominated by the most expensive mutation stage at $t=T$.

The chosen $\npart$, $\nstepe[\gamma_{\mathrm{H}}]$, and $\nstepe[1]$ are sufficient finite-sample conditions for the stated probability bound inherited from the coupling analysis of \cite{marion2023finite} and are not claimed to be optimal. In Section~\ref{sec:toy_problem}, we verify the assumptions of the above complexity analysis for a class of compact Bayesian inverse problems with additive stochastic solver noise, and illustrate the resulting cost reduction numerically.

\subsection{The general case}\label{sec:gener_is}

We now address how to perform the change of measure from $\postextml{t-1}$ to $\postextml{t}$ for $\nsamp_t>\nsamp_{t-1}$ when we do not have access to densities with respect to a common dominating measure. We decompose this change into two steps:
$
    \postextml{t-1}
    \longrightarrow
    \postextat{t}{\nsamp_{t-1}}
    \longrightarrow
    \postextml{t}.
$
The first step is the standard tempering change at fixed solver accuracy, with incremental weight
\begin{equation}\label{eq:tempering_substep_weight}
    \exp\bigl((\tp_{t-1}-\tp_t)\extend{\Phi}\extendarg\bigr).
\end{equation}
The second step changes the solver accuracy at fixed tempering parameter $\tp_t$ and is treated using a method based on generalized importance sampling; see also \cite{chopin2020introduction,del2006sequential}. No resampling or mutation is performed between the two steps. To formalize the use of generalized IS we use the following technical lemma.

Let us note that if $\kernel:(\spextended)\times\bsa{\noisesp}\rightarrow[0,1]$ is a Markov kernel, then also for every $u\in\paramsp$ the map $\kernel\big((u,\cdot),\cdot\big):\noisesp\times\bsa{\noisesp}\rightarrow[0,1]$ is a Markov kernel. Any measure on $(\noisesp,\bsa{\noisesp})$ can then be extended by $\kernel\big((u,\cdot),\cdot\big)$ in the sense of \eqref{eq:extendmeas}.

\begin{lemma}[Generalized IS]\label{lem:gener_is}
    Let $\kerforward[t]:(\spextended)\times\bsa{\noisesp}\rightarrow[0,1]$ and $\kerbackstage{t}:(\spextended)\times\bsa{\noisesp}\rightarrow[0,1]$ be Markov kernels. Assume that for all $u\in\paramsp$, the measure
    $\kernoise[t](u,\cdot)\otimes\kerbackstage{t}\big((u,\cdot),\cdot\big)$
    is absolutely continuous with respect to
    $\kernoise[t-1](u,\cdot)\otimes\kerforward[t]\big((u,\cdot),\cdot\big)$,
    with a jointly measurable Radon--Nikod\'ym derivative
    $R_t(u,e,\extend{e})$. Then
    \[
        \measxker{\postextml{t}}{\kerbackstage{t}}
        \ll
        \measxker{\postextat{t}{\nsamp_{t-1}}}{\kerforward[t]},
    \]
    and
    \begin{equation}\label{eq:gener_is_density}
        \frac{
            \td\bigl(\measxker{\postextml{t}}{\kerbackstage{t}}\bigr)
        }{
            \td\bigl(
                \measxker{\postextat{t}{\nsamp_{t-1}}}{\kerforward[t]}
            \bigr)
        }(u,e,\extend{e})
        \propto
        \frac{
            \exp\bigl(-\tp_t\extend{\Phi}(u,\extend{e})\bigr)
        }{
            \exp\bigl(-\tp_t\extend{\Phi}\extendarg\bigr)
        }
        R_t(u,e,\extend{e})
    \end{equation}
    almost surely with respect to
    $\measxker{\postextat{t}{\nsamp_{t-1}}}{\kerforward[t]}$.
\end{lemma}
\begin{proof}
    We first prove the claimed identity for measurable rectangles. 
    Let $A\in\bsa{\paramsp}$, let $B,C\in\bsa{\noisesp}$ and let $Z_t, Z_{t-1} > 0$ be the normalizing constants of $\postextml{t}$ and $\postextat{t}{\nsamp_{t-1}}$. We have
    \begin{equation*}
        (\measxker{\postextml{t}}{\kerbackstage{t}})(A\times C\times B) = \frac{1}{Z_t}\int_A\int_C\int_B \exp\bigl(-\tp_t\extend{\Phi}(u,\extend{e})\bigr) \kerbackstage{t}((u,\extend{e}),\td e) \kernoise[t](u,\td\extend{e})\prior(\td u),
    \end{equation*}
    where we have repeatedly used the definition of measure extension in \eqref{eq:extendmeas} and the density of $\postextml{t}$ from~\eqref{eq:inter_meas_ml}. Using the assumed absolute continuity with RN derivative $R_t(u,e,\extend{e})$ and Fubini's theorem to rearrange the order of integration, we obtain
    \begin{equation*}
        (\measxker{\postextml{t}}{\kerbackstage{t}})(A\times C\times B) = \frac{1}{Z_t}\int_A\int_B\int_C \exp\bigl(-\tp_t\extend{\Phi}(u,\extend{e})\bigr) R_t(u,e,\extend{e}) \kerforward[t]((u,e),\td\extend{e}) \kernoise[t-1](u,\td e)\prior(\td u).
    \end{equation*}
    Since $\exp\bigl(-\tp_{t}\extend{\Phi}\extendarg\bigr) > 0$ for all $(u,e)\in\paramsp\times\noisesp$ and $\postextat{t}{\nsamp_{t-1}}\ll\priorxker{\nsamp_{t-1}}$, we also have $\priorxker{\nsamp_{t-1}}\ll\postextat{t}{\nsamp_{t-1}}$ with RN derivative $$\rnder{\priorxker{\nsamp_{t-1}}}{\postextat{t}{\nsamp_{t-1}}}\propto\exp\bigl(-\tp_{t}\extend{\Phi}\extendarg\bigr)^{-1}.$$ Using this RN derivative and again \eqref{eq:extendmeas}, we finally obtain
    \begin{equation*}
        (\measxker{\postextml{t}}{\kerbackstage{t}})(A\times C\times B) = \frac{Z_{t-1}}{Z_t}\int_{A\times B\times C} \frac{\exp\bigl(-\tp_t\extend{\Phi}(u,\extend{e})\bigr)}{\exp\bigl(-\tp_{t}\extend{\Phi}\extendarg\bigr)} R_t(u,e,\extend{e}) (\measxker{\postextat{t}{\nsamp_{t-1}}}{\kerforward[t]})(\td u,\td e,\td\extend{e}).
    \end{equation*}
    The collection of sets in
    $\bsa{\paramsp \times \noisesp \times \noisesp}$ for which the above
    identity holds is a Dynkin system. Since it contains the measurable
    rectangles, which generate
    $\bsa{\paramsp \times \noisesp \times \noisesp}$ and are closed under
    finite intersections, the Dynkin-system theorem implies that the identity
    holds on the full product $\sigma$-algebra. This yields the stated
    Radon--Nikod\'ym derivative.
\end{proof}

This lemma recovers the IS step from \eqref{eq:inter_is_density_ml}. Suppose that $\kernoise\kernoisearg$ has a density $k_{\nsamp}$ with respect to a measure $\nu$, with
$\kernoise[t](u,\cdot)\ll\kernoise[t-1](u,\cdot)$. Choose
$
    \kerforward[t]((u,e),\cdot)=\delta_e(\cdot),
    \;
    \kerbackstage{t}((u,\extend{e}),\cdot)
    =
    \delta_{\extend{e}}(\cdot).
$
Then

\begin{equation}\label{eq:exact_ratio}
        R_t(u,e,\extend{e})
    =
    \frac{k_{\nsamp_t}(u,\extend{e})}
    {k_{\nsamp_{t-1}}(u,e)}
\end{equation}
almost surely. Since $\extend{e}=e$ a.s., the potential terms in
\eqref{eq:gener_is_density} cancel. This recovers Algorithm \ref{alg:gen_is}(A) as a special case of Lemma \ref{lem:gener_is}. To perform the complete change of measure from $\postextml{t-1}$ to $\postextml{t}$, we multiply \eqref{eq:exact_ratio} with \eqref{eq:tempering_substep_weight} to obtain
\[
    \exp\bigl((\tp_{t-1}-\tp_t)\extend{\Phi}\extendarg\bigr)
    \frac{k_{\nsamp_t}(u,e)}
    {k_{\nsamp_{t-1}}(u,e)},
\]
which is precisely \eqref{eq:inter_is_density_ml}.

To turn Lemma \ref{lem:gener_is} into a practical reweighting scheme, we need to specify the forward and backward kernels. We give two possible forward kernels.

\begin{example}\label{ex:gis_kernels}
    Fix $(u,e)\in\spextended$.
    \begin{enumerate}
        \item Let
        $\kerforward[t]((u,e),\cdot)=\kernoise[t](u,\cdot)$,
        so that $\extend{e}$ is sampled independently of $e$.
        \item Consider a Monte Carlo estimator using $\nsamp_{t-1}$ samples to obtain a realization $e\in\noisesp$ distributed according to $\kernoise[t-1](u,\cdot)$. Then $\kerforward[t]$ can be constructed by sampling $\extend{e}$ conditionally on $e$ by keeping $\nsamp_{t-1}$ samples forming $e$ and adding $\nsamp_t-\nsamp_{t-1}>0$ samples to form $\extend{e}$.
    \end{enumerate}
\end{example}

We will use Option 1 in Example \ref{ex:gis_kernels} from now on. For this forward kernel, the optimal backward kernel 
is
\[
    \kerbackopt{t}((u,\extend{e}),\td e)
    =
    \frac{
        \exp\bigl(-\tp_t\extend{\Phi}(u,e)\bigr)
    }{
        I_t(u)
    }
    \kernoise[t-1](u,\td e) 
\]
where $I_t(u)
    :=
    \int_\noisesp
    \exp\bigl(-\tp_t\extend{\Phi}(u,e^\prime)\bigr)
    \kernoise[t-1](u,\td e^\prime)$;
see \cite[Section 3.2.2]{del2006sequential} and
\cite[Section 17.4]{chopin2020introduction}. Hence,
\begin{equation}\label{eq:opt_dens}
    R_{t,\mathrm{opt}}(u,e,\extend{e})
    =
    \frac{
        \exp\bigl(-\tp_t\extend{\Phi}(u,e)\bigr)
    }{
        I_t(u)
    },
\end{equation}
and Lemma \ref{lem:gener_is} gives the weight
\begin{equation}\label{eq:opt_solver_weight}
    \frac{
        \exp\bigl(-\tp_t\extend{\Phi}(u,\extend{e})\bigr)
    }{
        I_t(u)
    }.
\end{equation}

Following \cite[Section 3.3.2.1]{del2006sequential}, approximating
$I_t(u)$ by
$\exp\bigl(-\tp_t\extend{\Phi}(u,e)\bigr)$
and multiplying by the tempering weight \eqref{eq:tempering_substep_weight} gives
\[
    \frac{
        \exp\bigl(-\tp_t\extend{\Phi}(u,\extend{e})\bigr)
    }{
        \exp\bigl(-\tp_{t-1}\extend{\Phi}\extendarg\bigr)
    }.
\]

Alternatively, using the \texttt{REPLACE} method of
\cite{botha2023automatically}, we can approximate $I_t(u)$ by
$\exp\bigl(-\tp_t\extend{\Phi}(u,\extend{e})\bigr)$. The weight~\eqref{eq:opt_solver_weight} is then approximated by one. Multiplying by \eqref{eq:tempering_substep_weight} yields
\begin{equation}\label{eq:replace_is_density}
    \exp\bigl((\tp_{t-1}-\tp_t)\extend{\Phi}\extendarg\bigr)
\end{equation}
as the approximate IS weights from $\postextml{t-1}$ to $\postextml{t}$. In our numerical experiments we use \eqref{eq:replace_is_density}, which we implement in Algorithm \ref{alg:gen_is}(B). Both approximations of $I_t(u)$ introduced above are heuristic, and we do not analyze theoretically the additional error they introduce, but we perform a numerical comparison in Section~\ref{sec:toy_problem}. Note that generalized IS is performed before resampling in Algorithm \ref{alg:pmsmc_specific}, whereas it is performed after resampling in \cite{botha2023automatically}.

\section{Hierarchical acceleration of mutation}\label{sec:mutation}

In practice, the majority of the cost of SMC samplers based on
invariant kernels typically comes from likelihood evaluations
during mutation. The mutation stage is necessary to mitigate the
errors arising from the empirical-measure approximation in
importance sampling and to decorrelate particles after resampling.

A well-known weakness of importance sampling is that the
importance weights tend to degenerate as the parameter dimension
increases \cite{bengtsson2008curse}. This means that only a few
particles have large weights, while the remaining particles have
very small weights. After resampling, the empirical measure
then contains only a few distinct particles. In a
practical tempering-based SMC sampler, two approaches can be used
to address this issue. Either the intermediate target
distributions are brought sufficiently close together, or the
particles are mutated more effectively. The former requires a
large number $T$ of tempering stages, while the
latter requires a large number $\nstepc[t]+\nstepe[t]$ of kernel transitions at each
mutation stage, thus both leading to a large number
of likelihood evaluations and high computational
cost.

We adopt the latter approach through the cheap and expensive
mutation blocks specified in Algorithm~\ref{alg:pmsmc_specific}.
Section~\ref{sec:mutation_analysis} provides an abstract analysis
of one mutation stage, while Section~\ref{sec:pm_smc} applies the
resulting conditions to SMC for BIPs with randomized likelihoods.

\subsection{Accelerated mutation analysis}\label{sec:mutation_analysis}

Throughout this subsection, we focus on one fixed mutation stage of
Algorithm~\ref{alg:pmsmc_specific} and suppress the corresponding
stage index $t\in\{1,\ldots,T\}$. 
Let $\msp=\spextended$, and for a fixed number of particles $\npart\in\N$ let
$\mspex$ be a product space equipped with its product Borel
$\sigma$-algebra $\bsa{\mspex}$.

We formulate the analysis in
terms of the cheap and expensive kernels used at this stage and
their $\npart$-particle product kernels. Our main assumption is
that these kernels are $V$-uniformly ergodic for a suitable function $V$. In an abstract setting, sufficient conditions for the
corresponding product-space ergodicity are collected in
Appendix~\ref{app:mutation}; the underlying arguments use standard
coupling and contraction techniques.

Let $V\colon\msp\to[1,\infty)$ be measurable and define
\[
\varrho_V (x,y) =
\begin{cases}
V(x) + V(y), &\text{if } x \neq y,\\
0, &\text{otherwise}.
\end{cases}
\]
The function $\varrho_V$ satisfies all axioms of a metric. The corresponding
$1$-Wasserstein distance is denoted by $\mathcal{W}_V$ and can be understood
as a weighted total variation distance \cite{hairer2011yet}. The (unweighted) total variation distance is given by $\norm{\mu_1 - \mu_2}{tv} = \sup_{A \in \bsa{\msp}} \vert \mu_1(A) - \mu_2(A) \vert$. It is known that for $V = \mathbf{1}$ this coincides, up to a factor 2, with the Wasserstein distance $\mathcal{W}_V$. On $\mspex$,
we define the metric
\[
\widetilde{\varrho}_V(\xi,\xi')
=
\sum_{j=1}^{\npart}\varrho_V(x_j,x_j'),
\quad \text{where} \ \xi=(x_1,\ldots,x_\npart) \ \text{and} \ \xi'=(x_1',\ldots,x_\npart'),
\]
and denote by
$\widetilde{\mathcal{W}}_V$ the corresponding $1$-Wasserstein distance on $\mspex$.

We use the following notion of geometric ergodicity. Let $\mker$ be a
Markov kernel on $(\msp,\bsa{\msp})$ admitting $\minvar$ as an invariant
probability measure and assume $V$ is $\minvar$-integrable. We say that $\mker$ is $V$-uniformly geometrically
ergodic if there exist constants $K<\infty$ and $\tau\in[0,1)$ such that
\begin{equation}\label{equ:V_uniform_convergence}
    \sup_{\substack{f:\msp\rightarrow\mathbb{R}\ \mathrm{meas.}\\
                    \vert f\vert\leq V}}
    \left\vert
        \int_\msp f(y)\mker^k(x,\td y)-\minvar(f)
    \right\vert
    \leq
    K V(x)\tau^k,
\end{equation}
for every $x\in\msp$ and $k\in\N$, where $\vert f\vert\leq V$ is
understood pointwise. Below, the cheap and expensive kernels are assumed
to satisfy \eqref{equ:V_uniform_convergence} with the same function $V$,
but possibly different constants.

Let $\minvarc$ and $\minvare$ be probability measures on $(\msp,\bsa{\msp})$. Assume
that $\mkerc$ and $\mkere$ are Markov kernels on $(\msp,\bsa{\msp})$
admitting $\minvarc$ and $\minvare$ as invariant
distributions, respectively. We interpret $\minvarc$ as an approximation of $\minvare$
which can be targeted by applying the ``cheap'' kernel $\mkerc$.

For $\xi=(x_1,\ldots,x_\npart)\in\mspex$, let the corresponding product
kernels be defined by
\[
\mkerexc(\xi,\cdot)
=
\bigotimes_{j=1}^{\npart}\mkerc(x_j,\cdot),
\quad \text{and} \quad
\mkerexe(\xi,\cdot)
=
\bigotimes_{j=1}^{\npart}\mkere(x_j,\cdot).
\]
As shown in Theorem~\ref{th:product_meas_conv}, they admit the product measures $\minvarexc=\minvarc^{\otimes\npart}$
and $\minvarexe=\minvare^{\otimes\npart}$
as their invariant measures, respectively.

\begin{theorem}\label{th:mut_error}
Assume that $\mkere$ is geometrically ergodic in the sense of \eqref{equ:V_uniform_convergence}
with a $\minvare$-integrable function $V \colon \msp \to [1, \infty)$ and constants $\tau<1$ and $K< \infty$.
If $V$ is lower semi-continuous (w.r.t. the metric which renders $\msp$ Polish), then 
    \begin{equation*}
\widetilde{\mathcal{W}}_V (\eta  \mkerexe^{\nstepe} , \minvarexe) \leq K \tau^{\nstepe} \widetilde{\mathcal{W}}_V(\eta , \minvarexe),
    \end{equation*}
    for any $\nstepe, \nstepc \in \N$ and any initial distribution $\eta$ with finite first moment\footnote{A probability measure $\eta$ on $\mspex$ has a finite first moment if there exists $\xi \in \mspex$ such that $\int_{\mspex} \widetilde{\varrho}_V(\xi,\xi') \eta(\td \xi') < \infty$.}.

    If $\mkerc$ is also geometrically ergodic in the sense of \eqref{equ:V_uniform_convergence}, with the same (finite and lower semi-continuous) function $V$, but potentially different constants $K_c< \infty$ and $\tau_c<1$, then  
    \begin{equation}\label{eq:2stage_mut_error}
  \widetilde{\mathcal{W}}_V (\eta \mkerexc^{\nstepc} \mkerexe^{\nstepe} , \minvarexe) \leq K \tau^{\nstepe} \left( K_c\tau_c^{\nstepc} \widetilde{\mathcal{W}}_V(\eta , \minvarexc) + \widetilde{\mathcal{W}}_V(\minvarexc, \minvarexe) \right) ,  
    \end{equation}
    for any $\nstepe, \nstepc$ and $\eta$, and for any $\minvarexc$ with finite first moment. 
\end{theorem}
\begin{proof} 
Combining \cite[Lemma 2.1]{hairer2011yet} with the Kantorovich-Rubinstein duality formula \cite[Chapter 5]{villani2008optimal} we obtain 
\[
\mathcal{W}_V (\mkere^{\nstepe}(x, \cdot ), \minvare) = \sup_{\vert f \vert \leq V} \left\vert \int_\msp f(y) \mkere^{\nstepe}(x, \td y) - \minvare(f) \right\vert \leq KV(x) \tau^{\nstepe}. 
\]
Hence, by Theorem \ref{th:product_meas_conv} it follows that for any $\xi= (x_1, \dots, x_\npart) \in \mspex$
\[
\widetilde{\mathcal{W}}_V (\mkerexe^{\nstepe}(\xi, \cdot ), \minvarexe) \leq K \tau^{\nstepe} \sum_{j=1}^\npart V(x_j).
\]
Thus, for $\xi= (x_1, \dots, x_\npart) \in \mspex$ and $ \xi'= (x_1', \dots, x_\npart') \in \mspex$ an application of triangle inequality implies 
\[
\widetilde{\mathcal{W}}_V (\mkerexe^{\nstepe}(\xi, \cdot ), \mkerexe^{\nstepe}(\xi', \cdot)) \leq K \tau^{\nstepe} \sum_{j=1}^\npart \left(V(x_j) + V(x_j')\right) = K\tau^{\nstepe} \widetilde{\varrho}_V(\xi, \xi'),
\]
if $x_i\neq x_i'$ for all $i\in\{1,\ldots,\npart\}$. If some coordinates coincide, the corresponding coordinate transitions are coupled identically and the preceding estimate only needs to be applied to indices $i$ with $x_i\neq x_i'$. Consequently,
\[
\sup_{\xi \neq \xi'} \frac{\widetilde{\mathcal{W}}_V (\mkerexe^{\nstepe}(\xi, \cdot ), \mkerexe^{\nstepe}(\xi', \cdot))}{\widetilde{\varrho}_V(\xi, \xi')} \leq K \tau^{\nstepe}.
\]
Thus, we have a contraction in the sense of \cite[Section 2]{rudolf2018perturbation} and
\[
\widetilde{\mathcal{W}}_V
\bigl(
    \eta\mkerexe^{\nstepe},
    \minvarexe
\bigr)
\leq
K\tau^{\nstepe}
\widetilde{\mathcal{W}}_V
\bigl(
    \eta,
    \minvarexe
\bigr)
\]
for any initial distribution $\eta$ with finite first moment. 

The second part follows by applying the same arguments to $\mkerexc$ instead of $\mkerexe$ and a straightforward application of the triangle inequality.
\end{proof}

We find this approach particularly efficient when employed within Algorithm \ref{alg:pmsmc_specific}. The utility of Theorem~\ref{th:mut_error} becomes apparent when the cost of applying $\mkere$ and $\mkerc$ is taken into consideration. At a fixed mutation stage $t$, let $\eta$ denote the distribution on $\mspex$ of the $N$-particle vector immediately after the resampling step in Algorithm~\ref{alg:pmsmc_specific}. Thus, $\eta$ describes the joint law of the particles entering the mutation stage. The measure $\minvarexe$ is interpreted as the target product measure of Algorithm~\ref{alg:pmsmc_specific} at stage~$t$. Assuming the distance $\widetilde{\mathcal{W}}_V(\eta,\minvarexe)$ to be large, $\minvarexe$ to be well approximated by $\minvarexc$, and the cost of applying $\mkerc$ to be small, we first move from the initial distribution $\eta$ to $\minvarexc$ using the cheap kernel $\mkerc$ before applying the expensive~$\mkere$ and moving to~$\minvarexe$. We now analyze this formally by formulating an optimization problem based on Theorem \ref{th:mut_error}.

We minimize the upper bound for $\widetilde{\mathcal{W}}_V (\eta \mkerexc^{\nstepc} \mkerexe^{\nstepe} , \minvarexe)$, assuming for simplicity that $\nstepe,\nstepc\in\R$, $\nstepe,\nstepc\geq 0$. That is, we want to solve the following constrained minimization problem:
\begin{equation}\label{eq:mut_opt_problem}
\begin{split}
  &\argmin_{\nstepe,\nstepc\geq 0}\; K \tau^{\nstepe} \left( K_c\tau_c^{\nstepc} \widetilde{\mathcal{W}}_V(\eta ,\minvarexc) + \widetilde{\mathcal{W}}_V(\minvarexc,\minvarexe) \right) ,\\
  &\text{such that}\quad C \nstepe+ C_c \nstepc=C \km,
\end{split}
\end{equation}
where $\km>0$ is a fixed budget given by a maximum number of transitions with the expensive target kernel. Here, $C_c$ and $C$ are the costs of one Markov transition for the cheap and expensive chain, respectively. 

We assume the following:
\begin{equation}\label{eq:as_ratios}
    \frac{C_c}{C}=:\pp\in(0,1), \qquad 
    \frac{\tau_c}{\tau}=:\qq\in(0,\infty), \qquad 
    \frac{\widetilde{\mathcal{W}}_V(\eta , \minvarexc)}{\widetilde{\mathcal{W}}_V(\minvarexc, \minvarexe)}=: \rr\in(0,\infty), 
\end{equation}
where $\tau$ is assumed to be bounded away from zero and $\tau\in(0,1)$. The constraint in \eqref{eq:mut_opt_problem} can now be more compactly written as
$\nstepe + \pp \nstepc=\km$.

The following theorem characterizes the optimal budget allocation.
Substituting the constraint, the proof reduces to a simple one-dimensional optimization problem, which is given in Appendix \ref{app:mutation}.

\begin{theorem}\label{th:budget_opt}
    Consider the optimization problem defined in \eqref{eq:mut_opt_problem}. Define
    \begin{equation}\label{eq:def_D}
        D = \frac{\pp \ln \tau}{K_c \rr 
        \big( (1-\pp)\ln \tau + \ln \qq \big)}.
    \end{equation}
    The objective function admits a unique minimum $(\nstepe^\star, \nstepc^\star)$ on the feasible set $(\nstepe,\nstepc) \in [0, \km]\times [0, \km/\pp]$. Specifically, the optimal budget allocation is determined by the following cases.
    \begin{enumerate}[(i)]
        \item\label{th:fine_only} (Expensive chain only.) If $\qq \geq \tau^{-(1-\pp)}$ or 
        $D \geq 1$, then 
            $\nstepc^\star = 0, \quad \nstepe^\star = \km.$
        \item\label{th:coarse_only} (Cheap chain only.) If $\qq < \tau^{-(1-\pp)}$ and $D \leq (\qq\tau)^{\km/\pp}$, then 
            $\nstepc^\star = \km/\pp, \quad \nstepe^\star = 0.$
        \item\label{th:chain_comb} (Chain combination.) If $\qq < \tau^{-(1-\pp)}$ and $(\qq\tau)^{\km/\pp} < D < 1$, then there exists a unique minimizer
        \begin{equation*}
            \nstepc^\star = \frac{\ln D}{\ln(\qq \tau)}, \quad 
            \nstepe^\star = \km - \pp \nstepc^\star. 
        \end{equation*}
    \end{enumerate}
\end{theorem}

The results in Theorem \ref{th:budget_opt} admit an intuitive interpretation. First, the condition $\qq < \tau^{-(1-\pp)}$, which is equivalent to $\tau_c < \tau^\pp$, establishes a cost-efficiency threshold. Recall that one cheap chain transition costs a fraction $\pp$ of one expensive chain transition. This condition therefore implies that one transition of the cheap chain yields a better error reduction than taking $\pp$ (fractional) expensive transitions.

The quantity $D$ acts as a relative bias penalty. By definition,
$D \propto (K_c \rr)^{-1}$, which 
is proportional to the discrepancy
$\widetilde{\mathcal{W}}_V(\minvarexc,\minvarexe)$ relative to
$\widetilde{\mathcal{W}}_V(\eta,\minvarexc)$. Notice that in
\eqref{eq:def_D}, a low cost of the cheap chain ($\pp\ll1$) can to some
extent compensate for a larger discrepancy between $\minvarexc$ and
$\minvarexe$. This suggests that even a bad (highly biased) cheap chain may be used if it is cheap enough.

Finally, let us examine regime \eqref{th:chain_comb} in Theorem \ref{th:budget_opt} more closely. Here, the optimal cheap chain length $\nstepc^\star$ is independent of the mutation budget $\km$. This has an important practical consequence: if our budget increases for a fixed problem, the number of cheap transitions should stay the same and all the additional budget should be spent on the expensive chain. Conversely, if for a fixed problem our budget decreases, 
then at some point $(\qq\tau)^{\km/\pp} \geq D$ and all budget should be spent on the cheap chain.

\subsection{Application to SMC for BIPs with randomized likelihoods}
\label{sec:pm_smc}

We now verify the abstract assumptions of Sections~\ref{sec:acc_is} and \ref{sec:mutation_analysis} for mutation kernels we use within density-tempered marginalized SMC. 
In particular, we impose boundedness and positivity conditions that yield uniform ergodicity independently of the solver accuracy and tempering stage. 

For each tempering index $t\in\{0,\ldots,T\}$ and solver accuracy $\nsamp>0$, let $\mkerpm_{t,\nsamp}$ and $\mkermc_{t,\nsamp}$ denote the pseudo-marginal Metropolis--Hastings (PMMH) and Monte Carlo within Metropolis (MCWM) kernels of \cite[Algorithms~1 and~2]{medina2016stability}, 
respectively. The PMMH kernels act on the extended space $\spextended$, have invariant measures $\postextsl{t}$ and marginal targets $\margmeas{t,\nsamp}$. The MCWM kernels act on $\paramsp$ and approximately target $\margmeas{t,\nsamp}$. Under the assumptions below, they admit unique invariant measures $\minvarmc{t}{\nsamp}$ that differ in general from~$\margmeas{t,\nsamp}$.
To apply the analysis of Section~\ref{sec:mutation_analysis}, we fix an arbitrary stage $t\in\{1,\ldots,T\}$ and accuracies $0<\nsamp_c<\nsamp$, taking $\mkere=\mkerpm_{t,\nsamp}$
and use a version of $\mkermc_{t,\nsamp_c}$ as the cheap kernel $\mkerc$. The corresponding invariant measures then take the roles of $\minvare$ and $\minvarc$ in Theorem~\ref{th:mut_error}. We make this precise below.

PMMH allows us to marginally target $\margmeas{t,\nsamp}$ exactly, but its mixing properties may deteriorate as $\nsamp$ decreases and the likelihood noise increases; see \cite{medina2016stability}. By contrast, MCWM only approximately targets $\margmeas{t,\nsamp}$, but its mixing properties have been observed to degenerate less when $\nsamp$ decreases. This motivates using PMMH 
as the expensive 
and MCWM 
as the cheap kernel.

One application of MCWM is twice as expensive as one application of PMMH when both use the same accuracy $\nsamp_c=\nsamp$; see \cite{medina2016stability}. Indeed, MCWM requires two likelihood evaluations at each transition, whereas PMMH requires only one. Thus MCWM is cheaper than PMMH only if $\nsamp_c<\nsamp/2$. However, in applications we use values as small as $\nsamp_c=\nsamp/20$, which still makes one MCWM transition significantly cheaper than one PMMH transition; see Section \ref{sec:numerics}. The following assumptions yield Doeblin minorization conditions that are uniform over the solver accuracy and over arbitrary tempering schedules.

\begin{assumption}\label{ass:bounded_setting}
    The parameter space $\paramsp\subset\R^d$ is bounded and satisfies $\paramleb(\paramsp)>0$, where $\paramleb$ denotes Lebesgue measure restricted to $\bsa{\paramsp}$. The prior admits a density, denoted again by $\prior$, with respect to $\paramleb$, and a version of this density satisfies
    \begin{equation}
        0<\priorl\leq \prior(u)\leq\priorh<\infty,
        \qquad u\in\paramsp.\tag{B1}\label{ass:bd_domain}
    \end{equation}
    The randomized potential is non-negative and uniformly bounded:
    \begin{equation}
        0\leq\extend{\Phi}(u,e)\leq\potbd<\infty,
        \qquad (u,e)\in\spextended.\tag{B2}\label{ass:bd_potential}
    \end{equation}
    The proposal kernel admits a density $Q$ with respect to $\paramleb$ satisfying\footnote{Here
$Q:\paramsp\times\paramsp\to[0,\infty)$ is such that, for every
$u\in\paramsp$, the map $v\mapsto Q(u,v)$ is a density of
$Q(u,\cdot)$ with respect to $\paramleb$.}
    \begin{equation}
        0<\propl\leq Q(u,v)\leq\proph<\infty,
        \qquad u,v\in\paramsp.\tag{B3}\label{ass:bd_proposal}
    \end{equation}
\end{assumption}

\begin{remark}\label{rem:assumptions_compact}
Sufficient conditions for Assumption~\ref{ass:bounded_setting} to hold are that $\paramsp$ and $\noisesp$ are compact, that the prior density is continuous and strictly positive on $\paramsp$, and that $\fwopext$ is continuous on $\spextended$. Furthermore, a Gaussian random-walk proposal conditioned to remain in $\paramsp$ has a continuous, strictly positive density on $\paramsp\times\paramsp$ and therefore satisfies Assumption~(\ref{ass:bd_proposal}). Symmetry of $Q$ is not required.
\end{remark}

\begin{theorem}\label{th:pm_geom_erg}
Let Assumption~\ref{ass:bounded_setting} be satisfied, let $T\in\N$, and let $0=\tp_0<\cdots<\tp_T=1$ be an arbitrary tempering schedule. Define
\begin{equation}\label{eq:def_minor_const}
    \minc
    \coloneqq
    \exp(-\potbd)\,
    \frac{\priorl}{\priorh}\,
    \frac{\propl^2}{\proph}\,
    \paramleb(\paramsp)
    \in(0,1].
\end{equation}
Then, for every $t\in\{0,\ldots,T\}$ and every $\nsamp>0$, the following statements are true.
\begin{enumerate}[(i)]
    \item\label{th:pm_erg_pmmh}
    The PMMH kernel $\mkerpm_{t,\nsamp}$ on $\spextended$ satisfies
    \begin{equation}\label{eq:minor_pmmh}
        \mkerpm_{t,\nsamp}\bigl((u,e),A\bigr)
        \geq \minc\,\minormeas{\nsamp}(A),
        \qquad
        (u,e)\in\spextended,
        \quad A\in\bsa{\spextended},
    \end{equation}
    where
    \[
        \minormeas{\nsamp}(A)
        =
        \frac{1}{\paramleb(\paramsp)}
        \int_\paramsp\int_\noisesp
        \mathbf{1}_A(v,e')\,\kernoise(v,\td e')\,\td v.
    \]
    Consequently, $\mkerpm_{t,\nsamp}$ is uniformly ergodic and admits $\postextsl{t}$ as unique invariant distribution.

    \item\label{th:pm_erg_mcwm}
    The MCWM kernel $\mkermc_{t,\nsamp}$ on $\paramsp$ satisfies
    \begin{equation}\label{eq:minor_mcwm}
        \mkermc_{t,\nsamp}(u,A)
        \geq
        \minc\,\frac{\paramleb(A)}{\paramleb(\paramsp)},
        \qquad
        u\in\paramsp,
        \quad A\in\bsa{\paramsp}.
    \end{equation}
    It is uniformly ergodic and admits a unique invariant probability measure $\minvarmc{t}{\nsamp}$.
\end{enumerate}
For either kernel $P$ and its corresponding invariant measure $\pi$,
\begin{equation}\label{eq:uniform_tv_pm_mc}
    \norm{P^k(x,\cdot)-\pi}{tv}
    \leq (1-\minc)^k,
    \qquad k\in\N.
\end{equation}
Hence both kernels satisfy \eqref{equ:V_uniform_convergence} with $V\equiv1$, $K=2$, and $\tau=1-\minc$. The constant $\minc$ depends only on the constants in Assumption~\ref{ass:bounded_setting}; in particular, it is independent of $\nsamp$, $t$, $T$, and $\tp_t$.
\end{theorem}
The proof is given in Appendix \ref{app:mutation}.

The complexity analysis of Section~\ref{sec:compl_anal}, and in particular Theorem~\ref{cor:st_vs_ad_complex}, is formulated in terms of the $L^2$-spectral gaps of the mutation kernels. The next proposition converts part~\eqref{th:pm_erg_pmmh} of Theorem~\ref{th:pm_geom_erg} into the required bound.

\begin{proposition}\label{prop:pm_gap}
Under the assumptions and notation of Theorem~\ref{th:pm_geom_erg}, for every $t\in\{0,\ldots,T\}$ and every $\nsamp>0$, the PMMH kernel $\mkerpm_{t,\nsamp}$ is reversible with respect to $\postextsl{t}$. Its $L^2(\postextsl{t})$ absolute spectral gap satisfies
\begin{equation}\label{eq:pm_gap}
    \rho\bigl(\mkerpm_{t,\nsamp}\bigr)\geq\minc.
\end{equation}
Consequently, for the mutation stages of Algorithm~\ref{alg:pmsmc_specific},
\[
    \inf_{\nsamp>0}
    \min_{t\in\{1,\ldots,T\}}
    \rho_t(\nsamp)
    \geq\minc>0.
\]
The lower bound is independent of $\nsamp$, $t$, $T$, and the choice of the tempering schedule.
\end{proposition}

\begin{proof}
By construction, $\mkerpm_{t,\nsamp}$ is the Metropolis--Hastings kernel on $\spextended$ with target $\postextsl{t}$, proposal kernel
\[
    Q(u,v)\,\td v\,\kernoise(v,\td e'),
\]
and acceptance probability given by \eqref{eq:acc_pmmh}. The standard Metropolis--Hastings detailed-balance identity therefore applies. Thus $\mkerpm_{t,\nsamp}$ satisfies detailed balance with respect to $\postextsl{t}$ and is therefore reversible; in particular, its unique invariant measure from part~\eqref{th:pm_erg_pmmh} of Theorem~\ref{th:pm_geom_erg} is $\postextsl{t}$.

For every probability measure $\eta$ on $\spextended$,
\[
    \norm{\eta(\mkerpm_{t,\nsamp})^k-\postextsl{t}}{tv}
    \leq (1-\minc)^k,
    \qquad k\in\N,
\]
which follows by integrating \eqref{eq:uniform_tv_pm_mc} with respect to $\eta$. In particular, this bound holds for every probability measure whose density belongs to $L^2(\postextsl{t})$. Since $\mkerpm_{t,\nsamp}$ is reversible, \cite[Theorem~2.1 and Remark~2.2]{roberts1997geometric} imply that its operator norm on the mean-zero subspace of $L^2(\postextsl{t})$ is at most $1-\minc$, which is equivalent to~\eqref{eq:pm_gap}. The final assertion follows from the uniformity of $\minc$.
\end{proof}

No reversibility property is required for the cheap MCWM kernel in
Theorems~\ref{th:mut_error} and \ref{th:budget_opt}. To apply Theorems~\ref{th:mut_error} and \ref{th:budget_opt}, it only remains to place it on the same extended state space as PMMH.

\begin{corollary}\label{cor:extended_ergodicity}
Let Assumption~\ref{ass:bounded_setting} hold, let $t\in\{1,\ldots,T\}$, and let $0<\nsamp_c<\nsamp$. Set $\mkere\coloneqq\mkerpm_{t,\nsamp}$ and define the extended MCWM kernel $\mkerc$ on $\spextended$ by
\begin{equation}\label{eq:extended_mcwm}
    \mkerc\bigl((u,e),A\bigr)
    \coloneqq
    \int_\paramsp\int_\noisesp
    \mathbf{1}_A(u',e')\,
    \kernoise(u',\td e')\,
    \mkermc_{t,\nsamp_c}(u,\td u'),
    \qquad A\in\bsa{\spextended}.
\end{equation}
Then, $\mkerc$ and $\mkere$ are uniformly ergodic Markov kernels on $\spextended$ with unique invariant measures
\[
    \minvarc
    =
    \measxker{\minvarmc{t}{\nsamp_c}}{\kernoise},
    \qquad
    \minvare
    =
    \postextsl{t},
\]
respectively. Both kernels satisfy \eqref{equ:V_uniform_convergence} with $K=2$, $\tau_c=\tau=1-\minc$ and common Lyapunov function $V\equiv1$. Thus, they satisfy the assumptions of Theorem~\ref{th:mut_error} concerning the kernels and their invariant measures.
\end{corollary}

\begin{proof}
The kernel $\mkerc$ is well defined by \eqref{eq:extended_mcwm}. Its stated invariant measure follows from the invariance of $\minvarmc{t}{\nsamp_c}$ under $\mkermc_{t,\nsamp_c}$ and the definition of the measure-kernel extension in \eqref{eq:extendmeas}. 

For $A\in\bsa{\spextended}$, part~\eqref{th:pm_erg_mcwm} of Theorem~\ref{th:pm_geom_erg} gives
\begin{equation*}
    \mkerc\bigl((u,e),A\bigr)
    \geq
    \frac{\minc}{\paramleb(\paramsp)}
    \int_\paramsp\int_\noisesp
    \mathbf{1}_A(u',e')\,
    \kernoise(u',\td e')\,\td u'
    =\minc\,\minormeas{\nsamp}(A).
\end{equation*}
Thus $\mkerc$ satisfies the same one-step Doeblin condition as $\mkere$. The asserted uniform ergodicity and the values $V\equiv1$, $K=2$, and $\tau_c=\tau=1-\minc$ follow from Theorem~\ref{th:pm_geom_erg} and the preceding argument.

Finally, $V\equiv1$ is finite, continuous, and integrable with respect to every probability measure. The associated metric satisfies $\varrho_V\leq2$ on $\spextended\times\spextended$, and its product-space counterpart satisfies $\widetilde{\varrho}_V\leq2\npart$ on $\mspex\times\mspex$. Hence all probability measures appearing in Theorem~\ref{th:mut_error} have finite first moment.
\end{proof}

\begin{remark}
\begin{enumerate}[(i)]
\item The extension \eqref{eq:extended_mcwm} is only a theoretical construction. Since the MCWM update of $u$ does not use the auxiliary coordinate $e$, an implementation need not sample from $\kernoise(u',\cdot)$ after every cheap transition; it suffices to generate the auxiliary variable once, at accuracy $\nsamp$, after the cheap block and before the first PMMH transition.
%
\item Corollary~\ref{cor:extended_ergodicity} verifies the mixing assumptions for Theorems~\ref{th:mut_error} and \ref{th:budget_opt}, but it does not control the discrepancy $\widetilde{\mathcal{W}}_V(\minvarexc,\minvarexe)$. Thus the budget regimes of Theorem~\ref{th:budget_opt} remain conditional on either a perturbation bound for the MCWM invariant measure or an empirical estimate of this discrepancy.
%
\item Assumption~\ref{ass:bounded_setting} is sufficient rather than necessary. On an unbounded parameter space, a random-walk proposal has no positive global lower bound, so the one-step Doeblin argument must be replaced by drift and small-set conditions. Applying Theorem~\ref{th:mut_error} would then require a common unbounded Lyapunov function for PMMH and the extended MCWM kernel, together with tail control of the randomized potential uniform in the solver accuracy. We do not pursue this extension here.
\end{enumerate}
\end{remark}

\section{Applications}\label{sec:numerics}

\subsection{A linear problem}\label{sec:toy_problem}
We consider a class of inverse problems with Gaussian observation noise for which the assumptions of Sections~\ref{sec:acc_is} and~\ref{sec:mutation} can be verified explicitly. The numerical experiments below use a simplified linear variant as a controlled benchmark.

\begin{example}
\label{ex:compact_gaussian}
Let the parameter space $\paramsp\subset\R^d$ be compact with positive Lebesgue measure, and let $\prior$ be the uniform distribution on $\paramsp$. Fix data $y\in\R^m$ and a symmetric positive-definite covariance matrix $\Gamma\in\R^{m\times m}$. Let the forward map $\fwop:\paramsp\to\R^m$ be continuous and define the potential by $\Phi(u)=\frac12\norm{y-\fwop(u)}{\Gamma^{-1}}^2$.

Fix $R>0$, set $\noisesp=\{e\in\R^m:\norm{e}{}\leq R\}$, and define the stochastic forward solver by $\fwopext\extendarg=\fwop(u)+e$ for $(u,e)\in\paramsp\times\noisesp$. The associated potential is $\extend{\Phi}(u,e)=\frac12\norm{y-\fwopext(u,e)}{\Gamma^{-1}}^2$.

For each $\nsamp>0$, suppose that the noise model satisfies $\kernoise(u,\td e)=k_\nsamp(u,e)\noiseleb(\td e)$, where $\noiseleb$ denotes Lebesgue measure restricted to $\noisesp$ and
\begin{equation}
\label{equ:compact_noise_density}
    k_\nsamp(u,e)
    =
    \frac{1}{Z_\nsamp(u)}
    \exp\left(
        -\frac{\nsamp}{2}e^\top\Sigma(u)^{-1}e
    \right),
    \qquad
    (u,e)\in\paramsp\times\noisesp.
\end{equation}
Here $Z_\nsamp(u)$ is the normalizing constant such that $\int_\noisesp k_\nsamp(u,e)\noiseleb(\td e)=1$. Assume that $u\mapsto\Sigma(u)$ is continuous on $\paramsp$ and that $\Sigma(u)\in\R^{m\times m}$ is symmetric positive definite for every $u\in\paramsp$.
\end{example}

The resulting marginal measure $\margmeas{\nsamp}$ is given by its RN derivative as 
\[
\td\margmeas{\nsamp}/\td\prior\propto\likmarg{\nsamp} \quad \text{with} \quad \likmarg{\nsamp}(u)=\int_\noisesp\exp(-\extend{\Phi}(u,e))k_\nsamp(u,e)\noiseleb(\td e).
\]
To construct the SMC mutation kernels we use PMMH and MCWM as in Section~\ref{sec:pm_smc}. We fix a symmetric positive-definite matrix $\Gamma_Q\in\R^{d\times d}$ and use the Gaussian random-walk proposal conditioned to remain in $\paramsp$: given $u\in\paramsp$, propose $v=u+\xi$, where $\xi\sim\mathcal N(0,\Gamma_Q)$, conditionally on $u+\xi\in\paramsp$. In this case, Assumption~\ref{ass:bounded_setting} is satisfied and therefore Proposition~\ref{prop:pm_gap} implies that the PMMH kernels $\mkerpm_{t,\nsamp}$ are reversible with respect to $\postextsl{t}$ and that their absolute spectral gaps satisfy $\inf_{\nsamp>0}\min_{t\in\{1,\ldots,T\}}\rho_t(\nsamp)\geq\minc>0$. Moreover, the assumptions of Theorem~\ref{th:mut_error} are satisfied by Corollary~\ref{cor:extended_ergodicity}. The precise mutation setup is stated when numerical experiments are presented.

We use the computational cost model in which generating one sample from $k_\nsamp(u,\cdot)$, and hence one realization of the stochastic solver at accuracy $\nsamp$, has cost $\nsamp$. This is motivated by Monte Carlo simulation, where $\nsamp$ represents the number of samples used to construct one solver realization. For the complexity analysis of Section~\ref{sec:acc_is} to apply, it remains to verify Assumptions~\eqref{eq:assump_W_ad}--\eqref{eq:assump_CB_ad}.

\begin{proposition}
\label{prop:example_complexity}
Consider the setting of Example~\ref{ex:compact_gaussian}. For every $T\in\N$, every tempering schedule $0=\tp_0<\cdots<\tp_T=1$, and every $\gamma\geq1$, Assumptions~\eqref{eq:assump_W_ad}--\eqref{eq:assump_CB_ad} are satisfied.
\end{proposition}

The proof is given in Appendix~\ref{app:numerics}.

We now demonstrate performance of the developed methods numerically on this problem.

\paragraph{Numerical Results.}
The benchmark is motivated by approximating a matrix--vector product $Au$ using a Monte Carlo solver.
We consider a linear Gaussian inverse problem based on
Example~\ref{ex:compact_gaussian}, but replace the compact parameter
space and uniform prior by \(\paramsp=\R^d\) and
\(\prior=\mathcal{N}(0,I_d)\). This modification is made because the
uniform prior in Example~\ref{ex:compact_gaussian} leads to poor
behavior in high dimensions. We take \(m=d\), generate a fixed matrix
\(A\in\R^{d\times d}\) with independent entries
\(A_{ij}\sim\mathcal{N}(0,1)\), and set \(\fwop(u)=Au\). The ground
truth \(u^\dagger\) is sampled from the prior, and the data are
generated according to
\[
    y=Au^\dagger+\eta,
    \qquad
    \eta\sim\mathcal{N}(0,\sigma_{\mathrm{obs}}^2I_d),
\]
where either \(\sigma_{\mathrm{obs}}^2=0.05\) and \(d=8\), or
\(\sigma_{\mathrm{obs}}^2=0.00625\) and \(d=64\). The stochastic solver
and its noise model are otherwise those of
Example~\ref{ex:compact_gaussian}, with \(R=1\) and
\(\Sigma=8I_d\). Since the exact model is linear Gaussian, the true
posterior \(\post\) is available in closed form. We use
\(N_{\mathrm{ref}}=10^4\) independent draws from \(\post\) as reference
samples.

In the hierarchical algorithm, we use the discrete solver hierarchy with
\(\nsamp_{0}=10\) and refinement factor~\(2\). The value of $\nsamp_{0}$ is fixed throughout the experiments in this subsection. For a prescribed
final accuracy \(\nsamp_T\), we set
\begin{equation}\label{eq:lin_solver_schedule}
    \nsamp_t=\nsamp_{0}2^{k_t},\qquad\text{where}\qquad
    k_t
    =
    \operatorname{round}\left(
        \log_2\left(
            \frac{\nsamp_T}{\nsamp_{0}}
        \right)
        \sqrt{\frac{t}{T}}
    \right),
    \qquad
    t=0,\ldots,T.
\end{equation}

In all numerical experiments we use the following error metric. For each independent run of an algorithm, we compute the empirical one-dimensional
\(W_1\) distance between the output particles of
Algorithm~\ref{alg:pmsmc_specific} and the reference posterior samples
marginally in each coordinate, and then average over the \(d\) coordinates. The distance is computed using \texttt{scipy.stats.wasserstein\_distance()}. The cost of an algorithm is the total cost of initialization,
importance sampling, auxiliary-variable replacement, and mutation.

In the first experiment, we compare standard SMC (Algorithm \ref{alg:pmsmc_specific} using \(\nsamp_t=\nsamp_T\) throughout) with hierarchical SMC using the solver schedule in \eqref{eq:lin_solver_schedule} and either the \texttt{Classical} IS (Algorithms~\ref{alg:pmsmc_specific}~and~\ref{alg:gen_is}(A)) or the
\texttt{Generalized} IS (Algorithms \ref{alg:pmsmc_specific}~and~\ref{alg:gen_is}(B)). For each algorithm, a tempering schedule is first constructed in a single reference SMC run using a relative ESS threshold of \(0.8\). At each stage, the next tempering parameter is chosen so that the relative ESS of the incremental importance weights equals the prescribed threshold, unless the terminal value $\tp=1$ is reached. This schedule is then held fixed across the three algorithms, all particle counts, and all independent repetitions.
%
We set \(\nstepc[t]=0\) in Algorithm \ref{alg:pmsmc_specific} at every stage. In each configuration, we mutate the particles using $\nstepe[t]=2d$ transitions of the PMMH kernel (see Section \ref{sec:pm_smc}). At each mutation stage, the Gaussian random-walk proposal covariance is $\frac{2.38^2}{d}\left(C+10^{-8}I_d\right)$, where \(C\) is the empirical covariance of the resampled particles. The proposal covariance is recalibrated between mutation stages.

\begin{figure}[!t]
    \centering
    \includegraphics[width=0.66\textwidth]
        {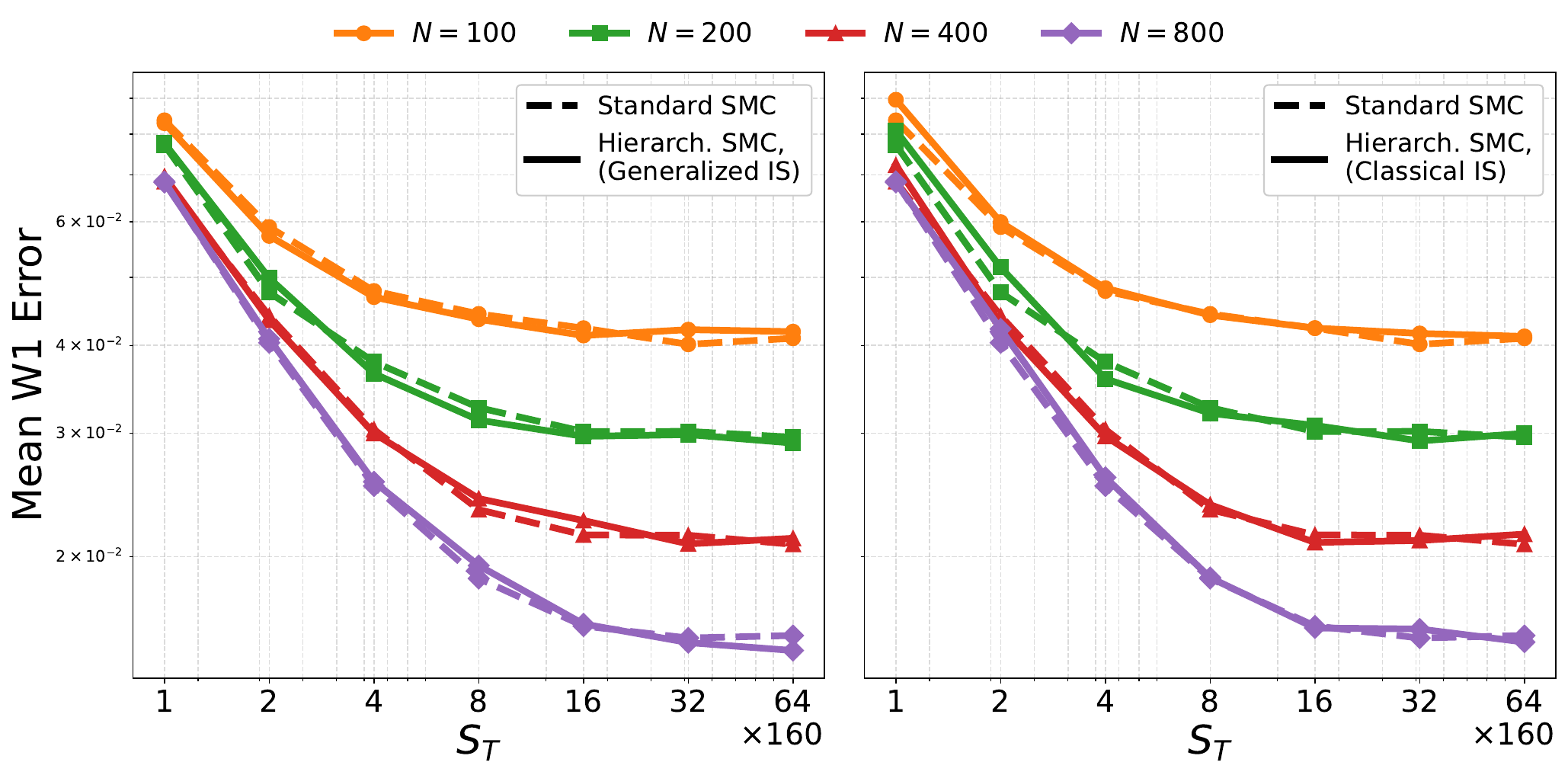}
    \hfill
    \includegraphics[width=0.32\textwidth]
        {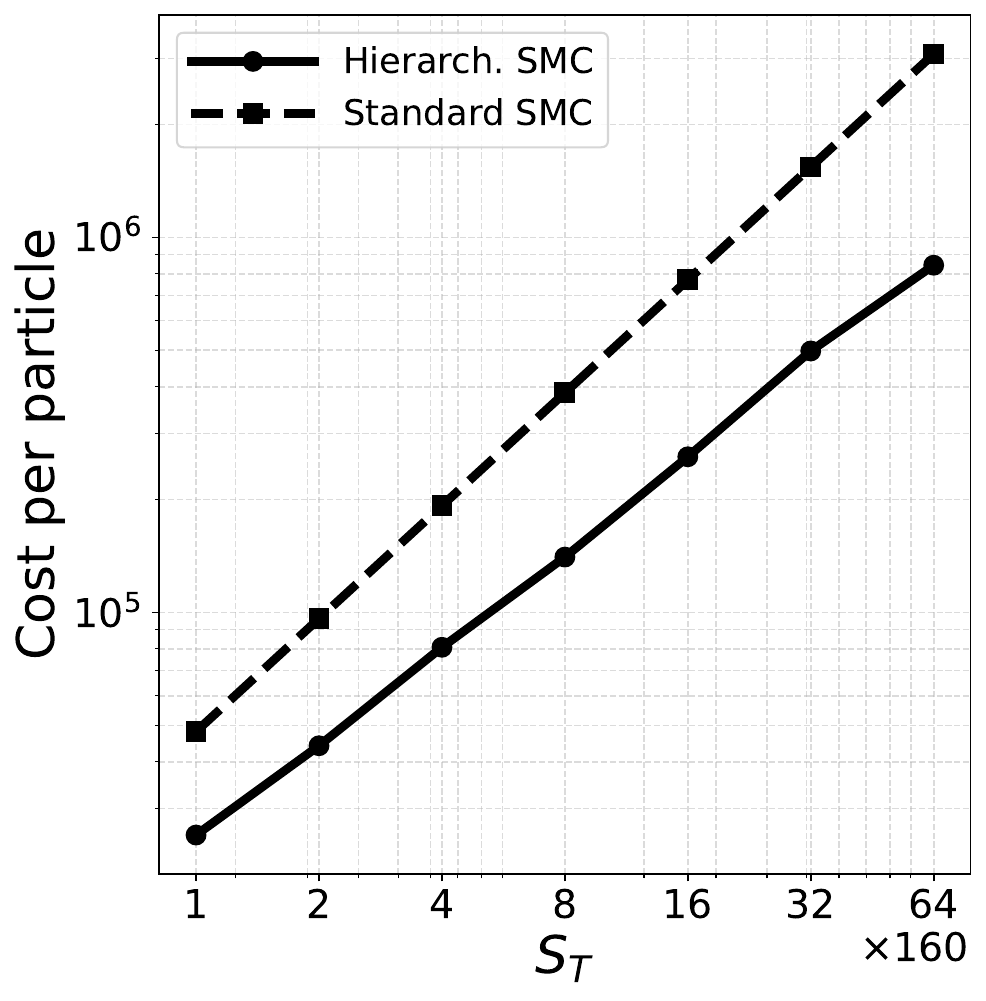}


    \includegraphics[width=0.66\textwidth]
        {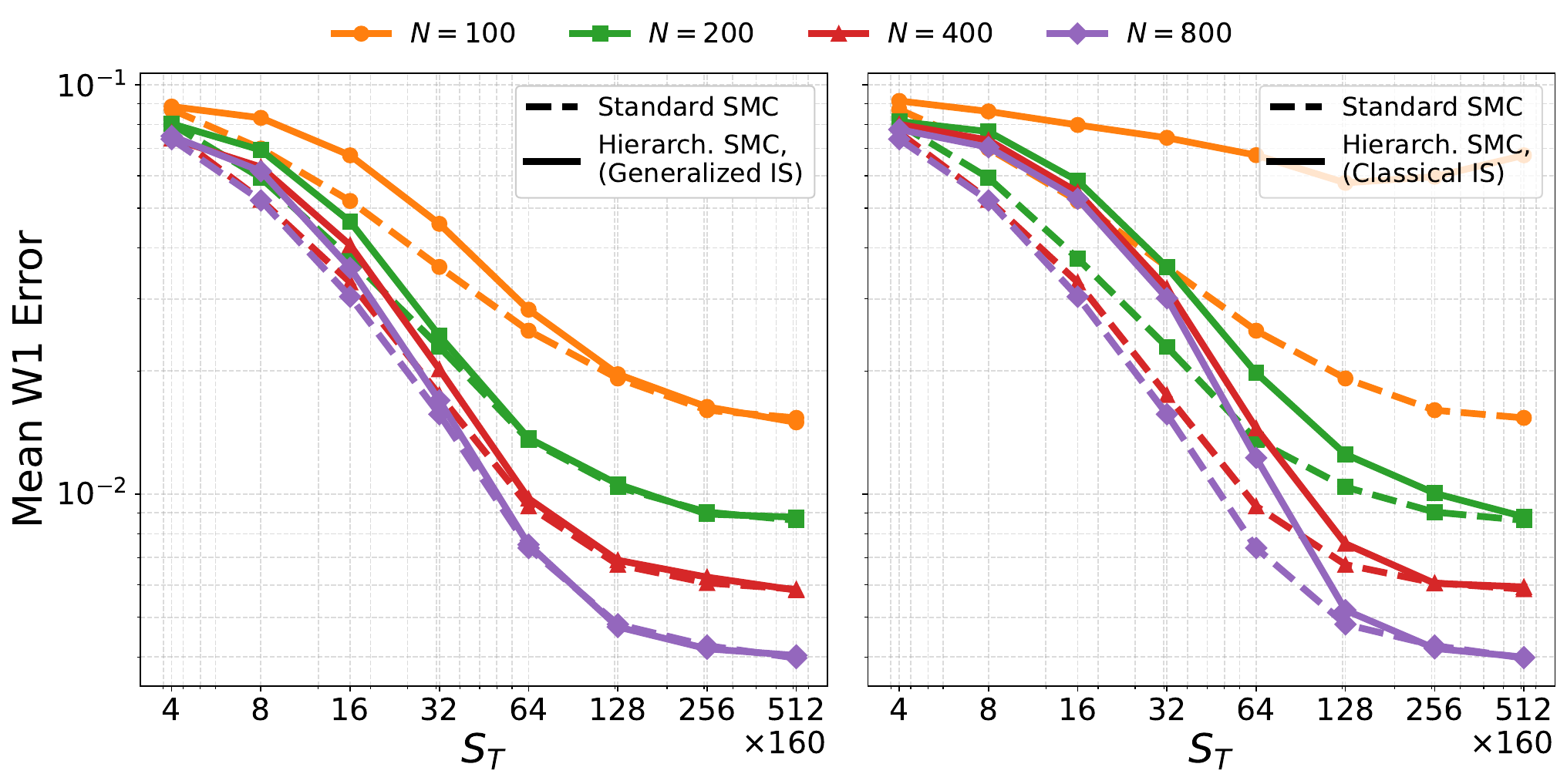}
    \hfill
    \includegraphics[width=0.32\textwidth]
        {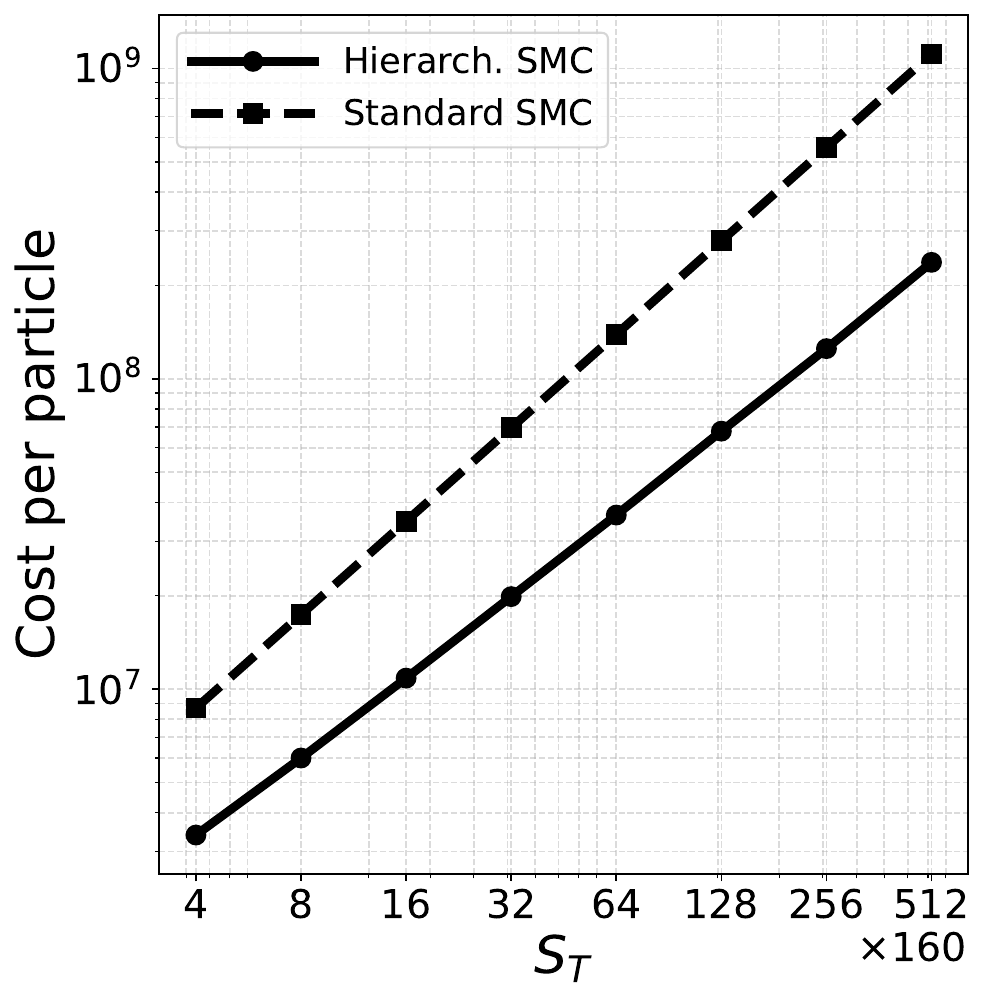}

    \vspace{-0.5em}

    \caption{Mean marginal \(W_1\) error and computational cost per particle
    for \(d=8\) (top row) and \(d=64\) (bottom row) as a function of the final solver accuracy \(\nsamp_T\). In each row, the left and middle panels compare standard SMC (dashed lines) with hierarchical SMC using \texttt{Generalized} IS and \texttt{Classical} IS (solid lines), respectively. The right-hand panels compare the corresponding costs of standard and hierarchical SMC; here ``Cost per particle = Total cost / $\npart$''.}
    \label{fig:gaussian_prior_experiment}
\end{figure}

The results are averaged over \(n_{\mathrm{run}}=100\) and \(n_{\mathrm{run}}=200\) runs for $d=8$ and $d=64$, respectively. Figure~\ref{fig:gaussian_prior_experiment} reports the mean of this quantity over the independent runs. For \(d=8\), all three algorithms produce closely matching errors over the full range of setups. We observe that for small values $\nsamp_T$ the low solver accuracy is the dominant source of error. This is due to the bias of the target posterior $\margmeas{\nsamp_T}$ with respect to the true posterior $\post$ and due to the sticking behavior of PMMH, which makes it difficult to mutate particles sufficiently for small values of $\nsamp_T$. In the region of $\nsamp_T$ where the Monte Carlo error of $\npart$ starts to dominate, the hierarchical algorithm achieves cost savings of a factor of approximately $3$.

For \(d=64\), the hierarchical algorithm with \texttt{Generalized} IS again achieves 
closely matching errors with the standard algorithm, with the maximum relative 
discrepancy being about $10\%$. The hierarchical algorithm with \texttt{Classical} IS 
breaks down for $\npart=100$. This is due to an importance weight 
collapse when the IS step 
$\nsamp_t\rightarrow\nsamp_{t+1}$, $\nsamp_{t+1}>\nsamp_t$, is performed, suggesting 
that the \texttt{Generalized} IS should be preferred even if the 
\texttt{Classical} IS can be implemented. In the regime where the Monte 
Carlo error of $\npart$ starts to dominate (large $\nsamp_T$), the hierarchical 
algorithm achieves cost savings of a factor of approximately $5$ while achieving 
the same accuracy. It is clear that the cost savings grow as $\nsamp_T$ increases 
and asymptotically we can achieve higher savings. However, in practice one aims 
to balance the error contributions of $\nsamp_T$ and $\npart$.

We now investigate the additional cost reduction
obtained by using a cheap approximate chain before the expensive chain in the mutation stage (i.e., additionally $\nstepc>0$ in Algorithm~\ref{alg:pmsmc_specific}). We refer to the case when $\nstepc>0$ as \emph{two-stage}. We use the same \(d=m=64\) linear inverse problem as above. Motivated by the mutation-budget analysis in Theorem \ref{th:budget_opt}, we investigate further cost reductions when total solver cost is constrained, and examine how these gains change with a larger budget. We use the solver refinement scheme \eqref{eq:lin_solver_schedule}, with \(\nsamp_0=10\) and final solver accuracy \(\nsamp_T=160\) for \(T=24,37\), and \(\nsamp_T=1280\) for \(T=50\).

In this experiment, we consider three fixed tempering schedules. Each is generated in a reference SMC run with \(1000\) particles and \(1000\) mutation transitions per stage by choosing successive tempering parameters to meet a prescribed relative ESS threshold (see \cite[Section 17.2.3]{chopin2020introduction}). The thresholds \(0.05\), \(0.2\), and \(0.4\) lead to \(T=24\), \(37\), and \(50\), respectively. Each schedule is then held fixed across all mutation strategies, particle counts, budgets, and independent repetitions.

The mutation kernels are constructed as follows. The expensive kernel consists of $\nstepe$ transitions of PMMH, each transition using accuracy $\nsamp_t$ in the likelihood evaluation. As the cheap kernel we use $\nstepc$ transitions of MCWM (see Section \ref{sec:pm_smc}). At stage \(t\) the cheap MCWM accuracy is
\(\nsamp_{c,t}=\nsamp_t/20\).
In contrast to the preceding experiment, we use the isotropic Gaussian
random-walk proposal $\mathcal{N}\big(u,\frac{2.38^2}{d}I_d \big)$ for both the PMMH and MCWM kernels. The empirical-covariance proposal
used above performs poorly in the present setting, since the shorter
tempering schedules and more concentrated importance weights lead to
unreliable covariance estimates.

To determine the chain lengths $\nstepc$ and $\nstepe$ in the two-stage method, we fix a mutation budget \(\km\) per particle and tempering stage, in units of one PMMH transition at the current solver accuracy. One MCWM transition requires two stochastic solver evaluations, whereas one PMMH transition requires one new evaluation. Since \(\nsamp_{c,t}=\nsamp_t/20\), one MCWM transition is $10\times$ cheaper than one PMMH transition in each mutation stage. Theorem \ref{th:budget_opt} tells us that, given an optimal number of cheap transitions $\nstepc^\star$, all remaining resources should be spent on the expensive chain. We follow this allocation principle using a prescribed cap \(\nstepc^{\max}\): we take \(\nstepc=\min\{\nstepc^{\max},\lfloor10\km\rfloor\}\) and \(\nstepe=\lfloor\km-\nstepc/10\rfloor\). If no budget is left, then $\nstepe=0$. The same transition counts are used at every tempering stage $t$.

\begin{figure}[!t]
    \centering
    \begingroup
    \setlength{\fboxrule}{0.3bp}
    \setlength{\fboxsep}{2bp}
    \newcommand{\mutationcolumn}[1]{%
        \fcolorbox{gray!45}{white}{%
            \vbox{%
                \hbox{\includegraphics{img/two_stage_total_cost_comparison_T#1_cap100.pdf}}%
                \kern2bp
                \hbox{\includegraphics{img/two_stage_total_cost_comparison_T#1_cap1000.pdf}}%
            }%
        }%
    }
    \resizebox{\textwidth}{!}{%
        \begin{tabular}{@{}c@{\hspace{3bp}}c@{}}
            & \includegraphics{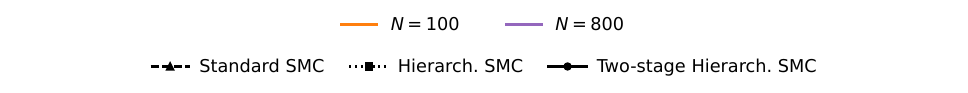} \\[2bp]
            \raisebox{-0.5\height}{\includegraphics{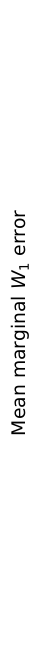}}
            & \raisebox{-0.5\height}{%
                \hbox{%
                    \mutationcolumn{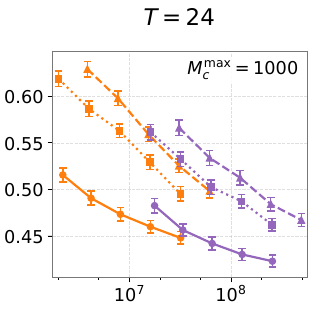}\hspace{8bp}%
                    \mutationcolumn{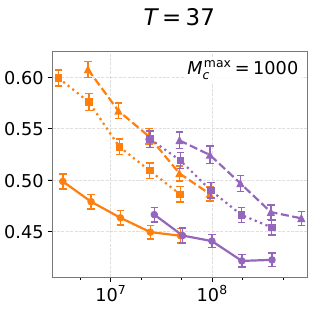}\hspace{8bp}%
                    \mutationcolumn{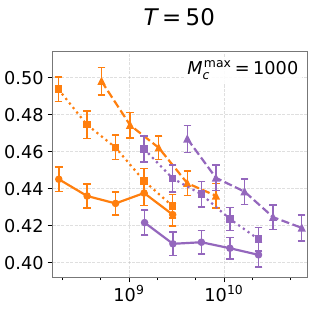}%
                }%
            } \\[2bp]
            & \includegraphics{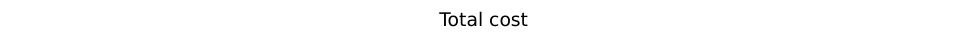}
        \end{tabular}%
    }
    \endgroup
    \caption{Mean marginal \(W_1\) error with approximate $95\%$ confidence intervals versus total cost for one SMC run. Columns correspond to 
    \(T=24,37,50\) tempering stages; the top and bottom rows use cheap-chain caps \(\nstepc^{\max}=100\) and \(1000\), respectively. Curves compare standard 
    SMC, hierarchical SMC without mutation acceleration, and two-stage hierarchical SMC for \(\npart\in\{100,800\}\); both hierarchical methods use 
    \texttt{Generalized} IS. Points along each curve correspond to increasing mutation budgets \(\km\). Results are averaged over \(400\) independent runs.}
    \label{fig:mutation-acceleration}
\end{figure}

We compare caps \(\nstepc^{\max}\in\{100,1000\}\), using mutation budgets \(\km\in\{10,20,40,80,160\}\) for \(T=24,37\) and \(\km\in\{80,160,320,640,1280\}\) for \(T=50\). Standard SMC and hierarchical SMC without mutation acceleration allocate the entire mutation budget to PMMH, with \(\nstepc=0\) and \(\nstepe=\km\). Figure~\ref{fig:mutation-acceleration} displays \(\npart\in\{100,800\}\); each curve is obtained by varying \(\km\) at a fixed particle count. The caps are prescribed choices, rather than estimates of the optimizer \(\nstepc^\star\) in Theorem~\ref{th:budget_opt}.

The results are averaged over \(400\) independent runs. 
Figure~\ref{fig:mutation-acceleration} shows the error versus total cost 
for \(\npart\in\{100,800\}\) and three algorithms: standard SMC 
(Algorithm \ref{alg:pmsmc_specific} with $\nsamp_t=\nsamp_T$ and $\nstepc=0$), 
hierarchical (Algorithms \ref{alg:gen_is}(B) and 
\ref{alg:pmsmc_specific} with $\nstepc=0$), and two-stage hierarchical (Algorithms \ref{alg:gen_is}(B) and \ref{alg:pmsmc_specific} 
with $\nstepc>0$). At a fixed total cost, hierarchical SMC reduces the error compared with standard SMC across all three tempering schedules. Adding two-stage mutation offers substantial additional error reduction at small budgets for \(T=24,37\). For \(T=50\), hierarchical tempering remains beneficial, while two-stage mutation with \(\nstepc^{\max}=100\) brings little additional improvement. Increasing the cap to \(\nstepc^{\max}=1000\) yields clear additional gains also in this regime.

Overall, hierarchical tempering with \texttt{Generalized} IS consistently reduces cost across the tested regimes. The additional gains from two-stage mutation depend on the mutation budget and the allocation to the cheap chain.

\subsection{A nonlinear application in quantitative photoacoustic imaging}\label{sec:nonlinear_problem}

Finally, we consider a nonlinear inverse problem arising from quantitative
photoacoustic imaging. In contrast to the toy problem of
Section~\ref{sec:toy_problem}, the stochastic forward solver is now a
Monte Carlo model for light transport in heterogeneous tissue. The inverse
problem is to infer a depth-dependent optical absorption profile from sparse
measurements of the resulting initial acoustic pressure. This provides a high-contrast medical-imaging example with a substantially
more expensive and nonlinear forward model. Due to the high number of simulated photons required for acceptable accuracies in each kernel transition, this is an extremely challenging inverse problem. We demonstrate that adapting the
solver accuracy during SMC can significantly reduce the computational cost while retaining posterior approximation accuracy.

Let \(\mathcal D\subset\R^3\) denote the domain,
\(r\in\mathcal D\) the spatial variable, and
\(\omega\in\mathbb S^2\) a propagation direction. We denote the optical
absorption and scattering coefficients by
\(\sigma_{\mathrm a}(r)\) and \(\sigma_{\mathrm s}(r)\), respectively.
Following the standard light-transport formulation used in quantitative
photoacoustic tomography~\cite{hanninen2023image}, the propagation of light
in a scattering medium is described by the radiative transfer equation. For
the time-integrated optical quantities considered here, we use its stationary
form
\begin{equation}\label{eq:rte-nonlinear-example}
    \omega\cdot\nabla I_{\mathrm{rad}}(r,\omega) +
    \bigl(
        \sigma_{\mathrm a}(r)
        +
        \sigma_{\mathrm s}(r)
    \bigr)
    I_{\mathrm{rad}}(r,\omega) =
    \sigma_{\mathrm s}(r)
    \int_{\mathbb S^2}
        \Theta_g(\omega\cdot\omega')
        I_{\mathrm{rad}}(r,\omega')
        \,\td\omega',
\end{equation}
where \(I_{\mathrm{rad}}\) denotes the radiance and
\(\Theta_g\) is the scattering phase function with anisotropy parameter
\(g\). The illumination is prescribed through the incoming boundary
condition
\[
    I_{\mathrm{rad}}(r,\omega)
    =
    \begin{cases}
        I_{\mathrm{in}}(r,\omega),
        & r\in\mathcal S,\ \omega\cdot n(r)<0,\\
        0,
        & r\in\partial\mathcal D\setminus\mathcal S,\
          \omega\cdot n(r)<0,
    \end{cases}
\]
where \(\mathcal S\subset\partial\mathcal D\) denotes the source region
and \(n(r)\) is the outward unit normal.

The photon fluence and absorbed optical energy density are
\[
    F_{\mathrm{opt}}(r)
    =
    \int_{\mathbb S^2}
        I_{\mathrm{rad}}(r,\omega)\,\td\omega,
    \qquad
    H(r)
    =
    \sigma_{\mathrm a}(r)F_{\mathrm{opt}}(r).
\]
The absorbed energy determines the initial acoustic pressure through
\begin{equation}\label{eq:initial-pressure-nonlinear-example}
    p_0(r)
    =
    \Upsilon(r)H(r)
    =
    \Upsilon(r)\sigma_{\mathrm a}(r)F_{\mathrm{opt}}(r),
\end{equation}
where \(\Upsilon\) denotes the Gr\"uneisen parameter. Since
\(F_{\mathrm{opt}}\) itself depends on \(\sigma_{\mathrm a}\)
through~\eqref{eq:rte-nonlinear-example}, the map from the absorption
coefficient to the initial pressure is nonlinear.

\begin{figure}[!t]
    \centering
    \includegraphics[width=\textwidth]
        {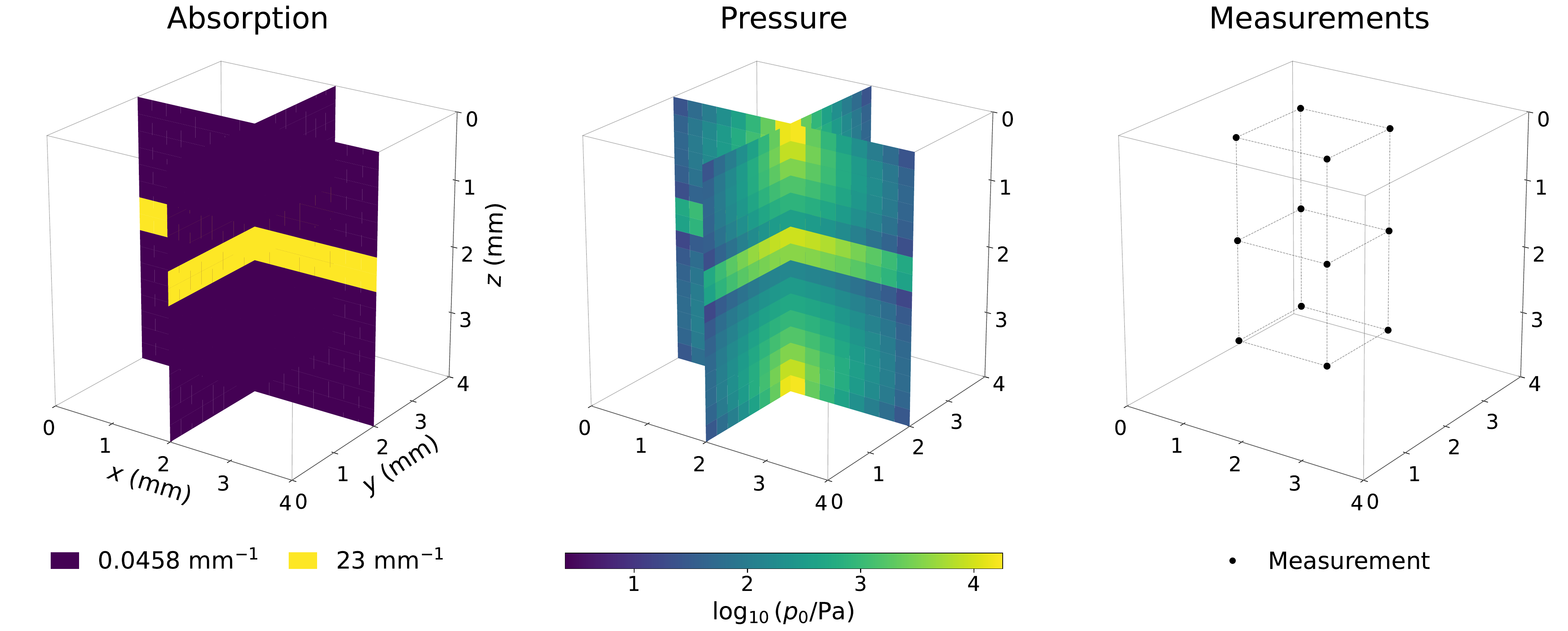}
    \caption{Reference nonlinear inverse problem. Left: ground-truth
    absorption field. Center: reference initial pressure generated by
    MCX using \(10^{11}\) photons for each of the two sources. Right:
    the twelve pressure-observation locations used to construct the data.}
    \label{fig:three-layer-reference}
\end{figure}

We consider light propagation in a layered tissue geometry; such geometries are commonly used as simplified models of biological tissue in biomedical optics \cite{wang1995mcml,zhu2013review}. The computational domain is a
\(4\,\mathrm{mm}\times4\,\mathrm{mm}\times4\,\mathrm{mm}\) cube,
discretized into a \(16\times16\times16\) voxel grid. We infer an
absorption coefficient that varies only in the depth direction. The
reference profile consists of a thin blood layer surrounded by muscle
tissue, as shown in Figure~\ref{fig:three-layer-reference}. The corresponding
absorption coefficients are
\[
    \sigma_{\mathrm a,\mathrm{muscle}}
    =
    0.0458\,\mathrm{mm}^{-1},
    \qquad
    \sigma_{\mathrm a,\mathrm{blood}}
    =
    23.0\,\mathrm{mm}^{-1},
\]
while the scattering coefficients are
\(35.6\,\mathrm{mm}^{-1}\) and \(9.4\,\mathrm{mm}^{-1}\),
respectively. The scattering coefficients and the remaining optical parameters are fixed throughout this section, with
anisotropy \(g=0.9\) and refractive index \(n=1\), and we only aim to recover the absorption coefficient $\sigma_{\mathrm a}(r)$ from measurements.

Illumination is provided by two pencil sources centered on opposing
\(z\)-faces. Figure~\ref{fig:three-layer-reference} shows the reference
absorption field, the corresponding initial pressure field, and the twelve
pressure-observation locations. 
We approximate light transport using MCX, a GPU-accelerated Monte Carlo
photon-transport simulator for heterogeneous three-dimensional
media~\cite{fang2009monte}. In this simulator, the transport is approximated by tracing
photon packets through successive scattering and absorption events. The
number of simulated photons therefore plays the role of the solver accuracy
\(\nsamp\) in the framework of Section~\ref{sec:bip}: increasing
\(\nsamp\) decreases the stochastic variability of the forward evaluation
but increases its computational cost. The reference pressure in Figure~\ref{fig:three-layer-reference} (middle) is generated using 
\(10^{11}\) MCX photons for each source, corresponding to
\(2\times10^{11}\) simulated photons in total.

Let $u$ denote a parametrization of the absorption $\sigma_{\mathrm a}$. 
In particular, we represent it through its logarithm using a truncated one-dimensional Haar-wavelet expansion, thus incorporating the structural prior information that the absorption only varies in depth. We further restrict this parametrization to the first six Haar coefficients, fixing all higher-order coefficients to zero, and incorporate further prior information through the prior mean. This defines a Gaussian prior on the six-dimensional Haar subspace, which may be viewed as a finite-dimensional truncation of the Gaussian wavelet construction associated with the Besov space \(B^s_{2,2}\), where \(s\) is the smoothness index. Specifically, we use
\[
    u
    \sim
    \mathcal N\left(
            [-7, 1,  -2,  0,  0,  -4]^{\mathsf T},
        \operatorname{diag}(4,4,2,2,1,1)
    \right)
\]
as the prior distribution for the coefficients of the Haar-wavelet expansion, where $\operatorname{diag}(\cdot)$ denotes a diagonal matrix. This implies a lognormal prior distribution on the absorption coefficient $\sigma_{\mathrm a}$. 

Let \(p_0(\,\cdot\,;u)\) denote the initial
pressure obtained from the corresponding absorption profile. The exact
forward map in the notation of Section~\ref{sec:bip} is
\[
    \fwop(u)
    =
    \bigl(
        p_0(r_1;u),
        \ldots,
        p_0(r_{12};u)
    \bigr)^{\mathsf T}
    \in\R^{12},
\]
where \(r_1,\ldots,r_{12}\) are the measurement locations shown in
Figure~\ref{fig:three-layer-reference} (right). An MCX evaluation using a smaller number of photons
provides the corresponding stochastic approximation \(\fwopext\) in our SMC algorithms. The \(2\times10^{11}\)-photon pressure field is treated as the reference
forward map $\fwop$ for the generation of synthetic observations.
The
observation model is
\[
    y
    =
    \fwop(u^\dagger)
    +
    \eta,
    \qquad
    \eta
    \sim
    \mathcal N(0,\Gamma),
    \qquad
    \Gamma
    =
    4000 I_{12}.
\]
The expansion of the reference absorption is
$u^\dagger = [-9.2244, 3.1095, -4.3975, 0, 0, -6.2190]^{\mathsf T}.$

We compare standard SMC with hierarchical SMC using adaptive solver
accuracy. Standard SMC is run once and serves as the reference, while
hierarchical SMC is run independently ten times. Algorithm~\ref{alg:pmsmc_specific} is formulated for an arbitrary tempering sequence. Rather than prescribing this sequence in advance, we determine it adaptively here: at each stage, the next tempering parameter is selected using a relative effective sample size threshold of \(0.7\); see \cite[Section 17.2.3]{chopin2020introduction}. This adaptive procedure resulted in $T=12$ tempering stages for standard SMC and an average of $T=12.9$ stages for hierarchical SMC.

In the notation of
Algorithm~\ref{alg:pmsmc_specific}, all runs use \(\npart=400\) particles
and \(\nstepe[t]=5\) PMMH transitions for all $t$. Accelerated mutation using a cheap kernel
is not employed, i.e., \(\nstepc[t]=0\) for all $t$. In this problem, further reducing the number of photons
in the MCX simulation (within a cheap mutation kernel) does not lead to a comparable reduction in wall-clock time. 
At sufficiently low photon counts, the GPU is no longer
fully utilized, so decreasing the number of photons further provides only
limited computational savings.
Nevertheless, selectively using two-stage mutation in some of the final tempering stages, where photon counts are higher, may yield additional gains if the cheap kernel provides sufficient wall-clock savings.

We set the maximum available number of photons (solver accuracy) to $\nsamp_{\max}=10^9$.
Standard SMC uses this accuracy throughout the complete tempering sequence.
The hierarchical SMC takes the solver accuracies
$\nsamp_0,\ldots,\nsamp_T$ as part of its input.
The initial MCX photon count is set to $\nsamp_0=10^5$ and, rather than fixing the
remaining values in advance, they are selected adaptively during the SMC run.
Whenever the solver accuracy changes, we use the \texttt{Generalized} update of
Algorithm~\ref{alg:gen_is}(B). After each mutation stage, the mean acceptance
rate is compared with the threshold \(0.15\). If it falls below this value, the
photon count used for the subsequent target is multiplied by \(10\), subject
to the upper bound $\nsamp_{\max}$. This adaptive rule therefore only
determines the solver accuracy sequence used in
Algorithm~\ref{alg:pmsmc_specific}; the SMC algorithm itself remains
unchanged. Four of the ten adaptive hierarchical SMC realizations reached the
maximum photon count of $\nsamp_{\max}=10^9$. (Similar adaptation rules have been used
in the context of SMC$^2$; see
\cite{duan2015density,chopin2020introduction}.)

\begin{figure}[!t]
    \centering
    \begin{subfigure}[t]{0.49\textwidth}
        \centering
        \includegraphics[width=\linewidth]
            {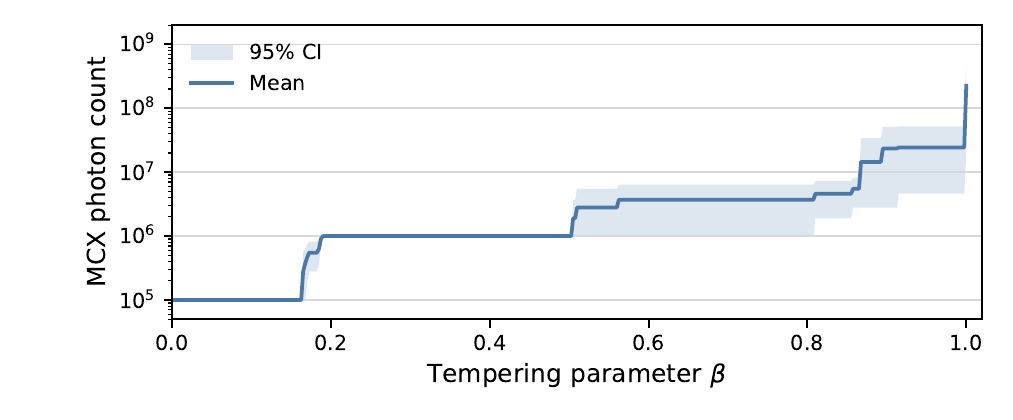}
        \caption{Mean MCX photon 
        count $S_t$ as a function of the tempering parameter (for ten hierarchical SMC
        realizations).}
        \label{fig:three-layer-photon-schedule}
    \end{subfigure}
    \hfill
    \begin{subfigure}[t]{0.49\textwidth}
        \centering
        \includegraphics[width=\linewidth]
            {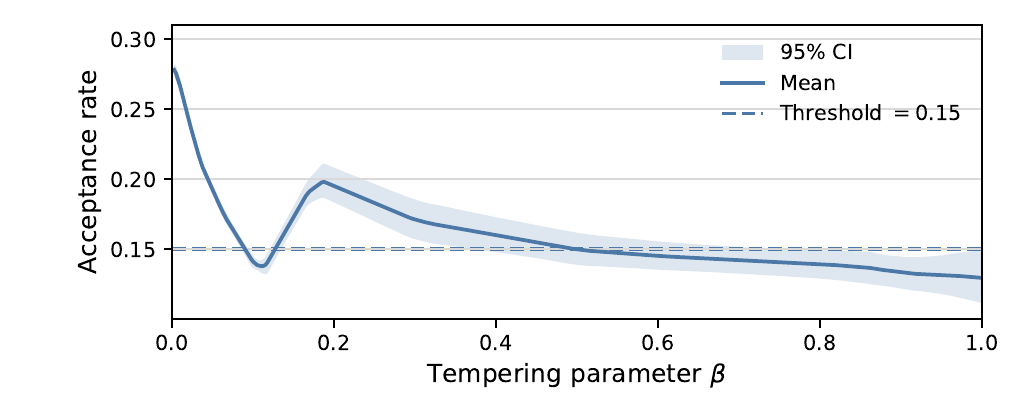}
        \caption{Mean mutation acceptance rate across the ten realizations,
        together with the adaptation threshold \(0.15\).}
        \label{fig:three-layer-acceptance-rate}
    \end{subfigure}
    \caption{Adaptive solver-accuracy diagnostics for hierarchical SMC.
    The individual realizations use different adaptive tempering schedules;
    the results are displayed against a common tempering parameter~$\beta$.}
    \label{fig:three-layer-diagnostics}
\end{figure}

Figure~\ref{fig:three-layer-diagnostics} summarizes the resulting hierarchical SMC
behavior across the ten independent realizations as a function of the
tempering parameter $\beta\in[0,1]$. The solid lines show the arithmetic mean across runs,
while the shaded regions are pointwise \(95\%\) bootstrap confidence
intervals. The mean photon count remains substantially below
\(\nsamp_{\max}\) over most of the tempering sequence and increases as $\beta$ approaches~$1$. The corresponding acceptance-rate summary remains close to the threshold \(0.15\) over most of the tempering sequence. Since the trajectories are aligned in \(\beta\) and averaged across realizations, the individual threshold crossings that trigger solver refinement are smoothed out and are not directly visible in the figure.

\begin{figure}[t]
    \centering
    \begin{subfigure}[t]{0.49\textwidth}
        \centering
        \includegraphics[width=\linewidth]
            {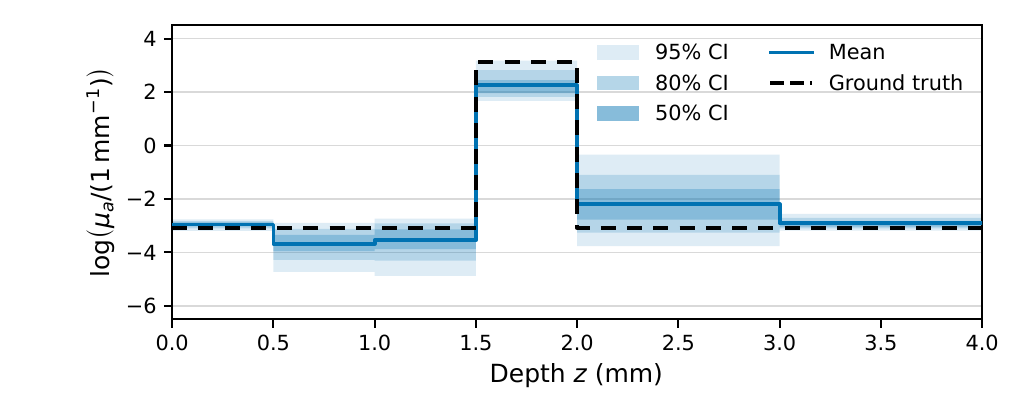}
        \caption{Reconstruction obtained with standard SMC.}
        \label{fig:three-layer-absorption-standard}
    \end{subfigure}
    \hfill
    \begin{subfigure}[t]{0.49\textwidth}
        \centering
        \includegraphics[width=\linewidth]
            {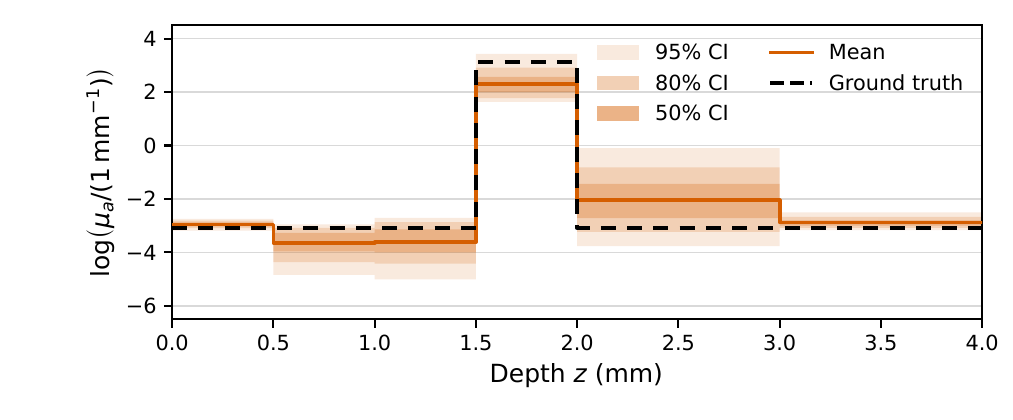}
        \caption{Reconstruction for the ``worst'' hierarchical SMC realization
        with largest mean-marginal \(W_1\) error.}
        \label{fig:three-layer-absorption-hierarchical}
    \end{subfigure}
    \caption{Posterior reconstruction of the log-absorption profile.} 
    \label{fig:three-layer-absorption-reconstruction}
\end{figure}

\begin{figure}[!t]
    \centering
    \begin{subfigure}[t]{0.49\textwidth}
        \centering
        \includegraphics[width=\linewidth]
            {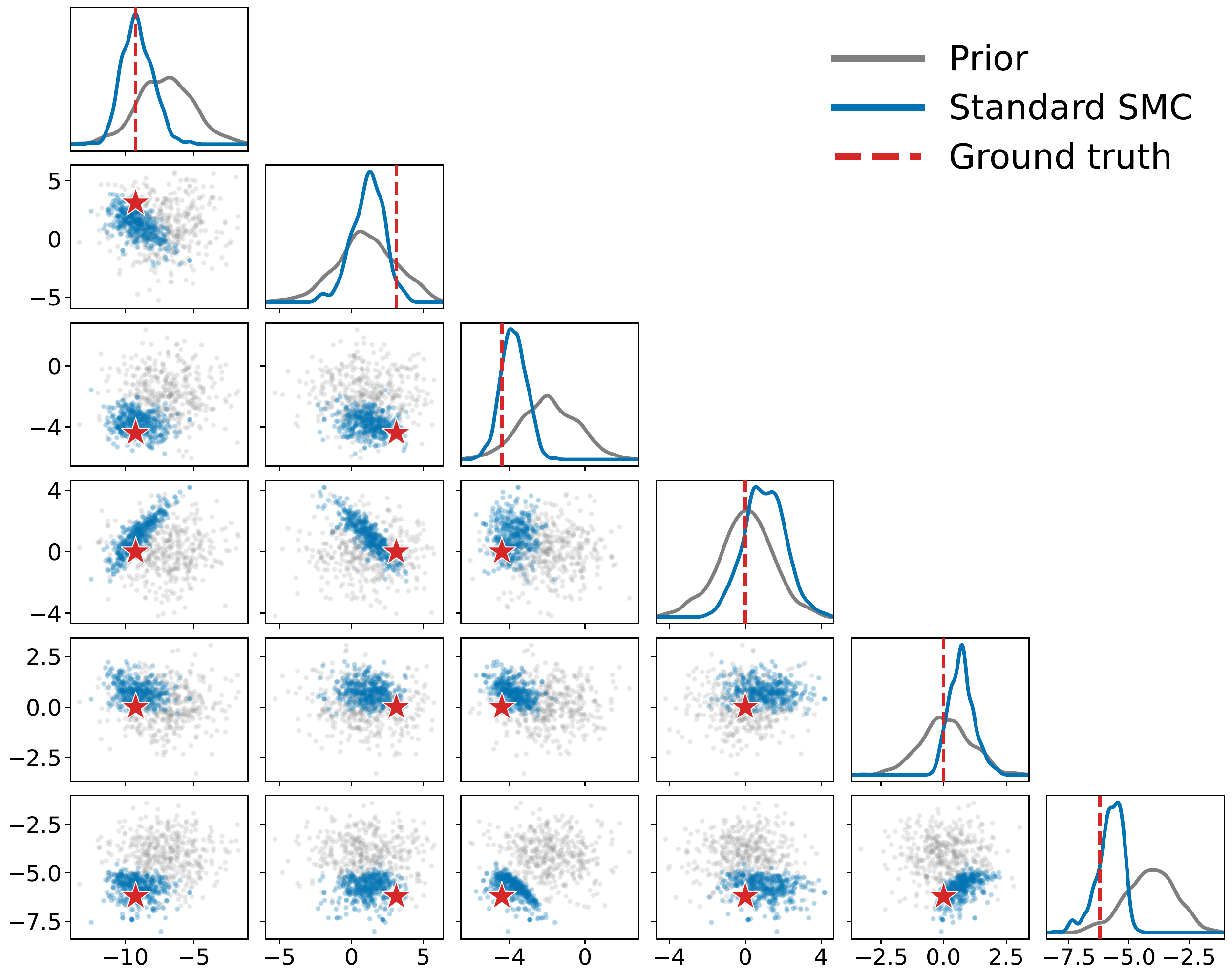}
        \caption{Prior and empirical posterior obtained with standard SMC (reference).}
        \label{fig:three-layer-prior-posterior}
    \end{subfigure}
    \hfill
    \begin{subfigure}[t]{0.49\textwidth}
        \centering
        \includegraphics[width=\linewidth]
            {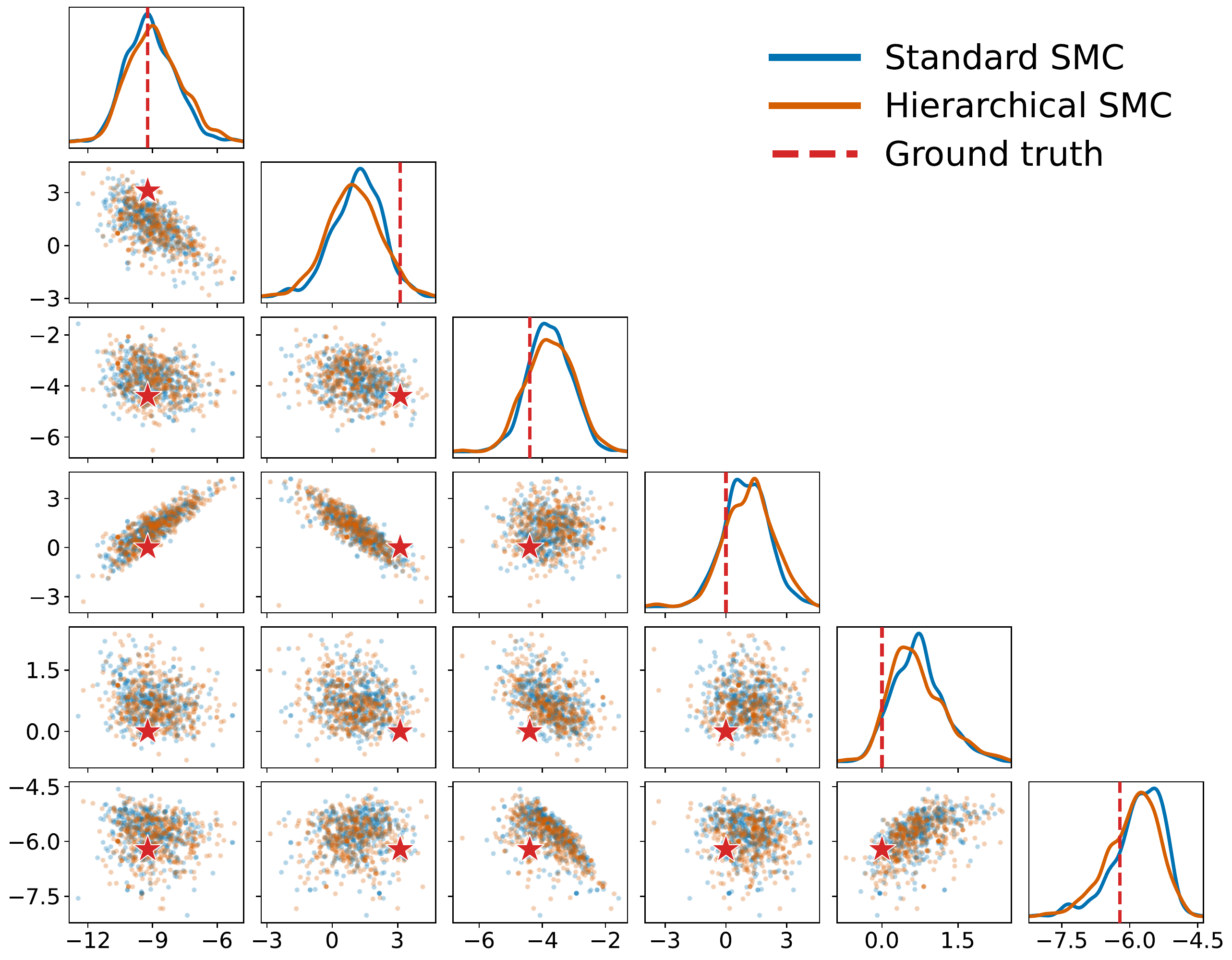}
        \caption{Standard SMC reference posterior and ``worst'' hierarchical SMC
        run (i.e., largest mean-marginal \(W_1\) error).}
        \label{fig:three-layer-posterior-comparison}
    \end{subfigure}
    \caption{Empirical posterior results for the nonlinear photoacoustic medical imaging inverse problem.}
    \label{fig:three-layer-posterior-results}
\end{figure}

The inferred log-absorption profiles are shown in
Figure~\ref{fig:three-layer-absorption-reconstruction}, with posterior summaries
obtained using standard SMC and the worst-case hierarchical SMC realization (attaining the
largest mean-marginal \(W_1\) error among the ten runs). We observe that the reconstruction
is best close to the domain boundaries, where the photon sources are located.
Figure~\ref{fig:three-layer-prior-posterior} illustrates the amount of
information gained from the observations. The diagonal panels compare the
marginal prior and reference-posterior distributions of the six Haar
coefficients, while the off-diagonal panels show their pairwise dependence.

We compare the adaptive posterior approximations with the standard-SMC
reference using two error metrics. As in
Section~\ref{sec:toy_problem}, we compute the empirical one-dimensional
\(W_1\) distance marginally in each coefficient and average over the six
coordinates. In addition, we consider the relative root mean squared error
on the reference posterior mean. Across the ten adaptive realizations, the mean of the
mean-marginal \(W_1\) errors is \(0.1016\), with maximum \(0.1413\).
The corresponding relative root mean squared errors have mean \(0.0177\)
and maximum \(0.0282\), so that the largest observed discrepancy in
the posterior mean is approximately \(2.8\%\) relative to the reference
posterior mean.
Figure~\ref{fig:three-layer-posterior-comparison} provides a
visual comparison by showing the adaptive realization with the largest
mean-marginal \(W_1\) error. Despite representing the worst-case comparison, the results remain visually close to those of the standard-SMC reference.

Relative to standard SMC using $10^9$ photons throughout, the hierarchical
algorithm achieved a speedup by a factor of $14.3$, reducing the mean inference wall-clock
time from about $4.6$ days to $7.7$ hours. The speedup is computed as the wall-clock runtime of the standard SMC realization divided by the mean wall-clock runtime across the ten hierarchical SMC realizations.
Given the limited number of realizations, we additionally report a speedup by a factor
of $6.4$ for the four realizations that reached the maximum number of photons
$\nsamp_{\max}=10^9$. All simulations were performed on an NVIDIA A40 GPU.

Recall that the standard SMC algorithm required only $T=12$ tempering stages.
We expect even larger computational gains for more challenging problems
requiring longer tempering sequences. Tempering parameters
typically increase approximately geometrically
\cite[Section 17.2.3]{chopin2020introduction}, resulting in many stages
at small tempering parameters, where hierarchical SMC offers the
greatest savings. This expectation is further supported by the
speedup proportional to $T$ established in
Theorem~\ref{cor:T_cost_gain}.

Overall, the experiment shows that the hierarchical algorithm can
substantially reduce the computational cost of SMC for a challenging nonlinear
inference problem while preserving accuracy.

\section{Conclusions}\label{sec:conclusion}

We developed a hierarchical SMC framework for Bayesian inverse problems with stochastic forward solvers, exploiting a hierarchy of solver accuracies in two ways. First, increasing solver accuracy along the tempering sequence yields a provable asymptotic cost reduction, with the speedup scaling proportionally to the number of tempering stages under the assumptions of Section~\ref{sec:acc_is}. Second, within each mutation stage, cheap approximate and expensive target-invariant kernels can be combined by allocating a fixed mutation budget between them, as analyzed in Section~\ref{sec:mutation}. The numerical experiments show consistent cost savings from hierarchical tempering with \texttt{Generalized} IS across the tested regimes. Two-stage mutation offers substantial additional savings in several regimes, but these depend on the mutation budget and the allocation to the cheap chain. In particular, on the nonlinear photoacoustic imaging problem, adaptive hierarchical SMC produced posterior approximations close to the standard-SMC reference while reducing the runtime from approximately \(4.6\) days to \(7.7\) hours, corresponding to a speedup of~\(14.3\).

Several extensions remain open. A principal theoretical direction is a joint complexity analysis of Algorithm~\ref{alg:pmsmc_specific} when both acceleration mechanisms are used simultaneously. More general multilevel constructions could employ a hierarchy of solver accuracies within one mutation stage, or combine the present framework with deterministic discretization or learned surrogates. The hierarchical methodology may also be useful beyond the present setting, including sequential inference in state-space models and SMC methods based on different likelihood models and approximations.

\subsection*{Acknowledgements} This work is supported by the Carl Zeiss-Stiftung through the project “Model-Based AI: Physical Models and Deep Learning for Imaging and Cancer Treatment” and by the Deutsche Forschungsgemeinschaft (German Research Foundation) under Germany’s Excellence Strategy EXC 2181/1 - 390900948 (the Heidelberg STRUCTURES Excellence Cluster). 
The authors acknowledge support by the state of Baden-Württemberg through bwHPC and the German Research Foundation through grant INST 35/1597-1 FUGG.

AMGC and JH are supported by the EPSRC programme grant EP/W026899/2 - MaThRad.
In addition, JM acknowledges financial support from the EPSRC Centre for Doctoral Training, SAMBa EP/S022945/1 and EPSRC Programme Grant MaThRad EP/W026899/2 for an extended visit at the University of Bath.

\printbibliography

@article{roberts1997geometric,
author = {G. O. Roberts and J. S. Rosenthal},
title = {Geometric ergodicity and hybrid {Markov} chains},
volume = {2},
journal = {Electronic Communications in Probability},
publisher = {Institute of Mathematical Statistics and Bernoulli Society},
pages = {13 -- 25},
year = {1997},
doi = {10.1214/ECP.v2-981},
OPTURL = {https://doi.org/10.1214/ECP.v2-981}
}

@book{dudley2002real, 
place={Cambridge},
 edition={2},
 series={Cambridge Studies in Advanced Mathematics}, 
title={Real Analysis and Probability}, 
publisher={Cambridge University Press}, 
author={Dudley, R. M.}, year={2002}, 
collection={Cambridge Studies in Advanced Mathematics},
  doi={10.1017/CBO9780511755347}
}

@article{gallegos2024equivalences,
  title={Equivalences of geometric ergodicity of Markov chains},
  author={Gallegos-Herrada, Marco A and Ledvinka, David and Rosenthal, Jeffrey S},
  journal={Journal of Theoretical Probability},
  volume={37},
  number={2},
  pages={1230--1256},
  year={2024},
  publisher={Springer},
  doi={10.1007/s10959-023-01240-1}
}

@article{rudolf2018perturbation,
author = {D. Rudolf and N. Schweizer},
title = {{Perturbation theory for Markov chains via Wasserstein distance}},
volume = {24},
journal = {Bernoulli},
number = {4A},
publisher = {Bernoulli Society for Mathematical Statistics and Probability},
pages = {2610 -- 2639},
year = {2018},
doi = {10.3150/17-BEJ938},
OPTURL = {https://doi.org/10.3150/17-BEJ938}
}

@inproceedings{hairer2011yet,
  title={Yet another look at Harris’ ergodic theorem for Markov chains},
  author={Hairer, Martin and Mattingly, Jonathan C},
  booktitle={Seminar on Stochastic Analysis, Random Fields and Applications VI: Centro Stefano Franscini, Ascona, May 2008},
  pages={109--117},
  year={2011},
  organization={Springer},
  doi={10.1007/978-3-0348-0021-1_7}
}

@article{eberle2016reflection,
  title={Reflection couplings and contraction rates for diffusions},
  author={Eberle, Andreas},
  journal={Probability Theory and Related Fields},
  volume={166},
  number={3},
  pages={851--886},
  year={2016},
  publisher={Springer},
  doi={10.1007/s00440-015-0673-1}
}

@book{villani2008optimal,
  title={Optimal Transport: Old and New},
  author={Villani, C{\'e}dric},
  OPTvolume={338},
  year={2008},
  publisher={Springer},
  doi={10.1007/978-3-540-71050-9}
}

@article{garegnani2021sampling,
  title={Sampling methods for Bayesian inference involving convergent noisy approximations of forward maps},
  author={Garegnani, Giacomo},
  journal={arXiv preprint},
  volume={arXiv:2111.03491},
  year={2021},
  doi={10.48550/arXiv.2111.03491}
}

@article{andrieu2015convergence,
  title={Convergence properties of pseudo-marginal Markov chain Monte Carlo algorithms},
  author={Andrieu, Christophe and Vihola, Matti},
  journal={The Annals of Applied Probability},
  volume={25},
  number={2},
  pages={1030--1077},
  year={2015},
  doi={10.1214/14-aap1022}
}

@article{lykkegaard2023multilevel,
  title={Multilevel delayed acceptance MCMC},
  author={Lykkegaard, Mikkel Bue and Dodwell, Tim J and Fox, Colin and Mingas, Grigorios and Scheichl, Robert},
  journal={SIAM/ASA Journal on Uncertainty Quantification},
  volume={11},
  number={1},
  pages={1--30},
  year={2023},
  publisher={SIAM},
  doi={10.1137/22m1476770}
}

@article{latz2018multilevel,
  title={Multilevel Sequential$^2$ Monte Carlo for Bayesian inverse problems},
  author={Latz, Jonas and Papaioannou, Iason and Ullmann, Elisabeth},
  journal={Journal of Computational Physics},
  volume={368},
  pages={154--178},
  year={2018},
  publisher={Elsevier},
  doi={10.1016/j.jcp.2018.04.014}
}

@book{chopin2020introduction,
  title={An Introduction to Sequential Monte Carlo},
  author={Chopin, Nicolas and Papaspiliopoulos, Omiros},
  OPTvolume={4},
  year={2020},
  publisher={Springer},
  doi={10.1007/978-3-030-47845-2}
}

@article{beskos2017multilevel,
  title={Multilevel sequential Monte Carlo samplers},
  author={Beskos, Alexandros and Jasra, Ajay and Law, Kody and Tempone, Raul and Zhou, Yan},
  journal={Stochastic Processes and their Applications},
  volume={127},
  number={5},
  pages={1417--1440},
  year={2017},
  publisher={Elsevier},
  doi={10.1016/j.spa.2016.08.004}
}

@article{jasra2017multilevel,
  title={Multilevel particle filters},
  author={Jasra, Ajay and Kamatani, Kengo and Law, Kody JH and Zhou, Yan},
  journal={SIAM Journal on Numerical Analysis},
  volume={55},
  number={6},
  pages={3068--3096},
  year={2017},
  publisher={SIAM},
  doi={10.1137/17m1111553}
}

@article{medina2016stability,
  title={Stability of noisy {M}etropolis--{H}astings},
  author={Medina-Aguayo, Felipe J and Lee, Anthony and Roberts, Gareth O},
  journal={Statistics and Computing},
  volume={26},
  pages={1187--1211},
  year={2016},
  publisher={Springer},
  doi={10.1007/s11222-015-9604-3}
}

@article{duan2015density,
  title={Density-tempered marginalized sequential Monte Carlo samplers},
  author={Duan, Jin-Chuan and Fulop, Andras},
  journal={Journal of Business \& Economic Statistics},
  volume={33},
  number={2},
  pages={192--202},
  year={2015},
  publisher={Taylor \& Francis},
  doi={10.1080/07350015.2014.940081}
}

@article{stuart2010inverse,
  title={Inverse problems: a Bayesian perspective},
  author={Stuart, Andrew M},
  journal={Acta Numerica},
  volume={19},
  pages={451--559},
  year={2010},
  publisher={Cambridge University Press},
  doi={10.1017/s0962492910000061}
}

@article{lie2018random,
  title={Random forward models and log-likelihoods in Bayesian inverse problems},
  author={Lie, Han Cheng and Sullivan, Timothy John and Teckentrup, Aretha L},
  journal={SIAM/ASA Journal on Uncertainty Quantification},
  volume={6},
  number={4},
  pages={1600--1629},
  year={2018},
  publisher={SIAM},
  doi={10.1137/18m1166523}
}

@article{andrieu2009pseudo,
  title={The pseudo-marginal approach for efficient Monte Carlo computations},
  author={Andrieu, Christophe and Roberts, Gareth O.},
  journal={The Annals of Statistics},
  volume={37},
  number={2},
  pages={697--725},
  year={2009},
  doi={10.1214/07-AOS574}
}

@article{beaumont2003estimation,
  title={Estimation of population growth or decline in genetically monitored populations},
  author={Beaumont, Mark A},
  journal={Genetics},
  volume={164},
  number={3},
  pages={1139--1160},
  year={2003},
  publisher={Oxford University Press},
  doi={10.1093/genetics/164.3.1139}
}

@article{del2006sequential,
  title={Sequential Monte Carlo samplers},
  author={Del Moral, Pierre and Doucet, Arnaud and Jasra, Ajay},
  journal={Journal of the Royal Statistical Society Series B: Statistical Methodology},
  volume={68},
  number={3},
  pages={411--436},
  year={2006},
  publisher={Oxford University Press},
  doi={10.1111/j.1467-9868.2006.00553.x}
}

@article{botha2023automatically,
  title={Automatically adapting the number of state particles in {SMC}$^2$},
  author={Botha, Imke and Kohn, Robert and South, Leah and Drovandi, Christopher},
  journal={Statistics and Computing},
  volume={33},
  number={4},
  pages={82},
  year={2023},
  doi={10.1007/s11222-023-10250-2}
}

@article{rohrbach2022multilevel,
  title={Multilevel simulation of hard-sphere mixtures},
  author={Rohrbach, Paul B and Kobayashi, Hideki and Scheichl, Robert and Wilding, Nigel B and Jack, Robert L},
  journal={The Journal of Chemical Physics},
  volume={157},
  number={12},
  year={2022},
  publisher={AIP Publishing},
  doi={10.1063/5.0102875}
}

@book{meyn2012markov,
  title={Markov Chains and Stochastic Stability},
  author={Meyn, Sean P and Tweedie, Richard L},
  year={2012},
  publisher={Springer Science \& Business Media},
  doi={10.1007/978-1-4471-3267-7}
}

@article{marion2023finite,
  title={Finite-sample complexity of sequential Monte Carlo estimators},
  author={Marion, Joe and Mathews, Joseph and Schmidler, Scott C},
  journal={The Annals of Statistics},
  volume={51},
  number={3},
  pages={1357--1375},
  year={2023},
  publisher={Institute of Mathematical Statistics},
  doi={10.1214/23-aos2295}
}

@article{fay2026complexity,
  title={On the complexity of standard and waste-free SMC samplers},
  author={Fay, Yvann Le and Chopin, Nicolas and Vihola, Matti},
  journal={arXiv preprint arXiv:2604.03352},
  year={2026},
  doi={10.48550/arXiv.2604.03352}
}

@book{hennig2022probabilistic,
  title={Probabilistic Numerics: Computation as Machine Learning},
  author={Hennig, Philipp and Osborne, Michael A and Kersting, Hans P},
  year={2022},
  publisher={Cambridge University Press},
  doi={10.1017/9781316681411}
}

@article{dodwell2019multilevel,
  title={Multilevel Markov chain Monte Carlo},
  author={Dodwell, Tim J and Ketelsen, Christian and Scheichl, Robert and Teckentrup, Aretha L},
  journal={SIAM Review},
  volume={61},
  number={3},
  pages={509--545},
  year={2019},
  publisher={SIAM},
  doi={10.1137/19m126966x}
}

@article{giles2015multilevel,
  title={Multilevel {M}onte {C}arlo methods},
  author={Giles, Michael B},
  journal={Acta Numerica},
  volume={24},
  pages={259--328},
  year={2015},
  publisher={Cambridge University Press},
  doi={10.1017/S096249291500001X}
}

@book{rasmussen2006gaussian,
  title={Gaussian Processes for Machine Learning},
  author={Rasmussen, Carl Edward and Williams, Christopher KI},
  year={2006},
  publisher={MIT Press, Cambridge, MA},
  doi={10.7551/mitpress/3206.001.0001}
}

@article{stuart2018posterior,
  title={Posterior consistency for {G}aussian process approximations of {B}ayesian posterior distributions},
  author={Stuart, Andrew M and Teckentrup, Aretha L},
  journal={Mathematics of Computation},
  volume={87},
  number={310},
  pages={721--753},
  year={2018},
  doi={10.1090/mcom/3244}
}

@book{lemieux2009monte,
  title={Monte {C}arlo and quasi-{M}onte {C}arlo Sampling},
  author={Lemieux, Christiane},
  year={2009},
  publisher={Springer Science \& Business Media},
  doi={10.1007/978-0-387-78165-5}
}

@article{cox2024bayesian,
  title={A Bayesian inverse approach to proton therapy dose delivery verification},
  author={Cox, Alexander MG and Hattam, Laura and Kyprianou, Andreas E and Pryer, Tristan},
  journal={Proceedings of the Royal Society A},
  volume={480},
  number={2301},
  pages={20230836},
  year={2024},
  OPTorganization={The Royal Society},
  doi={10.1098/rspa.2023.0836}
  
}

@article{kangasniemi2024stochastic,
  title={Stochastic Gauss-Newton method for estimating absorption and scattering in optical tomography with the Monte Carlo method for light transport},
  author={Kangasniemi, Jonna and Mozumder, Meghdoot and Pulkkinen, Aki and Tarvainen, Tanja},
  journal={Biomedical Optics Express},
  volume={15},
  number={8},
  pages={4925--4942},
  year={2024},
  publisher={Optica Publishing Group},
  doi={10.1364/boe.528666}
}

@article{lartaud2025uncertainty,
  title={Uncertainty quantification in Bayesian inverse problems with neutron and gamma time correlation measurements},
  author={Lartaud, Paul and Humbert, Philippe and Garnier, Josselin},
  journal={Annals of Nuclear Energy},
  volume={213},
  pages={111123},
  year={2025},
  publisher={Elsevier},
  doi={10.1016/j.anucene.2024.111123}
}

@article{conrad2017statistical,
  title={Statistical analysis of differential equations: introducing probability measures on numerical solutions},
  author={Conrad, Patrick R and Girolami, Mark and S{\"a}rkk{\"a}, Simo and Stuart, Andrew and Zygalakis, Konstantinos},
  journal={Statistics and Computing},
  volume={27},
  number={4},
  pages={1065--1082},
  year={2017},
  publisher={Springer},
  doi={10.1007/s11222-016-9671-0}
}

@incollection{bengtsson2008curse,
  title={Curse-of-dimensionality revisited: Collapse of the particle filter in very large scale systems},
  author={Bengtsson, Thomas and Bickel, Peter and Li, Bo},
  booktitle={Probability and statistics: Essays in honor of David A. Freedman},
  volume={2},
  pages={316--334},
  year={2008},
  publisher={Institute of Mathematical Statistics},
  doi={10.1214/193940307000000518}
}

@article{lovbak2025investigation,
  title={An Investigation into the Distribution of Ratios of Particle Solver-based Likelihoods},
  author={L{\o}vbak, Emil and Krumscheid, Sebastian},
  journal={arXiv preprint arXiv:2508.05303},
  year={2025},
  doi={10.48550/arXiv.2508.05303}
}

@inproceedings{hanninen2023image,
  author    = {H{\"a}nninen, Niko and Pulkkinen, Aki and Arridge, Simon and Tarvainen, Tanja},
  title     = {Image reconstruction in quantitative photoacoustic tomography using adaptive optical Monte Carlo},
  booktitle = {Photons Plus Ultrasound: Imaging and Sensing 2023},
  volume    = {12379},
  pages     = {1237916},
  publisher = {SPIE},
  year      = {2023},
  doi       = {10.1117/12.2647252}
}

@article{fang2009monte,
  author  = {Fang, Qianqian and Boas, David A.},
  title   = {Monte Carlo simulation of photon migration in 3D turbid media accelerated by Graphics Processing Units},
  journal = {Optics Express},
  volume  = {17},
  number  = {22},
  pages   = {20178--20190},
  year    = {2009},
  doi     = {10.1364/OE.17.020178}
}

@article{wang1995mcml,
  author  = {Wang, Lihong and Jacques, Steven L. and Zheng, Liqiong},
  title   = {{MCML}---Monte Carlo modeling of light transport in multi-layered tissues},
  journal = {Computer Methods and Programs in Biomedicine},
  volume  = {47},
  number  = {2},
  pages   = {131--146},
  year    = {1995},
  doi     = {10.1016/0169-2607(95)01640-F}
}

@article{zhu2013review,
  author  = {Zhu, Caigang and Liu, Quan},
  title   = {Review of Monte Carlo modeling of light transport in tissues},
  journal = {Journal of Biomedical Optics},
  volume  = {18},
  number  = {5},
  pages   = {050902},
  year    = {2013},
  doi     = {10.1117/1.JBO.18.5.050902}
}
\newpage
\begin{appendix}\label{sec:appendix}

\section{Proofs and Auxiliary Results of Section \ref{sec:mutation}}\label{app:mutation}

We collect here the abstract product-kernel and Wasserstein results used in
Section \ref{sec:mutation}, together with the proofs of Theorems \ref{th:budget_opt} and \ref{th:pm_geom_erg}.
 The arguments rely on standard coupling and contraction
techniques.

We begin with a general result applicable to any SMC sampler. To this end, we define the following notation. Let $(\msp,\bsa{\msp},\minvar)$ be a probability space with a Polish space $\msp$ and $\bsa{\msp}$ being the corresponding Borel $\sigma$-algebra. Note that this definition of $\msp$ is more general than in Section \ref{sec:mutation}. Let $\mker:\msp\times\bsa{\msp}\rightarrow[0,1]$ be an irreducible and aperiodic Markov kernel admitting $\minvar$ as its invariant distribution. 

For a fixed $\npart\in\N$ we investigate the behavior of the product kernel $\mkerex:\mspex\times\bsa{\mspex}\rightarrow[0,1]$, representing $\npart$ SMC particles. That is, for $\xi=(x_1,\ldots,x_\npart)\in\mspex$ the measure $\mkerex(\xi, \cdot )$ is defined as the product measure $\mker(x_1,\cdot)\otimes\cdots\otimes\mker(x_\npart,\cdot)$. Denote by $\minvarex$ the product measure on $\mspex$, i.e., $\minvarex = \minvar\otimes\cdots\otimes\minvar$. To analyze the convergence of the product kernel $\mkerex(\xi, \cdot) $ to $\minvarex$ our tools of choice are Wasserstein distances, based on assumptions about the convergence of the kernel $\mker$.

Let $\varrho \colon \msp \times \msp \to [0, \infty)$ be a metric, which need not be the metric rendering $\msp$ Polish.
Given two probability measures $\mu_1, \mu_2$ on $(\msp,\bsa{\msp})$ and $p\geq 1$ we define the $p$-Wasserstein distance as 
\[
\mathcal{W}_{p,\varrho}(\mu_1,\mu_2) = \inf_{\eta \in \mathcal{C}(\mu_1, \mu_2)} \mathbb{E}_{{(X, X') \sim \eta}} [\varrho(X, X')^p]^{1/p},
\]
where $\mathcal{C}(\mu_1, \mu_2)$ is the set of all couplings between $\mu_1$ and $\mu_2$.
Similarly, if $\widetilde{\varrho}$ is a metric on the product space $\msp^\npart$ we define
\[
\mathcal{W}_{p,\widetilde{\varrho}}(\mu_1,\mu_2) = \inf_{\eta \in \mathcal{C}(\mu_1, \mu_2)} \mathbb{E}_{{(X, X') \sim \eta}} [\widetilde{\varrho}(X, X')^p]^{1/p}.
\]

The upcoming result characterizes the convergence of $\mkerex$, based on assumptions about $\mker$. The (short) proof is based on a coupling construction which works for a suitably designed metric $\widetilde{\varrho}$. 
Similar constructions have been used earlier in the literature, e.g., in \cite{eberle2016reflection}.
\begin{theorem}\label{th:product_meas_conv}
Let $p\geq 1$ and assume the metrics $\varrho, \widetilde{\varrho}$ on $\msp$ and $\msp^\npart$ satisfy $\widetilde{\varrho}(\xi, \xi')^p \leq c \sum_{i=1}^\npart\varrho(x_i, x_i')^p$ for any $\xi=(x_1, \dots, x_\npart) \in \msp^\npart$ and $\xi'= (x_1', \dots , x_\npart') \in \msp^\npart$ and some constant $c>0$.  
Then 
    \begin{enumerate}
        \item The kernel $\mkerex$ admits $\minvarex$ as its invariant distribution. \label{point:extend_invar}
        \item Assume there exists $\tau\in[0,1)$ such that for all $x\in\msp$ there exists $C = C_x>0$ satisfying
    \begin{equation}\label{eq:geom_erg_ass}
        \mathcal{W}_{p, \varrho}(\mker^k(x,\cdot),\minvar)\leq C\tau^k
    \end{equation}
    for any $k\in\N$.     Then, for any $\xi\in\mspex$ there exists $K= K_\xi>0$ such that
    \begin{equation*}
        \mathcal{W}_{p, \widetilde{\varrho}} \,(\mkerex^k(\xi,\cdot),\minvarex)\leq K\tau^k
    \end{equation*}
    for any $k\in\N$. The constant $K$ can be chosen as $K=K_\xi= c^{1/p}\sum_{j=1}^\npart C_{x_j}$, for $\xi=(x_1, \dots, x_\npart)$. 
    \end{enumerate}
\end{theorem}
\begin{proof}
The first assertion follows by an application of the Dynkin-system theorem.

Now, for the sake of brevity denote  $\mathcal{W}_{p, \varrho}(\cdot , \cdot) = \was{\cdot}{\cdot}$ and $\mathcal{W}_{p, \widetilde{\varrho}} (\cdot , \cdot) = \widetilde{\mathcal{W}}(\cdot , \cdot)$.
Let $\varepsilon>0$ and choose a coupling $\eta^\varepsilon_x$ such that 
\[
\was{\mker^k(x,\cdot)}{\minvar} + \frac{\varepsilon}{c^{1/p}\npart} \geq  \mathbb{E}_{(X,X') \sim \eta^\varepsilon_x}[\varrho(X,X')^p]^{1/p} .
\]
Define the random variable $(Z;Z') = (X_1, \dots , X_\npart \,; \, X_1', \dots , X_\npart')$, such that the pairs $(X_j, X_j')$, $j\in\{1,\ldots,\npart\}$, are independent and the $j$-th pair has joint law $(X_j, X_j')\sim\eta^\varepsilon_{x_j}$. 
Note that $(Z;Z')$ defines a coupling\footnote{This can be seen by first considering sets of the form $A_1 \times \dots \times A_\npart \times \msp^\npart$ and using that product measures are unique if all involved spaces are $\sigma$-finite; cf. \cite{dudley2002real}.} for $\mkerex^k$ and $\minvarex$.
Hence
\begin{align*}
\widetilde{\mathcal{W}}(\mkerex^k(\xi, \cdot), \minvarex) &\leq \mathbb{E}[\widetilde{\varrho}(Z,Z')^p]^{1/p} \leq c^{1/p} \sum_{i=1}^\npart \mathbb{E}[\varrho(X_i, X_i')^p]^{1/p} \\
&\leq c^{1/p} \sum_{i=1}^\npart \was{\mker^k(x_i, \cdot )}{\minvar} + \varepsilon \leq c^{1/p} \tau^k \sum_{i=1}^\npart C_{x_i} + \varepsilon.
\end{align*}
Since $\varepsilon>0$ was arbitrary, the claimed result follows. 
\end{proof}

The total variation distance is one of the most commonly used distances in the MCMC literature; recall $\norm{\mu_1 - \mu_2}{tv} = \sup_{A \in \bsa{\msp}}\vert \mu_1(A) - \mu_2(A)\vert$.
By choosing $\varrho$ and $\widetilde{\varrho}$ to be the trivial metrics\footnote{That is $\varrho(x,x')=0$ if $x=x'$ and $\varrho(x,x')=1$ if $x\neq x'$. Similarly, we define the trivial metric $\widetilde{\varrho}$ on $\mspex$.} on $\msp$ and $\mspex$,  the Wasserstein distances $\mathcal{W}_{1,\varrho}$ and $\mathcal{W}_{1,\widetilde{\varrho}}$ coincide with the total variation distances on $\msp$ and $\mspex$, respectively.
We refer e.g. to \cite{rudolf2018perturbation} for a proof of this fact. 
In this case Theorem \ref{th:product_meas_conv} implies the following.

\begin{corollary}
Assume there exists $\tau<1$ such that for any $x \in \msp$ there exists $C_x < \infty$ such that 
\[
\norm{\mker^k(x, \cdot ) - \minvar}{tv} \leq C_x \tau^k,
\]
for any $k \in \N$. Then, for any $\xi \in \mspex$ there exists $ K_\xi$ such that 
\[
\norm{\mkerex^k(\xi, \cdot) - \minvarex}{tv} \leq K_\xi \tau^k
\]
for any $k \in \N$.
The constant $K_\xi$ can be chosen as $K_\xi= \sum_{j=1}^\npart C_{x_j}$, for $\xi=(x_1, \dots, x_\npart)$.
\end{corollary}

This result is the key for understanding  SMC samplers based on invariant kernels. It says that when we apply a geometrically ergodic Markov kernel $\mker$ to individual particles, the resulting product kernel $\mkerex$ acting on $\npart$ particles inherits geometric ergodicity and converges as desired to the product measure $\minvarex$.

The upcoming auxiliary lemma is an immediate consequence of \cite{gallegos2024equivalences}, where various equivalent formulations of geometric ergodicity for Markov kernels can be found. For the sake of exposition, we focus on the setting where $\mker$ admits exponential convergence in total variation.

\begin{lemma}\label{lem:V_convergence}
Assume that there exists $\alpha<1$ such that for any $x \in \msp$ there exists $C=C_x < \infty$ such that 
\[
\norm{\mker^k(x, \cdot) - \minvar}{tv} \leq C_x \alpha^k
\]
for any $k\in \N$.
Then, there exists a $\minvar$-integrable function $V \colon \msp \to [1, \infty]$, finite for $\pi$-almost all $x \in \msp$, and constants $K < \infty$, $\tau<1$ such that $\mker$ is $V$-uniformly geometrically ergodic in the sense of \eqref{equ:V_uniform_convergence}.
\end{lemma}
We conclude the section with proofs of Theorems \ref{th:budget_opt} and \ref{th:pm_geom_erg}.

\begin{proof}[Proof of Theorem \ref{th:budget_opt}]
To simplify the notation we introduce the constant $\rr_c = K_c \rr$.
    Substituting the constraint $\nstepe = \km - \pp \nstepc$ and using the assumptions \eqref{eq:as_ratios} leads to a constrained minimization problem for 
    \begin{equation*}
        J(\nstepc) = \rr_c (\qq\tau)^{\nstepc} \tau^{\km - \pp \nstepc} + \tau^{\km - \pp \nstepc} = \tau^{\km} \Big( \rr_c (\qq \tau^{1-\pp})^{\nstepc} + \tau^{-\pp \nstepc} \Big)
    \end{equation*}
    with $\nstepc\in[0,\frac{\km}{\pp}]$. Since $J$ is continuous, it attains a minimum on the compact set $[0,\frac{\km}{\pp}]$.
    
    Differentiating $J$ with respect to $\nstepc$ yields
    \begin{equation*}
        J'(\nstepc) = \tau^{\km} \Big( \rr_c (\qq \tau^{1-\pp})^{\nstepc} \ln(\qq \tau^{1-\pp}) - \pp \tau^{-\pp \nstepc} \ln \tau \Big).
    \end{equation*}
    To find the stationary point, we solve $J'(\nstepc) = 0$ in $(0,\km/\pp)$. This is equivalent to
    \begin{equation}\label{aux:J'eq0}
        \rr_c (\qq \tau)^{\nstepc} \ln(\qq \tau^{1-\pp}) - \pp \ln \tau = 0.
    \end{equation}
    Since $\pp \ln \tau < 0$, we have that $\ln(\qq \tau^{1-\pp}) < 0$ is a necessary condition for \eqref{aux:J'eq0} to hold. Further, $\ln(\qq \tau^{1-\pp}) < 0$ is equivalent to $\qq < \tau^{-(1-\pp)}$. Assuming $\qq < \tau^{-(1-\pp)}$, we solve for $\nstepc$ to obtain the stationary point $\nstepc[s] = \frac{\ln D}{\ln(\qq \tau)}$. To verify optimality let us compute the second derivative of $J$ with respect to $\nstepc$:
    \begin{equation*}
        J''(\nstepc) = \tau^{\km} \Big( \rr_c (\qq \tau^{1-\pp})^{\nstepc} \big(\ln(\qq \tau^{1-\pp})\big)^2 + \tau^{-\pp \nstepc} (-\pp \ln \tau)^2 \Big).
    \end{equation*}
    It follows that $J''(\nstepc) > 0$ for all $\nstepc \in \R$. Consequently, the stationary point $\nstepc[s]$ is the unique global minimizer $\nstepc^\star$ of $J$ on $\R$. Substituting $\nstepc^\star$ into the budget constraint we obtain $\nstepe^\star = \km - \pp \nstepc^\star$. It remains to employ the domain constraint $\nstepc \in (0, \km/\pp)$. A straightforward calculation shows that we have $0 < \nstepc^\star < \km/\pp$ if and only if $(\qq\tau)^{\km/\pp} < D < 1$. 
    
    Let us now analyze the cases where the stationary point $\nstepc[s]$ either does not exist, i.e., $\qq \geq \tau^{-(1-\pp)}$, or falls outside $(0, \km/\pp)$, i.e., $D \leq (\qq\tau)^{\km/\pp}$ or $D\geq1$:
\begin{itemize}    
    \item In the case where $\qq \geq \tau^{-(1-\pp)}$, we have $J'(\nstepc) > 0$ for all $\nstepc \in [0, \km/\pp]$. The function $J$ is increasing on $[0, \km/\pp]$, and therefore the minimizer satisfies $\nstepc^\star = 0$ and $\nstepe^\star=\km$.

    \item Assuming $\qq < \tau^{-(1-\pp)}$ and $D \geq 1$ implies $\ln D \geq 0$. In this case, 
    $J'(\nstepc[s])=0$ and thus $\nstepc[s]\leq 0$. Since $J''>0$ on $\R$, we have $J'(\nstepc) > 0$ for all $\nstepc > 0$ and therefore again $\nstepc^\star = 0$ and $\nstepe^\star = \km$.

    \item If $\qq < \tau^{-(1-\pp)}$ and $D \leq (\qq\tau)^{\km/\pp}$, we obtain $J'(\nstepc) < 0$ for all $\nstepc < \km/\pp$, and therefore $\nstepc^\star = \km/\pp$. Substituting this into the budget constraint yields $\nstepe^\star = 0$, completing the proof. \qedhere
    \end{itemize}
\end{proof}

\begin{proof}[Proof of Theorem \ref{th:pm_geom_erg}]
We first consider PMMH. Given the current state $(u,e)\in\spextended$, draw $v$ from the density $Q(u,\cdot)$ and then draw $e'$ from $\kernoise(v,\cdot)$. The proposal $(v,e')$ is accepted with probability
\begin{equation}\label{eq:acc_pmmh}
    a\bigl((u,e),(v,e')\bigr)
    =
    \min\left\{
        1,
        \frac{\prior(v)\exp\bigl(-\tp_t\extend{\Phi}(v,e')\bigr)Q(v,u)}
        {\prior(u)\exp\bigl(-\tp_t\extend{\Phi}(u,e)\bigr)Q(u,v)}
    \right\}.
\end{equation}
Assumption~(\ref{ass:bd_potential}) and $\tp_t\in[0,1]$ imply
\[
    \exp\bigl(-\tp_t\extend{\Phi}(v,e')\bigr)\geq\exp(-\potbd),
    \qquad
    \exp\bigl(-\tp_t\extend{\Phi}(u,e)\bigr)\leq1.
\]
Together with Assumptions~(\ref{ass:bd_domain}) and~(\ref{ass:bd_proposal}), this gives
\begin{equation}\label{aux:acc_floor_pmmh}
    a\bigl((u,e),(v,e')\bigr)
    \geq
    \exp(-\potbd)\,
    \frac{\priorl}{\priorh}\,
    \frac{\propl}{\proph} \ \in (0,1].
\end{equation}
Writing out the PMMH transition kernel and discarding the non-negative term corresponding to rejected proposals, then applying \eqref{aux:acc_floor_pmmh} and $Q(u,v)\geq\propl$, we obtain
\begin{align*}
    \mkerpm_{t,\nsamp}\bigl((u,e),A\bigr)
    &\geq
    \int_\paramsp\int_\noisesp
    \mathbf{1}_A(v,e')
    a\bigl((u,e),(v,e')\bigr)
    \kernoise(v,\td e')Q(u,v)\,\td v
    \\
    &\geq
    \exp(-\potbd)\,
    \frac{\priorl}{\priorh}\,
    \frac{\propl^2}{\proph}
    \int_\paramsp\int_\noisesp
    \mathbf{1}_A(v,e')\,\kernoise(v,\td e')\,\td v =
    \minc\,\minormeas{\nsamp}(A),
\end{align*}
which proves \eqref{eq:minor_pmmh}.

We next consider MCWM. Given $u\in\paramsp$, draw $v$ from the density $Q(u,\cdot)$ and independently draw $e'\sim\kernoise(v,\cdot)$ and $e''\sim\kernoise(u,\cdot)$. The conditional acceptance probability is
\[
    a^{\mathrm{MC}}(u,v;e',e'')
    =
    \min\left\{
        1,
        \frac{\prior(v)\exp\bigl(-\tp_t\extend{\Phi}(v,e')\bigr)Q(v,u)}
        {\prior(u)\exp\bigl(-\tp_t\extend{\Phi}(u,e'')\bigr)Q(u,v)}
    \right\}.
\]
The argument leading to \eqref{aux:acc_floor_pmmh} is pointwise in $(e',e'')$ and therefore yields the same lower bound for $a^{\mathrm{MC}}(u,v;e',e'')$. Hence, for $A\in\bsa{\paramsp}$,
\begin{align*}
    \mkermc_{t,\nsamp}(u,A)
    &\geq
    \int_A Q(u,v)
    \int_\noisesp\int_\noisesp
    a^{\mathrm{MC}}(u,v;e',e'')
    \kernoise(v,\td e')\kernoise(u,\td e'')\,\td v
    \\
    &\geq
    \exp(-\potbd)\,
    \frac{\priorl}{\priorh}\,
    \frac{\propl^2}{\proph}\,
    \paramleb(A)
    =
    \minc\,\frac{\paramleb(A)}{\paramleb(\paramsp)},
\end{align*}
which proves \eqref{eq:minor_mcwm}.

Since $Q(u,\cdot)$ is a probability density, $\propl\paramleb(\paramsp)\leq1$, and therefore $\minc\in(0,1]$. Both inequalities are one-step Doeblin minorization conditions. Hence the corresponding kernels admit unique invariant probability measures and are uniformly ergodic. Moreover, \cite[Theorem~16.2.4]{meyn2012markov} gives \eqref{eq:uniform_tv_pm_mc}; recall that we use the convention $\norm{\mu_1 - \mu_2}{tv} = \sup_{A \in \bsa{\msp}}\vert \mu_1(A) - \mu_2(A)\vert$.
Since
\[
    \sup_{\abs{f}\leq1}
    \abs{P^k f(x)-\pi(f)}
    =2\norm{P^k(x,\cdot)-\pi}{tv},
\]
we obtain \eqref{equ:V_uniform_convergence} with $V\equiv1$, $K=2$, and $\tau=1-\minc$. The asserted uniformity follows directly from \eqref{eq:def_minor_const}.
\end{proof}

\section{Proofs of Section \ref{sec:numerics}}\label{app:numerics}

\begin{proof}[Proof of Proposition \ref{prop:example_complexity}]
~\medskip

\noindent
\textbf{Assumption~\eqref{eq:assump_W_ad}.}
Let $t\in\{1,\ldots,T\}$, $\nsamp>0$, $\gamma\geq1$, and $(u,e)\in\spextended$. Since $\tp_{t-1}\leq\tp_t$, $\extend{\Phi}\extendarg\geq0$, and $\Sigma(u)^{-1}$ is positive definite,
\begin{align*}
w_{t,\nsamp,\gamma\nsamp}(u,e)
&=
\exp\bigl((\tp_{t-1}-\tp_t)\extend{\Phi}\extendarg\bigr)
\frac{Z_\nsamp(u)}{Z_{\gamma\nsamp}(u)}
\exp\left(
-\frac{(\gamma-1)\nsamp}{2}
e^\top\Sigma(u)^{-1}e
\right) 
\leq
\frac{Z_\nsamp(u)}{Z_{\gamma\nsamp}(u)}.
\end{align*}
Applying the change of variables $x=\sqrt{\nsamp}\,e$ in the numerator and $x=\sqrt{\gamma\nsamp}\,e$ in the denominator gives
\begin{align*}
\frac{Z_\nsamp(u)}{Z_{\gamma\nsamp}(u)}
&=
\gamma^{m/2}
\frac{
\displaystyle
\int_{\norm{x}{}\leq\sqrt{\nsamp}R}
\exp\left(-\frac12x^\top\Sigma(u)^{-1}x\right)\td x
}{
\displaystyle
\int_{\norm{x}{}\leq\sqrt{\gamma\nsamp}R}
\exp\left(-\frac12x^\top\Sigma(u)^{-1}x\right)\td x
}
\leq
\gamma^{m/2},
\end{align*}
where the final inequality follows from the inclusion
$\{x:\norm{x}{}\leq\sqrt{\nsamp}R\}
\subseteq
\{x:\norm{x}{}\leq\sqrt{\gamma\nsamp}R\}$.
Consequently, Assumption~\eqref{eq:assump_W_ad} holds with
$W_\gamma=\gamma^{m/2}$.\medskip

\noindent
\textbf{Assumption~\eqref{eq:assump_Z_ad}.}
Let $\nsamp>0$ and $\gamma\geq1$. By continuity of $\fwopext$ and compactness of $\paramsp\times\noisesp$, there exists $K_\Phi<\infty$ such that $0\leq\extend{\Phi}(u,e)\leq K_\Phi$ for every $(u,e)\in\paramsp\times\noisesp$. Consequently, since $\tp_t\in[0,1]$, we have $\exp(-K_\Phi)\leq\exp(-\tp_t\extend{\Phi}(u,e))\leq1$.
Using that $\prior$ and $\kernoise(u,\cdot)$ are probability measures, we obtain
\[
z_{t-1,\nsamp}
=
\int_{\paramsp\times\noisesp}
\exp\bigl(-\tp_{t-1}\extend{\Phi}(u,e)\bigr)
k_\nsamp(u,e)
\prior(\td u)\noiseleb(\td e)
\leq 1
\]
and
\[
z_{t,\gamma\nsamp}
=
\int_{\paramsp\times\noisesp}
\exp\bigl(-\tp_t\extend{\Phi}(u,e)\bigr)
k_{\gamma\nsamp}(u,e)
\prior(\td u)\noiseleb(\td e)
\geq \exp(-K_\Phi).
\]
Therefore,
\[
\frac{z_{t-1,\nsamp}}{z_{t,\gamma\nsamp}}
\leq
\exp(K_\Phi),
\]
uniformly in $t$, $\nsamp$, and $T$. Hence Assumption~\eqref{eq:assump_Z_ad} holds with $Z_\gamma=\exp(K_\Phi)$.\medskip

\noindent
\textbf{Assumption~\eqref{eq:assump_CB_ad}.}
The exact posterior $\post$ and the marginal target $\margmeas{\nsamp}$ admit the unnormalized densities $\likexact(u)=\exp(-\extend{\Phi}(u,0))$ and $\likmarg{\nsamp}(u)=\int_\noisesp\exp(-\extend{\Phi}(u,e))k_\nsamp(u,e)\td e$ with respect to $\prior$, respectively.

The map $e\mapsto\exp(-\extend{\Phi}(u,e))$ is twice continuously differentiable. Since $\fwop$ is continuous and $\paramsp\times\noisesp$ is compact, its Hessian with respect to $e$ is uniformly bounded. Taylor's theorem therefore gives
\begin{equation}\label{aux:taylor}
    \exp(-\extend{\Phi}(u,e))
    =
    \exp(-\extend{\Phi}(u,0))
    -
    \exp(-\extend{\Phi}(u,0))
    \nabla_e\extend{\Phi}(u,0)^\top e
    +
    \mathcal R(u,e),
\end{equation}
where $|\mathcal R(u,e)|\leq C_R\norm{e}{}^2$ uniformly in $(u,e)\in\paramsp\times\noisesp$.

The set $\noisesp$ is symmetric and $e\mapsto k_\nsamp(u,e)$ is even. Hence $\int_\noisesp e\,k_\nsamp(u,e)\td e=0$, and using \eqref{aux:taylor} we obtain
\begin{equation*}
    \left|
        \likmarg{\nsamp}(u)-\likexact(u)
    \right|
    \leq
    C_R
    \int_\noisesp
    \norm{e}{}^2
    k_\nsamp(u,e)
    \td e.
\end{equation*}

Since $u\mapsto\Sigma(u)$ is continuous and $\Sigma(u)$ is symmetric positive definite for every $u\in\paramsp$, the maps $u\mapsto\lambda_{\min}(\Sigma(u))$ and $u\mapsto\lambda_{\max}(\Sigma(u))$ are continuous. Compactness of $\paramsp$ therefore implies that both maps attain their extrema. Since $\lambda_{\min}(\Sigma(u))>0$ for every $u\in\paramsp$, its minimum over $\paramsp$ is positive. Consequently, there exist constants $0<\lambda_-\leq\lambda_+<\infty$ such that $\lambda_-\leq\lambda_{\min}(\Sigma(u))\leq\lambda_{\max}(\Sigma(u))\leq\lambda_+$ and $\lambda_+^{-1}\leq\lambda_{\min}(\Sigma(u)^{-1})\leq\lambda_{\max}(\Sigma(u)^{-1})\leq\lambda_-^{-1}$, uniformly in $u\in\paramsp$. Applying the change of variables $x=\sqrt{\nsamp}\,e$ yields\vspace{-3ex}
\[
    \int_\noisesp
    \norm{e}{}^2
    k_\nsamp(u,e)
    \td e \ = \
    \nsamp^{-1}
    \frac{
        \displaystyle
        \int_{\norm{x}{}\leq\sqrt{\nsamp}R}
        \norm{x}{}^2
        \exp\left(
            -\tfrac12x^\top\Sigma(u)^{-1}x
        \right)
        \td x
    }{
        \displaystyle
        \int_{\norm{x}{}\leq\sqrt{\nsamp}R}
        \exp\left(
            -\tfrac12x^\top\Sigma(u)^{-1}x
        \right)
        \td x
    }.
\]
For $\nsamp\geq R^{-2}$, the denominator is bounded below, uniformly in $u$ and $\nsamp$, by
\begin{equation*}
    \int_{\norm{x}{}\leq1}
    \exp\left(
        -\frac{\norm{x}{}^2}{2\lambda_-}
    \right)
    \td x
    >0,
\end{equation*}
whereas the numerator is bounded above by
\begin{equation*}
    \int_{\R^m}
    \norm{x}{}^2
    \exp\left(
        -\frac{\norm{x}{}^2}{2\lambda_+}
    \right)
    \td x
    <\infty.
\end{equation*}
It follows that there exists $C_E<\infty$ such that
$\int_\noisesp\norm{e}{}^2k_\nsamp(u,e)\td e\leq C_E\nsamp^{-1}$ uniformly in $u\in\paramsp$ and $\nsamp\geq R^{-2}$. Thus, for some $C_L<\infty$,
\begin{equation*}
    \sup_{u\in\paramsp}
    \left|
        \likmarg{\nsamp}(u)-\likexact(u)
    \right|
    \leq
    C_L\nsamp^{-1},
    \qquad
    \nsamp\geq R^{-2}.
\end{equation*}

Let $z_\nsamp=\int_\paramsp\likmarg{\nsamp}(u)\prior(\td u)$ and
$z=\int_\paramsp\likexact(u)\prior(\td u)$. Since
$0\leq\extend{\Phi}(u,e)\leq K_\Phi$, we have
$\likmarg{\nsamp}(u)\geq\exp(-K_\Phi)$ and hence
$z_\nsamp\geq\exp(-K_\Phi)$. Therefore, for every $f\in\mathcal F_b$, using
\[
    \frac{\likmarg{\nsamp}}{z_\nsamp}-\frac{\likexact}{z}
    =
    \frac{\likmarg{\nsamp}-\likexact}{z_\nsamp}
    +
    \likexact\left(\frac{1}{z_\nsamp}-\frac{1}{z}\right),
\]
we obtain
\begin{align*}
    \left|
        \int_\paramsp f(u)\margmeas{\nsamp}(\td u)
        -
        \int_\paramsp f(u)\post(\td u)
    \right|
    &\leq
    \frac{1}{z_\nsamp}
    \int_\paramsp
    \left|\likmarg{\nsamp}(u)-\likexact(u)\right|
    \prior(\td u)
    +
    \frac{|z-z_\nsamp|}{z_\nsamp} \\
    &\leq
    \frac{2}{z_\nsamp}
    \int_\paramsp
    \left|\likmarg{\nsamp}(u)-\likexact(u)\right|
    \prior(\td u)
    \leq
    2\exp(K_\Phi)C_L\nsamp^{-1},
\end{align*}
where we used
$|z-z_\nsamp|
\leq
\int_\paramsp
|\likmarg{\nsamp}(u)-\likexact(u)|
\prior(\td u)$.
Hence Assumption~\eqref{eq:assump_CB_ad} holds with $C_{\mathrm B}=2\exp(K_\Phi)C_L$ and any $\nsamp_{\mathrm B}\geq R^{-2}$.
\end{proof}
\end{appendix}

\end{document}